\documentclass[prb,aps,twocolumn,superscriptaddress,floatfix,citeautoscript,longbibliography]{revtex4-2}
\usepackage{graphicx,rotating,subfigure,amsmath,amsfonts,amssymb,delarray,color,hyperref}
\usepackage{dsfont}
\usepackage{bbold}
\usepackage[T1]{fontenc}
\usepackage{multirow}
\usepackage{comment}
\renewcommand{\vec}[1]{\boldsymbol #1}
\newcommand{\e}{\text{e}}

\newcommand{\tr}{\text{tr}}

\def\12{\frac{1}{2}}

\usepackage{placeins}
\usepackage{mciteplus}
\usepackage{lineno}

\begin{document}

\title{Thermodynamics based on Neural Networks}

\author{Dennis Wagner}
\affiliation{Department of Physics and Astronomy, University of Manitoba, Winnipeg R3T 2N2, Canada}
\affiliation{Department of Physics, University of Wuppertal, Gaussstraße 20, 42119 Wuppertal, Germany}

\author{Andreas Kl\"umper}
\affiliation{Department of Physics, University of Wuppertal, Gaussstraße 20, 42119 Wuppertal, Germany}

\author{Jesko Sirker}
\email{sirker@physics.umanitoba.ca}
\affiliation{Department of Physics and Astronomy, University of Manitoba, Winnipeg R3T 2N2, Canada}
\affiliation{Manitoba Quantum Institute, University of Manitoba, Winnipeg R3T 2N2, Canada}

\date{\today}

\begin{abstract}
We present three different neural network (NN) algorithms to calculate thermodynamic properties as well as dynamic correlation functions at finite temperatures for quantum lattice models. The first method is based on purification, which allows for the exact calculation of the operator trace. The second one is based on a sampling of the trace using minimally entangled states, whereas the third one makes use of quantum typicality. In the latter case, we approximate a typical infinite-temperature state by wave functions which are given by a product of a projected pair and a NN part and evolve this typical state in imaginary time.  

\end{abstract}

\maketitle
\section{Introduction}
While being able to obtain precise results for the ground states of quantum lattice Hamiltonians is essential to understand different states of matter and quantum phase transitions between such states, phase transitions can also occur at finite temperatures $T$ for dimensions $d\geq 2$. Furthermore, experiments are always performed at $T>0$. The influence of thermal fluctuations on the physical properties of a quantum system are therefore also of fundamental interest. Experiments on solid state systems, for example, often probe thermodynamic quantities such as the specific heat, the susceptibility, or the compressibility. Also of interest are linear response functions at finite temperatures such as the spin structure factor, measured in neutron scattering experiments, or the spin-lattice relaxation rate, measured in nuclear magnetic resonance. \\
Except for exactly solvable quantum models such as Gaussian fermionic and bosonic models and Bethe ansatz integrable models in one dimension---which share the common property of having an infinite set of local conservation laws---approximate analytical or numerical methods are needed to study the thermodynamic properties of a quantum system. For a closed system, we are usually interested in calculating the partition function $Z$ which gives access to the free energy $f$ as well as in expectation values of operators $A$. These quantities are obtained in the canonical ensemble as
\begin{equation}
    \label{Z}
Z = \tr\, \e^{-\beta H}, \; f = -\frac{T}{N}\ln Z ,  \; \langle A\rangle =\frac{\tr \{A\e^{-\beta H}\}}{Z} \ ,
\end{equation}
where $T=1/\beta$ denotes the temperature, $H$ is the Hamiltonian, $N$ the number of lattice sites, and we set $k_B=1$. Important numerical methods to obtain approximations of the partition function and expectation values include various Quantum Monte Carlo (QMC) techniques \cite{GubernatisQMC,SandvikSSE}. While QMC can be applied to problems in any dimension, it suffers from a sign problem in the sampling weights for fermionic models and frustrated spin models \cite{Nightingale1999QuantumMC,becca_sorella_2017}. In such cases, it is usually not possible to reach low temperatures. Furthermore, dynamical quantities at finite temperatures can only be obtained in imaginary time (Matsubara frequency) with the analytical continuation of the discrete numerical data to real times being an ill-posed problem with no unique solution \cite{SilverSivia}. An alternative approach to calculate the thermodynamic properties of quantum systems are matrix product state/density-matrix renormalization group (MPS/DMRG) algorithms \cite{Schollwock_review, Schollwock_MPS_review}. This approach was originally invented to obtain the ground state properties of one-dimensional quantum systems \cite{Schollwock_review}. Two different ideas were then pursued to extend such algorithms to finite temperatures: On the one hand, one can map any one-dimensional quantum lattice model at finite temperatures to a two-dimensional classical model. 
For the classical model, one can then define a so-called quantum transfer matrix (QTM) which evolves along the spatial direction and whose largest eigenvalue and corresponding eigenstate completely determine the properties of the system in the {\it thermodynamic limit}. The QTM can then be approximated as a matrix product operator \cite{WangXiang,BursillXiang,SirkerKluemperPRB,SirkerKluemperEPL}. In this approach, it is also possible to calculate dynamical correlation functions directly in real time \cite{SirkerKluemperDTMRG,EnssSirker}. \\
An alternative approach is based on introducing an ancilla site for each real site of the lattice and to obtain the partition function by purification \cite{FeiguinWhite}. Geometrically, this corresponds to replacing the cylinder which represents the partition function in the two-dimensional classical model by two sheets (formed by the real sites and ancilla sites) connected at the two edges by a state which is a direct product of maximally entangled local states between a site and its corresponding ancilla. Like the QTM algorithm, one can also include a real-time evolution and thus access dynamical correlations at finite temperatures \cite{SirkerKluemperDTMRG,KarraschMoore,KarraschBardarson,BabenkoGohmann,GohmannKozlowski}. While fermionic and frustrated spin models in one dimension do not pose any difficulties for MPS-type algorithms in contrast with QMC, they can only be applied to small clusters in dimensions higher than one in which case calculations are usually restricted to the ground state \cite{WhiteScalapinoStripes1,Zheng1155}.
\\
Recently, variational QMC approaches, where the parameters of the variational wave function are optimized using a neural network (NN), have attracted renewed attention \cite{TroyerScience,CarleoCirac,CarleoNomura,SchmittHeyl} as a numerical approach possibly able to overcome some of these issues. Variational QMC approaches have, at least in principle, the advantage that they are applicable in any dimension, that sign problems can be circumvented, and that real-time dynamics can be studied directly. However, the choice of the variational wave function biases the obtained results. It is expected that by using ansatz wave functions based on NNs with many variational parameters, the system can learn by itself which parameters are most important instead of fixing the form of the wave function by physical intuition. Mathematically, the basis for this approach are universal approximation theorems which guarantee that any smooth real function can be approximated to any desired precision by a NN with a non-polynomial activation function (for example, the sigmoid function) \cite{Cybenko,Hornik}. Such approximation theorems can also be generalized to complex-valued NNs \cite{Voigtlaender} which are needed to approximate complex wave functions. In practice, some form of gradient descent in the parameter space of the variational wave function is then used to optimize the ground state energy. The variational principle guarantees that the obtained energy is always larger than or equal to the true ground state energy. This approach has been applied successfully to investigate the ground state properties of a number of different low-dimensional quantum lattice models. However, for fermionic and frustrated spin systems where a Marshall sign rule \cite{LiebSchultzMattis} does not exist, learning the sign structure has remained challenging at least in ansatz wave functions which are explicitly translationally invariant \cite{SzaboCastelnovo,PP-NN1,PP-NN2}. \\
In this paper, we want to expand such NN approaches to allow for the calculation of static and dynamic finite-temperature properties. We will discuss three possible approaches: Like MPS/DMRG algorithms, we can introduce an additional ancilla site for each real lattice site and obtain traces via purification. The second approach is based on sampling the trace using minimally entangled states \cite{deRaedt}. The third approach is based on the notion of quantum typicality. Properly implemented, a {\it single pure state} can reproduce the expectation values of observables in the Gibbs ensemble for large system sizes \cite{Gemmer2009,SteinigewegGemmer,Sugiura}. We note that the first approach was already considered for deep Boltzmann machines in Refs.~\cite{nys2023realtime,PurifyingDBM}. However, the method to obtain the initial, infinite temperature state is different from our approach. We will discuss this point in more detail below. The second approach has also very recently been studied in the NN context in Ref.~\cite{RBM-pure}. One of the main goals of our paper is to compare these three different approaches instead of focusing on only one of them.

In all cases, we start out by constructing an appropriate infinite-temperature state $|\Psi(\infty)\rangle$ and then adjust the parameters of the NN by solving the Schr\"odinger equation for the imaginary-time evolution
\begin{equation}
    \label{Schrodinger}
    |\Psi(\beta)\rangle = \exp \left(-\frac{\beta H}{2} \right) |\Psi(\infty)\rangle \, . 
\end{equation}
We note that the task here is technically more demanding than in the ground state case. While any gradient descent through parameter space which is successively lowering the energy of the system is acceptable if we are only interested in obtaining an approximation for the ground state, we want to follow here as precisely as possible the thermodynamic path so that the state describes the physical properties of the system at any given temperature $T=1/\beta$. \\
Our paper is organized as follows: In Sec.~\ref{sec.Purification}, we present the NN algorithm based on purification and test the algorithm by calculating thermodynamic and dynamic properties at finite temperatures for the Heisenberg model. In Sec~.\ref{sec.Sampling}, we instead sample the trace using minimally entangled states. Thermodynamic properties are computed by evolving these states in imaginary time. In Sec.~\ref{sec.Quantum Typicality}, we then introduce, as a third alternative, an algorithm based on quantum typicality and compare the obtained results with those obtained by purification and the sampling of the trace. The last section is devoted to a short summary and a critical discussion of the power of NN algorithms for thermodynamic simulations and of some of the remaining open issues.

\section{Purification}
\label{sec.Purification}
To exactly calculate the partition function $Z$ as well as thermodynamic expectation values, a trace over all basis states must be performed. Since we do not want to diagonalize the Hamiltonian itself but rather express wave functions in the Hilbert space by a variational ansatz, we cannot perform the trace directly. Let us assume that $|\{s\}\rangle = |\{s_1^z...s_N^z\}\rangle$ represents the local $s^z$-basis for a quantum lattice model with $N$ sites. Then, one way to obtain thermodynamic quantities solely based on wave functions and the action of operators on these wave functions is to introduce additional ancilla degrees of freedom ${s'}$ by 
\begin{equation}
    \label{ancillas}
  |\{s\}\rangle \to |\{s\}\rangle \otimes |\{s'\}\rangle \equiv |\{s,s'\}\rangle \ .  
\end{equation}
 While the Hamiltonian will act only on the real lattice sites, the state of the ancillas remains unchanged. We can therefore think of the ancillas as an \lq index\rq \ keeping track of the initial state of the real sites before we have acted on this state with some operator. We will consider the initial state
\begin{equation}
    \label{ancillas2}
 |\Psi(\infty)\rangle = \bigotimes_{i=1}^N \sum_{s=-s'} |s,s'\rangle_i \ ,
\end{equation}
where $N$ is the number of lattice sites. $|\Psi(\infty)\rangle$ represents the (not normalized) infinite-temperature state and an expectation value or a dynamical correlation function can be obtained by 
\begin{eqnarray}
    \label{ancillas3}
 &&\langle\hat O_1(0) \hat O_2 (t)\rangle_{T=\infty} \\
 &=& \frac{\langle\Psi(\infty)|(\hat O_1(0) \otimes \mathbb{1})(\hat O_2(t) \otimes \mathbb{1}) |\Psi(\infty)\rangle}{\langle\Psi(\infty)|\Psi(\infty)\rangle}  \ , \nonumber
\end{eqnarray}
with $\langle\Psi(\infty)|\Psi(\infty)\rangle =d^N$ where $d$ is the number of local degrees of freedom. A finite temperature state is then obtained by imaginary time evolution
\begin{equation}
    \label{finiteT_state}
|\Psi(\beta)\rangle = \left[ \exp \left(-\frac{\beta H}{2} \right) \otimes \mathbb{1} \right] |\Psi(\infty)\rangle    
\end{equation}
and the partition function is given by $Z(\beta)=\langle\Psi(\beta)|\Psi(\beta)\rangle$. Note that the two sheets---consisting of the real spins and the ancilla spins, respectively---become connected when performing a partial trace over the ancillas and thus together form a cylindrical geometry. In the numerical calculations, we therefore find it advantageous to move half of the imaginary-time evolution to the ancillary spins and to apply $\exp(-\beta H/4)$ on both sheets. This makes the problem more symmetric which is helpful for an efficient variational calculation. Similarly, the energy per site is then obtained in a more symmetric manner as  
\begin{equation}
    \label{inner}
    e = \frac{1}{2N}\frac{\langle\Psi(\beta)| (H\otimes\mathbb{1} + \mathbb{1}\otimes H) |\Psi(\beta)\rangle}{\langle\Psi(\beta)| \Psi(\beta)\rangle} \, .
\end{equation}
The description of the purification algorithm so far is general and can be performed with any variational ansatz wave function, provided that the initial infinite-temperature state \eqref{ancillas2} can be represented accurately. As a proof of principle, we consider in the following a simple implementation using a modified restricted Boltzmann machine (mRBM). We note that the performance of the algorithm discussed below might possibly be improved by using deep  NNs (DNNs) as, for example, convolutional NNs (CNNs), along similar lines as in ground state algorithms \cite{SchmittHeyl}. Here the focus is on demonstrating how thermodynamic calculations can be performed in principle and what the fundamental limitations are. For convenience, we label both the real and the ancilla spins as $s_i$ with  $i=1,\cdots, N$ for the real spins and $i=N+1,\cdots,2N$ for the ancillas. For the components of the wave function, we then choose the following ansatz
\begin{eqnarray}
    \label{ancilla_ansatz}
&& \Psi(s)=\langle s | \Psi\rangle \nonumber \\
&=& \exp\left(-\sum_{i=1}^{2N} a_i s_i\right)  \prod_{j=1}^M \cosh\left(b_j+\sum_{i=1}^{2N} W_{ij}s_i\right) \nonumber \\
&\times& \exp\left(-\sum_{i=1}^{2N} c_i (s_i + s_{(i+N) \textrm{mod}(2N)})^2\right)  \ .
\end{eqnarray}
Here $\vec{a},\vec{b}$ and $\vec{c}$ are complex vectors with $2N$ and $M$ components, respectively, while $\mathbf{W}$ is a complex matrix of dimension $2N\times M$. The second line in Eq.~\eqref{ancilla_ansatz} is the standard RBM ansatz for a system with $2N$ sites \cite{TroyerScience}. In NN theory, the RBM is classified as a single-layer feed-forward neural network and the imaginary-time evolution can be understood as a reinforcement learning algorithm \cite{CarleoCirac}. We modify this ansatz by multiplying with the term in the last line, which couples every spin with its respective ancilla. The (not normalized) infinite-temperature state $|\Psi(\infty)\rangle$ can now formally be obtained by setting $\vec{a},\vec{b},\vec{W}=0$ and $c_i\to +\infty$, $i=1,\cdots,2N$. This way, the term in the last line exponentially suppresses all components where real site and ancilla site do not have opposite signs and gives all components where the signs are opposite equal weight in the wave function. In practice, however, this limit cannot be performed exactly because numerical gradient methods---required to change the parameters of the wave function during the imaginary-time evolution---then can not find the proper descent path. We therefore set $c_i = c =\mbox{const}>0$ and choose the real and imaginary parts of $\vec{a}$ and $\vec{W}$ from a random normal distribution centered around zero with a very small width $w\sim 10^{-2}$ and keep $\vec{b}=0$. We note that in Refs.~\cite{nys2023realtime,PurifyingDBM} purification was used as well but without introducing the term in the last line of Eq.~\eqref{ancilla_ansatz}. Instead, the parameters $\mathbf{W}$ were chosen appropriately to realize the infinite-temperature state. We note that in our approach the suppression of unwanted contributions to the wave function is exponential while realizing the infinite-temperature state using the $\mathbf{W}$ parameters results in a suppression which is only quadratic if some noise is added to these parameters. In our approach, noise can be added to the $\mathbf{W}$ parameters while keeping the unwanted contributions exponentially suppressed.

To time evolve this approximation of the infinite-temperature state, we use the time-dependent variational principle (TDVP) which is a simple consequence of Eq.~\eqref{finiteT_state}. Considering that the temperature is encoded in the variational parameters $\mathbb{W}=\mathbb{W}[\beta]\equiv\{\vec{c}[\beta],\vec{a}[\beta],\vec{b}[\beta],\vec{W}[\beta]\}$ we obtain the Schrödinger equation
\begin{equation}
    \label{var_principle}
\frac{d}{d\beta}|\Psi_{\mathbb{W}}[\beta]\rangle = \sum_\alpha \frac{\partial \mathbb{W}_\alpha}{\partial\beta}|\partial_{\mathbb{W}_\alpha}\Psi_{\mathbb{W}}[\beta]\rangle 
= - P_{\mathbb{W}} \frac{H}{2} |\Psi_{\mathbb{W}}[\beta]\rangle \ ,
\end{equation}
where $P_{\mathbb{W}}$ is the projector onto the variational manifold. We will ignore this projection for now. We will return to this point and discuss the differences between the full time evolution and the one in the variational manifold in detail in Sec.~\ref{subsec.Beta Modification}.
We can now multiply with the vectors $\langle \partial_{\mathbb{W}_{\alpha'}}\Psi|$ from the left to obtain
\begin{equation}
    \label{var_principle2}
    \sum_\alpha \underbrace{\frac{\langle \partial_{\mathbb{W}_{\alpha'}}\Psi|\partial_{\mathbb{W}_\alpha}\Psi\rangle}{\langle\Psi|\Psi\rangle}}_{\vec{G}} \frac{\partial \mathbb{W}_\alpha}{\partial\beta} = -\frac{1}{2}\underbrace{\frac{\langle \partial_{\mathbb{W}_{\alpha'}}\Psi|H |\Psi\rangle}{\langle\Psi|\Psi\rangle}}_{\vec{H}} \ ,
\end{equation}
where we have defined the matrix $\vec{G}$ and the vector $\vec{H}$. Since the parameters $\mathbb{W}_\alpha$ are not necessarily linearly independent, the matrix $\vec{G}$ is, in general, not regular. We can, however, always compute the Moore-Penrose pseudoinverse which we denote by $\vec{G}^{-1}$ if we introduce a small cutoff. Throughout this paper, the cutoff is set to be between $10^{-8}$ and $10^{-12}$. We have tested that the differences between time evolutions using different cutoffs in this range are orders of magnitude smaller than the precision of this variational method except for very low temperatures $T$, see also Sec.~\ref{subsec.Numerical results for system sizes beyond exact diagonalizations}. This leads to the following matrix equation describing the time evolution of the state $|\Psi(\mathbb{W}[\beta]\rangle$ via a change of the variational parameters $\mathbb{W}$
\begin{equation}
    \label{var_principle3}
    \frac{\partial\mathbb{W}}{\partial\beta} = -\frac{1}{2}\vec{G}^{-1}\vec{H} \Rightarrow \mathbb{W}[\beta+\Delta\beta] \approx \mathbb{W}[\beta] -\frac{1}{2}\Delta\beta\vec{G}^{-1}\vec{H} ,
\end{equation}
where $\Delta\beta$ is the imaginary-time step (learning rate) and on the right-hand side of Eq.~\eqref{var_principle3} we have written, for simplicity, the approximate solution of the differential equation in the form of a Newton forward method. In the numerical calculations presented below, we will use instead the Heun method which is more stable, particularly if the states are Monte Carlo sampled. The elements of the matrix $\vec{G}$ and the vector $\vec{H}$ can be expressed in terms of the probability distribution $|\Psi(s)|^2$, for example,
\begin{equation}
    \label{var_principle4}
    \frac{\langle \partial_{\mathbb{W}_{\alpha'}}\Psi|\partial_{\mathbb{W}_\alpha}\Psi\rangle}{\langle\Psi|\Psi\rangle} = \frac{\sum_s |\Psi(s)|^2 \left(\frac{\partial_{\mathbb{W}_{\alpha'}}\Psi(s)}{\Psi(s)} \right)^* \frac{\partial_{\mathbb{W}_{\alpha}}\Psi(s)}{\Psi(s)}}{\sum_s |\Psi(s)|^2} \ .
\end{equation}

\subsection{Adaptive Heun Method}
\label{subsec.Adaptive Heun Method}
The adaptive Heun method is used throughout this paper for the imaginary-time evolution described in Eqs.~\eqref{finiteT_state} and \eqref{var_principle3}. In the following, we will use $\tau$ to denote the imaginary time instead of $\beta$. The reason for this change in notation will become clear in Sec.~\ref{subsec.Beta Modification}. The adaptive Heun method is an iterative method where in every time step the error of the integration is estimated and the length of the time step $\Delta \tau$ adjusted in a way that the error stays constant. We describe this predictor-corrector method in the following for a second-order integration for which the standard notation for the derivative $\dot{\mathbb{W}} = f(\mathbb{W})$ and the exact solution $\mathbb{W}[\tau+\Delta \tau]$ at time $\tau+\Delta \tau$, where $\mathbb{W}$ is a $P$-component vector and $P$ the number of parameters, is used \cite{SchmittHeyl}. We define
\begin{align}
    k_1 &= f(\mathbb{W}_n) \nonumber \\
    k_2 &= f(\mathbb{W}_n + \Delta \tau k_1) \nonumber \\
    \mathbb{W}_{n+1} &= \mathbb{W}_n + \frac{\Delta \tau}{2} \left(k_1+k_2 \right) ,
\end{align}
which leads to a cubic error in the integration step 
\begin{equation}
    \mathbb{W}_{n+1} = \mathbb{W}[\tau+\Delta \tau] + \alpha (\Delta \tau)^3,
\end{equation}
where $\alpha$ is a constant vector. This is now compared to the error for two time steps of length $\frac{\Delta \tau}{2}$
\begin{equation}
    \mathbb{W}'_{n+1} = \mathbb{W}[\tau+\Delta \tau] + \underbrace{2\alpha \left(\frac{\Delta \tau}{2} \right)^3}_{\delta},
\end{equation}
where $\delta$ is the integration error. The difference of these two results
\begin{equation}
    ||\mathbb{W}_{n+1}-\mathbb{W}'_{n+1}|| = \frac{3}{4} ||\alpha|| (\Delta \tau)^3 \equiv 3\delta
\label{eq.dist}
\end{equation}
allows to adjust the time step
\begin{equation}
    \Delta \tau' = \Delta \tau \left( \frac{\varepsilon}{\delta} \right)^{\frac{1}{3}}
\end{equation}
to obtain a desired accuracy $\varepsilon$. The norm in Eq.~\eqref{eq.dist} is defined by $ ||x|| = \frac{1}{P} \sqrt{x^\dagger \mathbf{G} x}$. Note that the tolerance $\varepsilon$ and the initial step size $\Delta \tau$ are in principle parameters which need to be specified at the beginning of the calculation. The adjusted step size $\Delta \tau'$ is always used as the proposed step size $\Delta \tau$ in the next step. By applying a pretempering at $\tau=0$ of a few steps in which parameters are not changed, $\Delta \tau'$ converges to a value that is used as the initial step size. Thus, the only remaining parameter to be fixed is the desired tolerance $\varepsilon$ which defines the accuracy of the method. In the following, we will denote the number of iterative steps in the Heun algorithm by $R$.

\subsection{\boldmath \texorpdfstring{$\beta$}{} Modification}
\label{subsec.Beta Modification}
The variational imaginary-time evolution in Eq.~\eqref{var_principle} needs a more thorough discussion that considers the projection onto the variational manifold $\mathcal{M}$, which is a submanifold of the projective Hilbert space. We start from the projected Schrödinger equation by using the projector $\mathtt{P}_{\Psi_{\mathbb{W}}[\tau]} $ which restricts the change of the variational state to be in the tangent space $\mathcal{T}_{\Psi_{\mathbb{W}}[\tau]} \mathcal{M}$ of the variational manifold $\mathcal{M}$ \cite{Geometry,RBM-pure}:
\begin{align}
\label{eq.ProjectedImaginarySE}
\frac{d |\Psi_{\mathbb{W}}[\tau] \rangle }{d\tau}
=& -\frac{1}{2} \mathtt{P}_{\Psi_{\mathbb{W}}[\tau]} H |\Psi_{\mathbb{W}}[\tau]\rangle \\
=& -\frac{1}{2} \mathtt{P}_{\Psi_{\mathbb{W}}[\tau]}(H-E[\tau]) |\Psi_{\mathbb{W}}[\tau]\rangle  \nonumber 
\end{align}
This is accomplished by using the orthogonal projector
\begin{equation}
\mathtt{P}_{\Psi_{\mathbb{W}}[\tau]} = \frac{(\vec{G}^{-1})^{^{\mu \eta}}}{\langle \Psi_{\mathbb{W}}[\tau] | \Psi_{\mathbb{W}}[\tau] \rangle} |V_{\mathbb{W},\mu}[\tau] \rangle \langle V_{\mathbb{W}, \eta}[\tau]| \ ,
\end{equation}
where 
\begin{equation}
|V_{\mathbb{W}, \mu}[\tau]\rangle =  \mathtt{Q}_{\Psi_{\mathbb{W}}[\tau]} \partial_{\mathbb{W}_{\mu}} |\Psi_{\mathbb{W}} [\tau]\rangle
\label{eq.V}
\end{equation}
is a local basis of the projected tangent space $\mathcal{T}_{\Psi_{\mathbb{W}}[\tau]} \mathcal{M}$ at the state $|\Psi_{\mathbb{W}}[\tau]\rangle$, $\mathtt{Q}_{\Psi_{\mathbb{W}}[\tau]} = \left(\mathds{1}- \frac{|\Psi_{\mathbb{W}}[\tau]\rangle \langle \Psi_{\mathbb{W}}[\tau]|}{ \langle \Psi_{\mathbb{W}}[\tau]|\Psi_{\mathbb{W}}[\tau]\rangle} \right)$ and
\begin{equation}
\vec{G}= \frac{ \langle V_{\mathbb{W},\mu}[\tau] | V_{\mathbb{W}, \eta}[\tau] \rangle}{\langle \Psi_{\mathbb{W}}[\tau] | \Psi_{\mathbb{W}}[\tau] \rangle} 
\end{equation}
the metric tensor, which was already obtained in Eq.~\eqref{var_principle2}. Due to the projection, the time traveled in the manifold and the imaginary time are not equal. This is quantified by the following ratio \cite{RBM-pure}
\begin{align}
\begin{split}
\gamma =& \frac{||\mathtt{P}_{\Psi_{\mathbb{W}}[\tau]} (H-E[\tau]) |\Psi_{\mathbb{W}}[\tau]\rangle||^2}{||(H-E[\tau]) |\Psi_{\mathbb{W}}[\tau] \rangle||^2} \\
=& -\frac{1}{\sigma^2[\tau]} \left(\frac{d E[\tau]}{d \tau} \right) \ ,
\end{split}
\label{eq.gamma}
\end{align}
where the two quantities of interest on the right hand side are easily calculated at every time step
\begin{align}
E[\tau] =& \frac{\langle \Psi_{\mathbb{W}}[\tau]| H |\Psi_{\mathbb{W}}[\tau]\rangle }{\langle\Psi_{\mathbb{W}}[\tau] |  \Psi_{\mathbb{W}}[\tau]\rangle}
\\
\sigma^2[\tau] =& \frac{\langle \Psi_{\mathbb{W}}[\tau]| (H-E[\tau])^2 |\Psi_{\mathbb{W}}[\tau]\rangle}{\langle\Psi_{\mathbb{W}}[\tau] |  \Psi_{\mathbb{W}}[\tau]\rangle} \ .
\end{align}

For greater clarity, the action of the projector in Eq.~\eqref{eq.gamma} is calculated step by step
\begin{align}
\begin{split}
&\frac{dE[\tau]}{d\tau} =  \frac{ \langle\Psi_{\mathbb{W}}[\tau]| H \left( \frac{d}{d\tau} | \Psi_{\mathbb{W}}[\tau] \rangle \right)}{\langle\Psi_{\mathbb{W}}[\tau] |  \Psi_{\mathbb{W}}[\tau]\rangle} + h.c. 
\\
=&- \frac{1}{2}\left[ \frac{ \langle\Psi_{\mathbb{W}}[\tau]| H \left(  \mathtt{P}_{\Psi_{\mathbb{W}}[\tau]}(H-E[\tau]) | \Psi_{\mathbb{W}}[\tau] \rangle \right)}{\langle\Psi_{\mathbb{W}}[\tau] |  \Psi_{\mathbb{W}}[\tau]\rangle} + h.c. \right]
\\
=& -\frac{ \langle\Psi_{\mathbb{W}}[\tau]| (H-E[\tau]) \mathtt{P}_{\Psi_{\mathbb{W}}[\tau]}^2(H-E[\tau]) | \Psi_{\mathbb{W}}[\tau] \rangle}{\langle\Psi_{\mathbb{W}}[\tau] |  \Psi_{\mathbb{W}}[\tau]\rangle} \ .
\end{split}
\end{align}
In the final line, we employed Eq.~\eqref{eq.ProjectedImaginarySE} and leveraged a property of projection operators, namely, \mbox{$\mathtt{P}_{\Psi_{\mathbb{W}}[\tau]}^2= \mathtt{P}_{\Psi_{\mathbb{W}}[\tau]}$}. The imaginary time $\tau$ is therefore adjusted and the modified inverse temperature $\beta[\tau]$ computed  by integrating Eq.~\eqref{eq.gamma} 
\begin{equation}
\gamma = \frac{d\beta}{d\tau} \quad \Rightarrow \quad \beta[\tau] = - \int_0^{\tau} \frac{1}{\sigma^2} \frac{d E}{d \tau'} \Big|_{\mathbb{W}[\tau']} d\tau' \ .
\label{eq.Gamma}
\end{equation}
We emphasize that the temperature adjustment does not fully correct the errors due to the restriction of the time evolution to the variational manifold $\mathcal{M}$. The proper state in the full Hilbert space is not known and, in general, different from the variational state.

\subsection{Results for Small Systems}
\begin{figure}[!htp]
    \centering
    \includegraphics*[width=0.99\columnwidth]{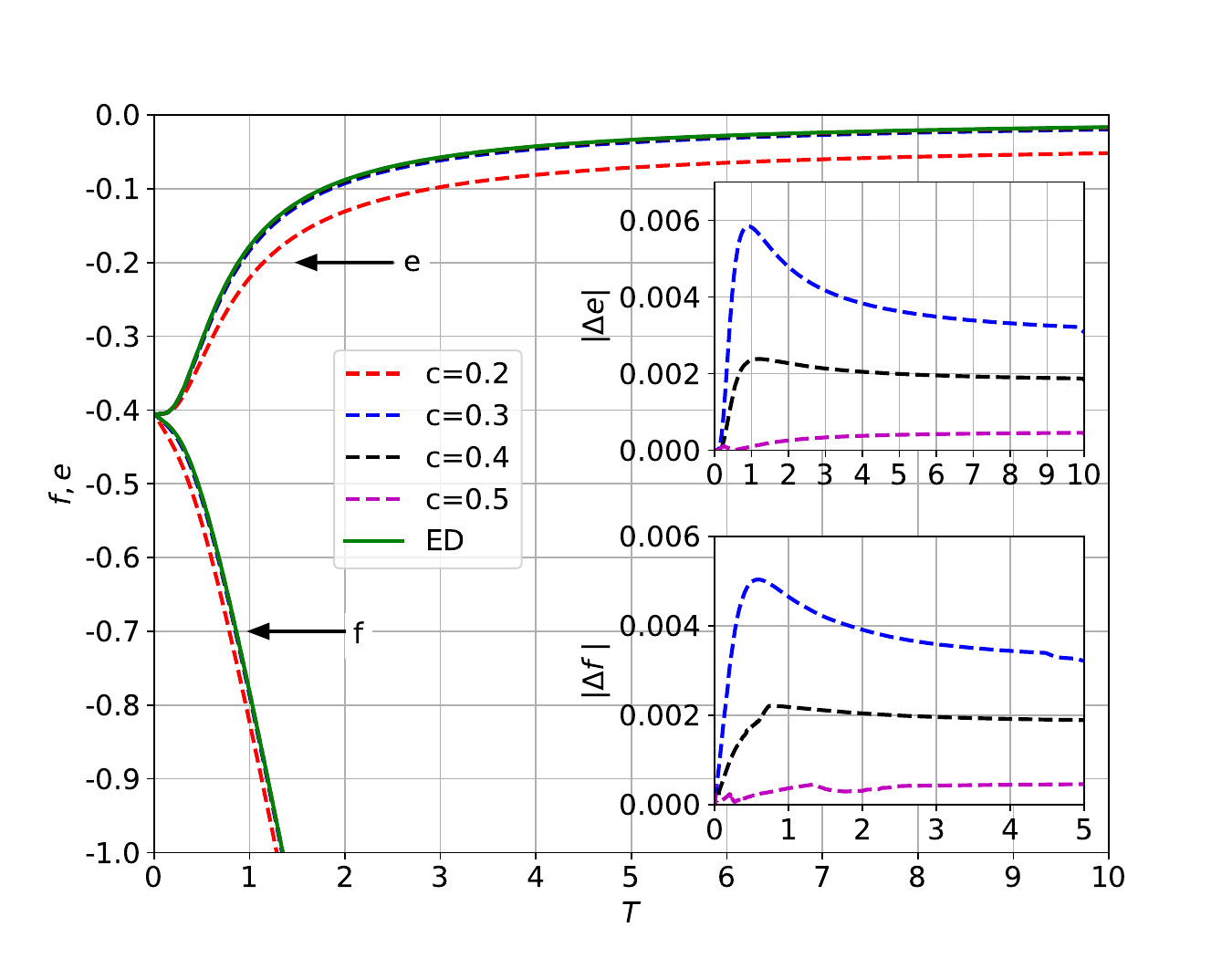}
    \caption{Inner and free energy for a Heisenberg chain with $N=7$ sites, a desired tolerance of $\epsilon = 10^{-9}$ and  \mbox{$R=2\cdot 10^3$} iterative steps obtained using the purification algorithm without the $\beta$ modification. In the mRBM ansatz, $M=28$ hidden units are used. The insets show the absolute errors for $c=0.3,0.4,0.5$.}
    \label{fig.Energy-7}
\end{figure}

\begin{figure}[!htp]
    \centering
    \includegraphics*[width=0.99\columnwidth]{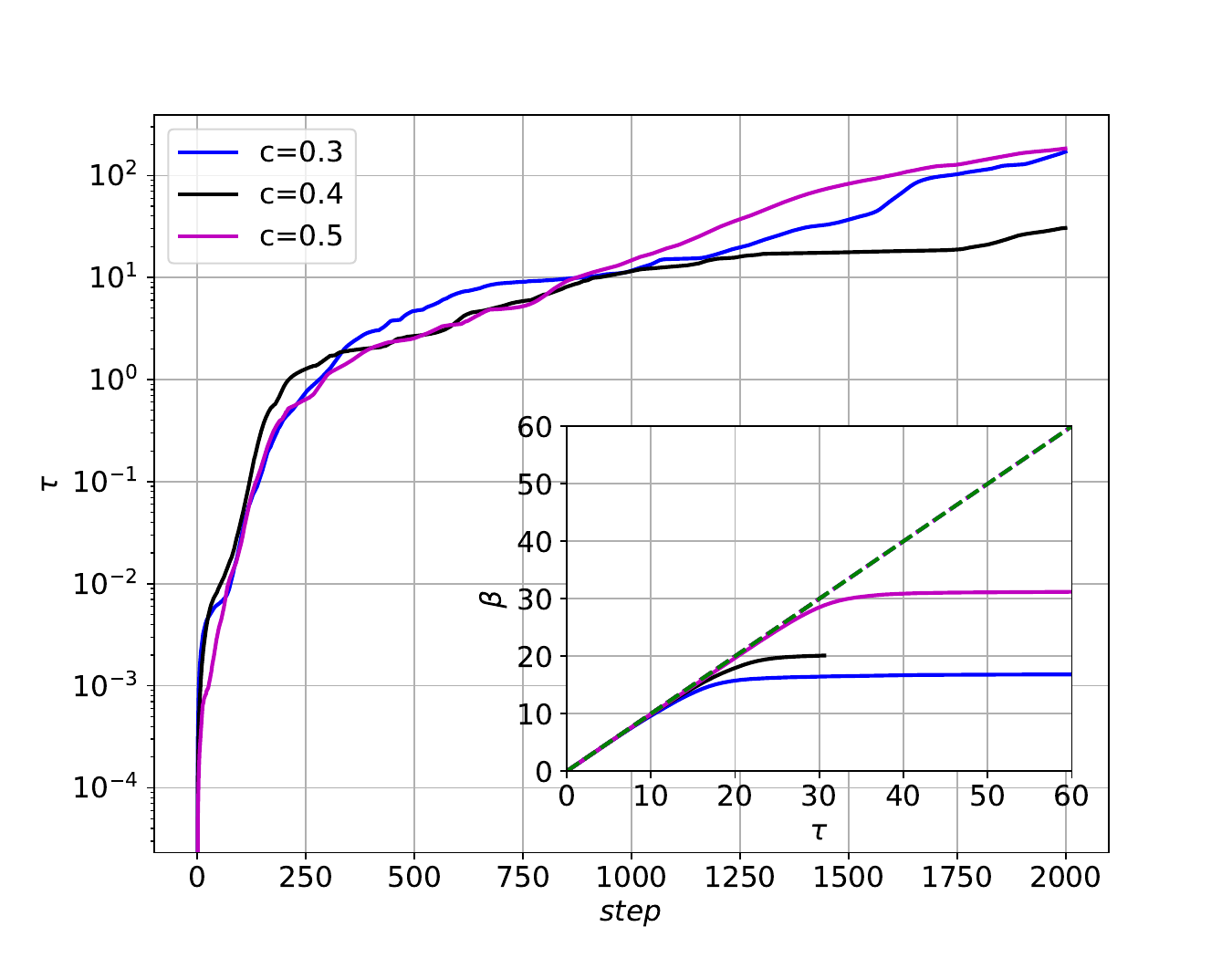}
    \caption{Imaginary time $\tau$ in the Heun algorithm vs number of iterative steps for a Heisenberg chain with $N=7$ sites. The parameters are the same as in Fig.~\ref{fig.Energy-7}. The $\beta$ modification is depicted in the inset.}
    \label{fig.Beta-7}
\end{figure}

\begin{figure}[!htp]
    \centering
    \includegraphics*[width=0.99\columnwidth]{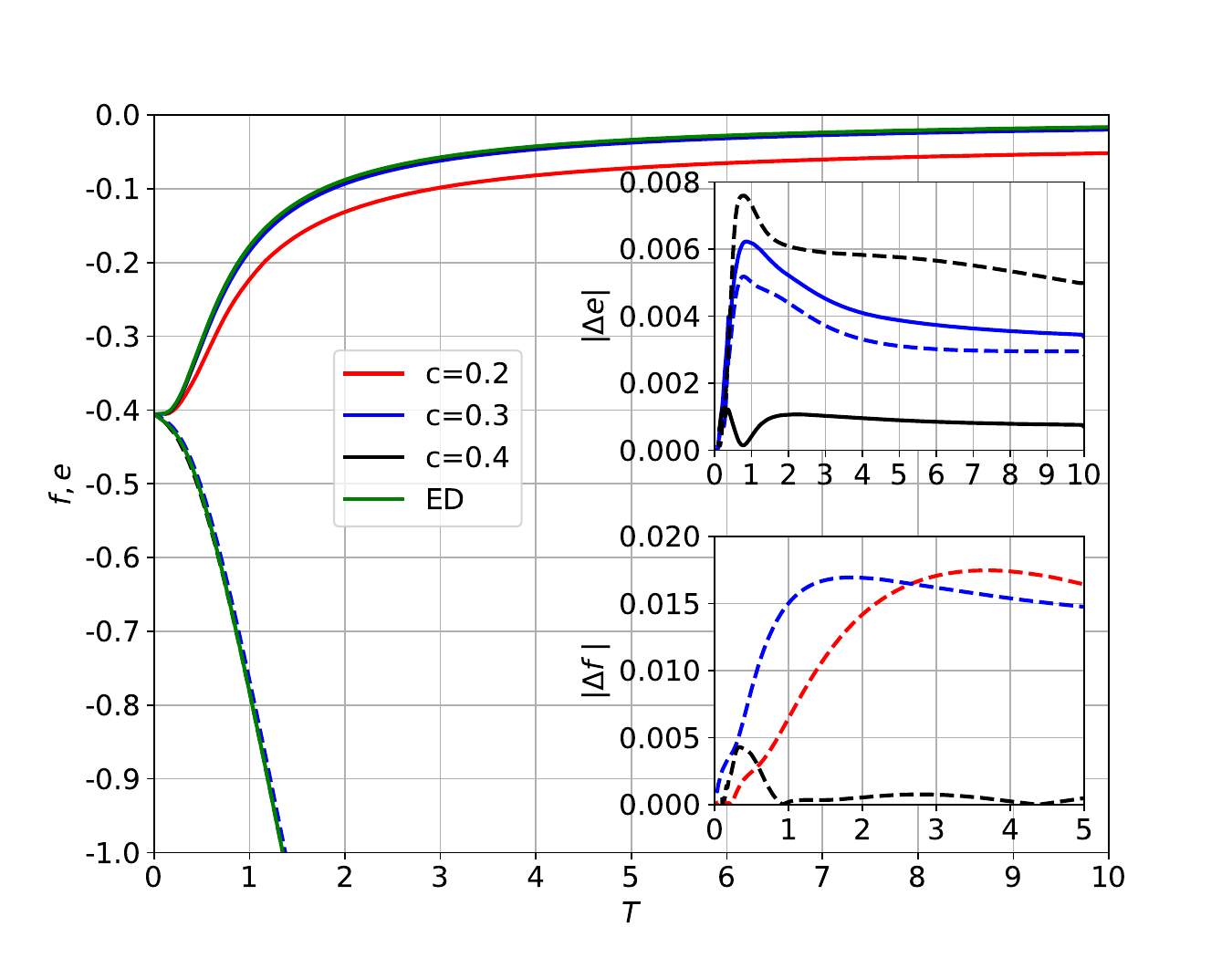}
    \caption{Inner and free energy for a Heisenberg chain with $N=7$ sites using the purification algorithm and a Monte Carlo sampling of the Hilbert space ($M=28$, $4\cdot 10^4$ samples, $\epsilon = 10^{-7}$, $R=2\cdot 10^3$). The results with $\beta$ modification are shown as solid lines, the ones without as dashed lines. The error in the energy for the case $c=0.2$ is not shown in the upper inset.}
    \label{fig.MC-Energy-7}
\end{figure}

\begin{figure}[!htp]
    \centering
    \includegraphics*[width=0.99\columnwidth]{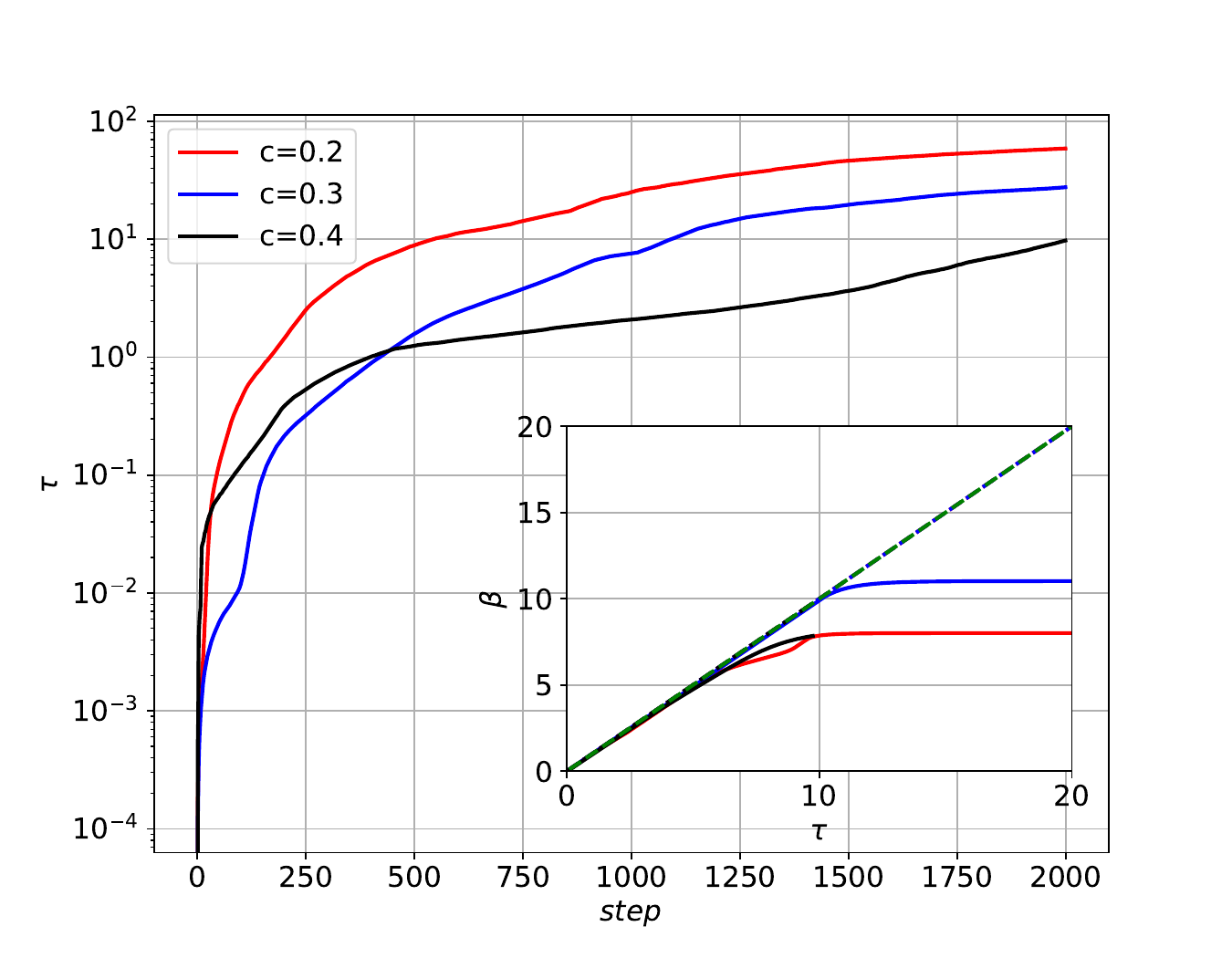}
    \caption{Imaginary time $\tau$ in the Heun algorithm vs number of iterative steps for a Heisenberg chain with $N=7$ sites. The parameters are the same as in Fig.~\ref{fig.MC-Energy-7}. The $\beta$ modification is depicted in the inset.}
    \label{fig.MC-Beta-7}
\end{figure}

To investigate the performance of the proposed algorithm, we use the spin-$\frac{1}{2}$ Heisenberg chain with Hamiltonian
\begin{equation}
    \label{XXZ}
    H=\sum_{j=1}^{N-1} \left\{-\frac{1}{2}(S^+_jS^-_{j+1}+h.c.)+S^z_jS^z_{j+1}\right\}
\end{equation}
as a test case. We use open boundary conditions. We note that while for periodic boundary conditions the number of variational parameters in the NN can be reduced due to translational invariance, this creates a new numerical problem for systems where there is no Marshall sign rule as, for example, for the Hamiltonian in Eq.~\eqref{XXZ} with a sign flip in the first term. This, however, is a separate issue which we do not wish to discuss here.
To show that the purification algorithm can indeed properly describe the thermodynamics of the spin-$\frac{1}{2}$ Heisenberg chain, we start by considering small systems where we can work with the full Hilbert space and can directly compare to exact diagonalization results. These calculations will also already indicate some issues and differences to zero-temperature calculations. In Fig.~\ref{fig.Energy-7}, we show results for the isotropic Heisenberg model with $N=7$ sites using the mRBM in the purification algorithm, where we keep all basis states and use no $\beta$ modification. Note that due to the doubling of the number of sites when adding the ancillas, this already corresponds to a Hilbert space dimension of $2^{14}$. 
The results show that for $M=28$ hidden units in the mRBM ansatz \eqref{ancilla_ansatz}, the exact results for the energy $e(T)$ and the free energy $f(T)$ are reproduced with absolute errors $\sim 10^{-3}-10^{-4}$. Note that we fix the normalization by setting $\langle\Psi(\infty)|\Psi(\infty)\rangle = 2^N$ which gives the proper free energy at infinite temperatures. If the parameter $c$ in the mRBM ansatz is chosen too small, then we already start with a state whose energy is strongly deviating from zero---the infinite-temperature value of the energy for the chosen form of the Hamiltonian in Eq.~\eqref{XXZ}---and thus a bad approximation of the infinite-temperature state. 
The course of the imaginary time in the Heun algorithm is different depending on the respective initial state and the topology of the area that is traversed in the time evolution (see Fig.~\ref{fig.Beta-7}).
The effect of the $\beta$ modification depicted in the lower right inset is negligible for temperatures larger than the finite size gap and is smaller than the inherent error of the approach. Thus, it is not visible in Fig.~\ref{fig.Energy-7}. The lower inset in Fig.~\ref{fig.Energy-7} shows that the free energy is obtained with a similar accuracy. 
For larger systems, it will no longer be possible to use the full Hilbert space. Instead, a Monte Carlo sampling using the Metropolis algorithm is employed. To better understand potential additional issues which may arise due to the sampling, we start by showing the results for the same system size, $N=7$, in  Fig.~\ref{fig.MC-Energy-7}.
Here, we find that the results for the energy have a similar accuracy as those for the full Hilbert space simulations if sufficiently many Monte Carlo samples are used and provided that the initial state is carefully chosen. The latter turns out to be more intricate here. First, a certain amount of noise in the $a$ and $W$ parameters is required for the imaginary time evolution to work properly. We have chosen to draw the real and imaginary parts of both sets of parameters from independent normal distributions with a width $w=0.02$. Second, the initial parameters $c_i=c=\mbox{const}>0$ have to be chosen just large enough to obtain $\langle\Psi(\infty)|H|\Psi(\infty)\rangle \sim 10^{-3}-10^{-4}$. If the initial $c$ parameters are chosen too large, then the algorithm remains stuck close to $|\Psi(\infty)\rangle$ for a large number of steps. Additionally, the sampling fluctuations become large and the algorithm shows instabilities for low temperatures. By decreasing the tolerance in the adaptive Heun algorithm and using the $\beta$ modification, acceptable results can still be obtained, but come with the cost of an increase in the number of iterations (see Fig.~\ref{fig.MC-Beta-7}).
Furthermore, the approximation of the infinite-temperature state is bounded by the sampling error. Thus, it is not feasible in this case to use values $c>0.4$. Making $c$ too small, on the other hand, leads to an initial error given by the non-zero initial energy which remains present down to low temperatures. \\ 
In practice, we proceed as follows: With $w$ small and fixed we obtain results for a series of calculations with increasing initial values for $c$ and different values for the desired tolerance $\epsilon$. The results are in general comparable for a certain range of initial values for $c$. Furthermore, we note that the error is dominated by the sampling error which is why an increase in the number of hidden units $M$ does not have a measurable effect. This can be deduced as well from the $\beta$ modification having no substantial effect up until low temperatures. Making $M$ too large, generally seems to lead to additional instabilities. For the considered model, $M\in[N,4N]$ seems to be optimal. While we find that this approach to find the right initial parameters and the right number of hidden units works, we note that it requires to run multiple simulations with different initial parameters $c$ and $\epsilon$. We note also, that without being able to compare with other methods it might not always be easy to decide what the optimal initial parameters are. Furthermore, we note that the free energy is more sensitive than the energy to the quality of approximation of the initial state and to the accuracy of the time evolution (see Fig.~\ref{fig.MC-Energy-7}).

\subsection{Results for System Sizes Beyond Exact Diagonalization}
\label{subsec.Numerical results for system sizes beyond exact diagonalizations}
\begin{figure}[!htp]
    \centering
    \includegraphics*[width=0.99\columnwidth]{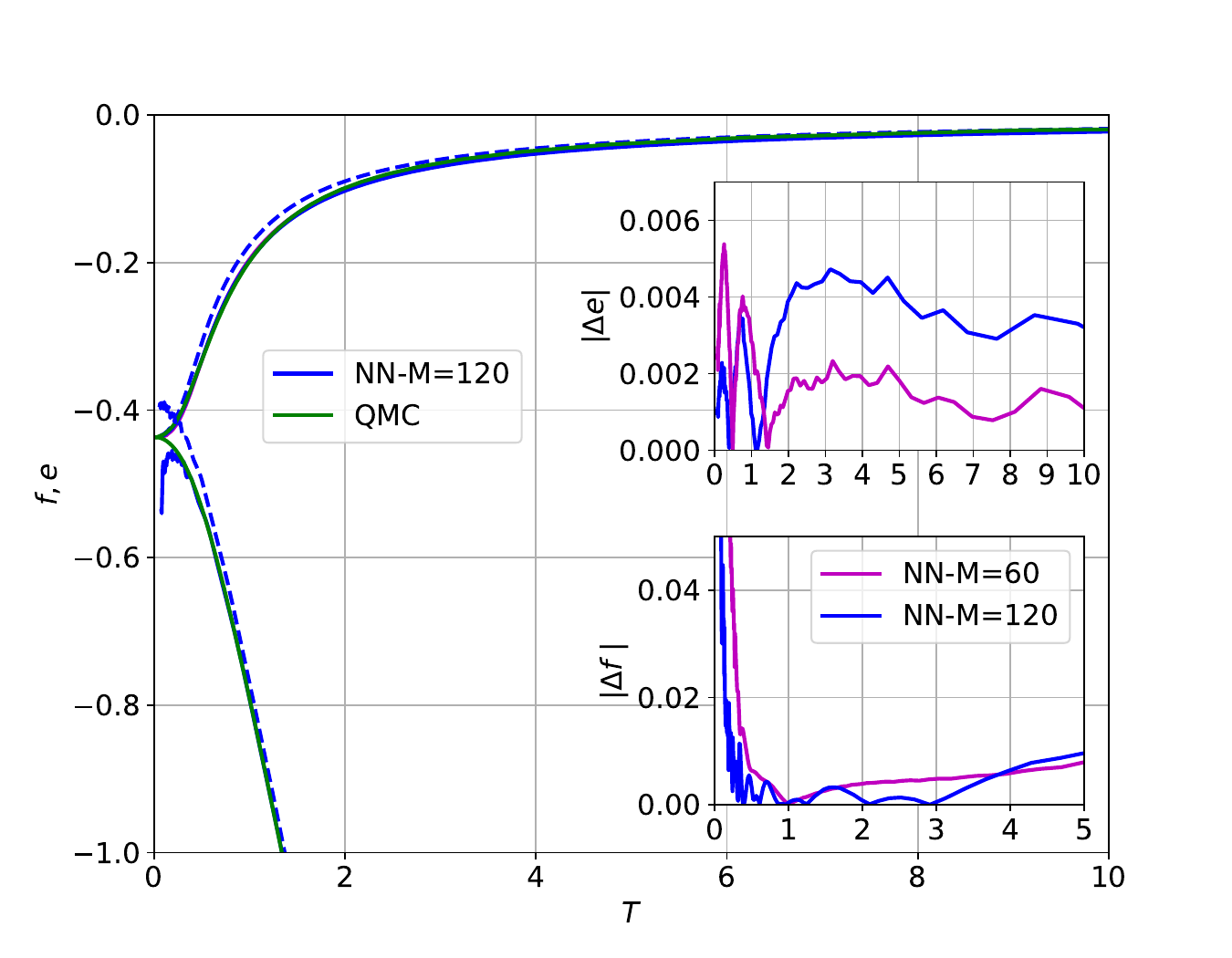}
    \caption{Inner and free energy for an open Heisenberg chain with $N=30$ sites using $4\cdot 10^4$ samples and $R=2\cdot 10^3$ iteration steps. We choose $c=0.34$ and $\epsilon=10^{-6}$. Results without the $\beta$ modification are represented by dashed lines. The results for $M=60$ are omitted in the main plot for visual clarity.}
    \label{fig.MC-Energy-30}
\end{figure}

\begin{figure}[!htp]
    \centering
    \includegraphics*[width=0.99\columnwidth]{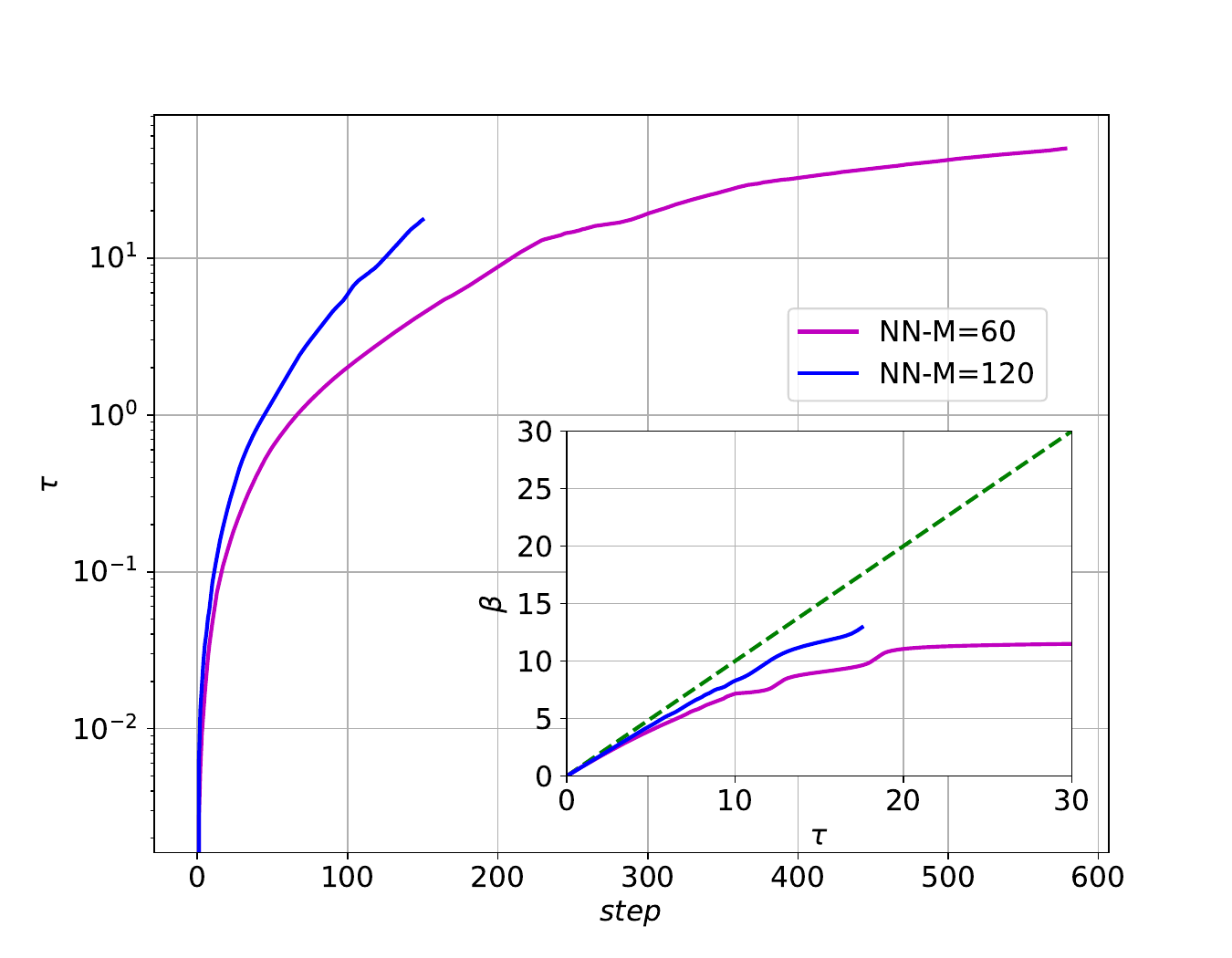}
    \caption{Imaginary time $\tau$ in the Heun algorithm vs number of iterative steps for an open Heisenberg chain with $N=30$ sites. The parameters are the same as in Fig.~\ref{fig.MC-Energy-30}. The effect of the $\beta$ modification is depicted in the inset.}
    \label{fig.MC-Beta-30}
\end{figure}

In this section, we want to briefly show that the purification algorithm can be scaled up and used to access the thermodynamics of quantum lattice models for system sizes outside the reach of exact diagonalization. We want to stress again that the purpose of this paper is a proof of principle demonstration of using NNs to investigate thermodynamic properties and the discussion of certain generic obstacles. There are a number of directions in which the algorithms presented here could possibly be improved, including but not limited to the use of more complicated network structures such as CNNs or modifications to the imaginary time evolution by using, for example, implicit iterative methods. \\
As an example, we present in Fig.~\ref{fig.MC-Energy-30} results obtained using the purification algorithm for a Heisenberg chain with 30 sites and compare the results with QMC results for the energy and the free energy, where the latter is obtained by the respective integration of the former. We used the stochastic series expansion code available at \cite{Sandvik-Webpage} which is described in Ref.~\cite{Sandvik}.  \\
As for the smaller system sizes considered in the previous section, a postselection of data is still required and can be done in a systematic manner. We find again that the parameter range $c=0.3...0.4$ gives initial energies $|e|\sim 10^{-3}$ and allows for an accurate imaginary-time evolution. The algorithm is executed for different tolerances. For $c=0.34$ and $M=60$, the results for the energy appear to be converged sufficiently for the $\beta$ modified results to stay within the mentioned accuracy for all temperatures. Increasing the number of hidden units results in a small improvement for low temperatures, but also leads to instabilities that occur earlier in imaginary time $\tau$. However, the algorithm can proceed faster to reach lower temperatures. We want to emphasize that the $\beta$ modification depicted in Fig.~\ref{fig.MC-Beta-30} saves us from increasing the number of hidden units to an extent that is not numerically feasible anymore. The results without $\beta$ modification shown as dashed lines are about one order of magnitude less accurate than the modified data. The $\beta$ modification is thus a useful tool to achieve a good accuracy without having to increase the number of hidden units to the point where numerical instabilities become more and more problematic. The lowest temperatures reached are limited by the progress of the Heun method and the $\beta$ modification, which at some point result in a temperature which is not decreasing anymore. To overcome this, increasing the number of hidden units and the number of sampled states is ultimately inevitable. To obtain a better approximation of the ground state, the regularization can be adjusted by adding a diagonal matrix with entries being $\sim 10^{-4}-10^{-6}$ (see Ref.\cite{TroyerScience} and \cite{PP-NN1}) to the matrix $\vec{G}$ defined in Eq.~\eqref{var_principle2} . This, however, does not work well for the computation of the finite-temperature properties of a system since such a regulation distorts the imaginary-time evolution itself. We find that the free energy is more sensitive to the sampling error and that the $\beta$ modification is not helping for low temperatures leading to the large deviations observed in Fig.~\ref{fig.MC-Energy-30}. Additional issues are jumps in the inner and free energy at temperatures close or below the finite size gap. The simulation must be stopped in this case. Related problems were also observed in Ref.~\cite{SpuriousModes}, where it was suggested that they can be mediated by modifying the respective loss function. 
We also note that very recently, the authors of Ref.~\cite{nys2023realtime} proposed a Hadamard transformation of the infinite-temperature state to circumvent some of the issues related to small gradients at high temperatures. While all purifications are unitarily equivalent, it is possible that some provide a better starting point for numerical computations than others. We have not systematically studied the effect of such unitary transformations of the initial state here and leave such investigations for future work. The general problem of having states which overlap with only a small number of basis states and are thus difficult to represent in a Monte Carlo sampling but could, potentially, yield important contributions to the gradient expansion has also been considered for small model systems in Ref.~\cite{Sinibaldi2023unbiasingtime}.

\subsection{Correlation Functions}
\label{subsec.Numerical results for correlation functions with finite support}
\begin{figure}[!htp]
    \centering
   \includegraphics*[width=0.99\columnwidth]{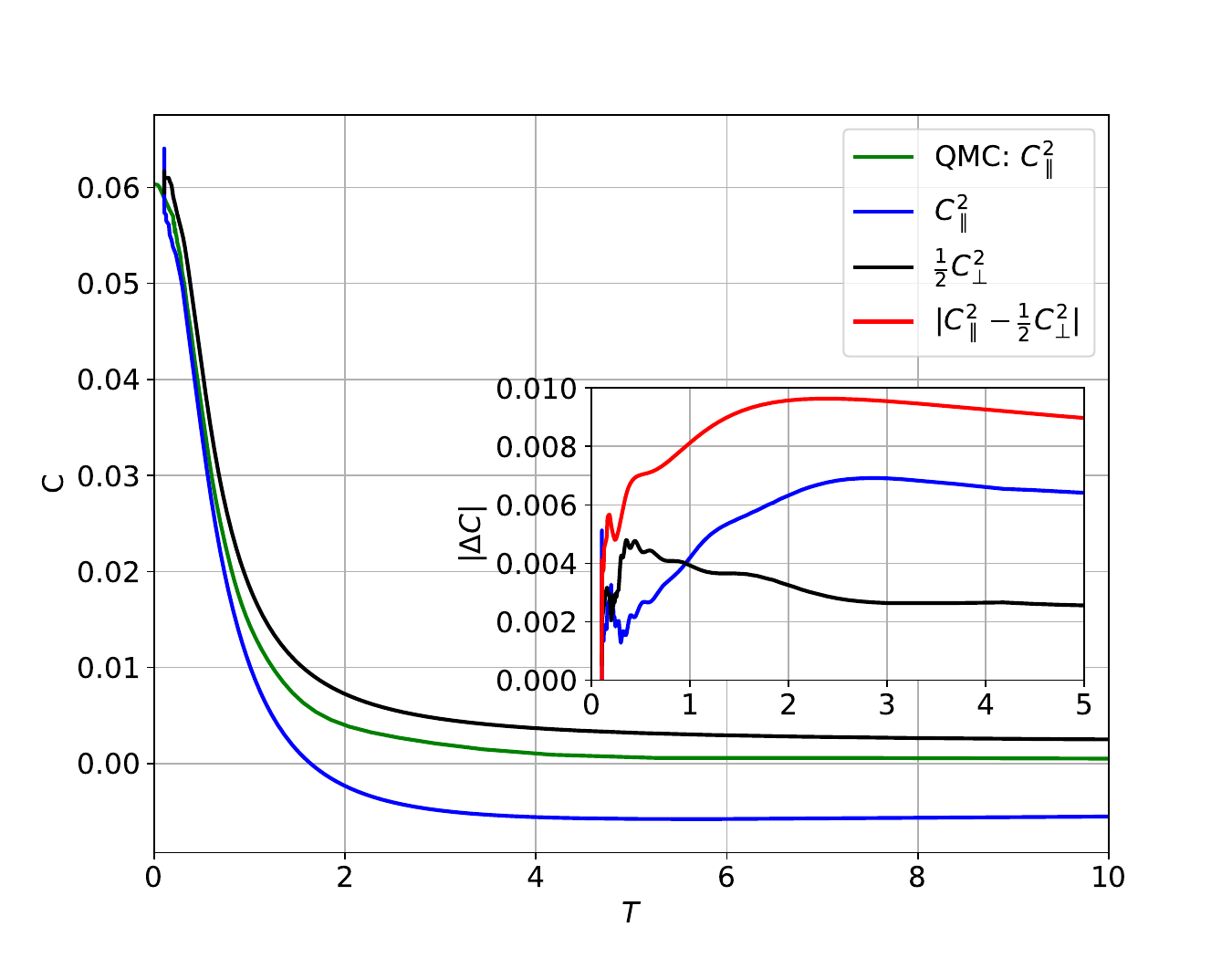}
    \caption{Correlation functions $C^2_{\parallel,\perp}$ for a chain of length $N=30$ obtained using the purification method are compared to the QMC result. The parameters are the same as in Fig.~\ref{fig.MC-Energy-30} and $M=60$. The inset shows $|C^2_\parallel -\frac{1}{2}C^2_\perp|$ and the deviations of the two correlation functions from the QMC result.}
    \label{fig.SzSz-AC-30}
\end{figure}

To check how well the proper thermodynamics is approximated beyond the energy, we compute also the correlation functions $C^d_\parallel=\langle S_k^zS_{k+d}^z\rangle$ and \mbox{$C^d_\perp=-\frac{1}{2}\langle S_k^+S_{k+d}^- + S_k^-S_{k+d}^+\rangle$}. In Fig.~\ref{fig.SzSz-AC-30}, the results for $k=10$ and $d=2$ are compared to the QMC results and agree to an accuracy of order $10^{-3}$.
The deviations observed in Fig.~\ref{fig.SzSz-AC-30} are dominated by the Monte Carlo sampling error. We note that the SU(2) symmetry is broken, in general, but that it is restored at low temperatures.

\subsection{Real-time Dynamics at Finite Temperatures}
\begin{figure}[!htp]
    \centering
    \includegraphics*[width=0.99\columnwidth]{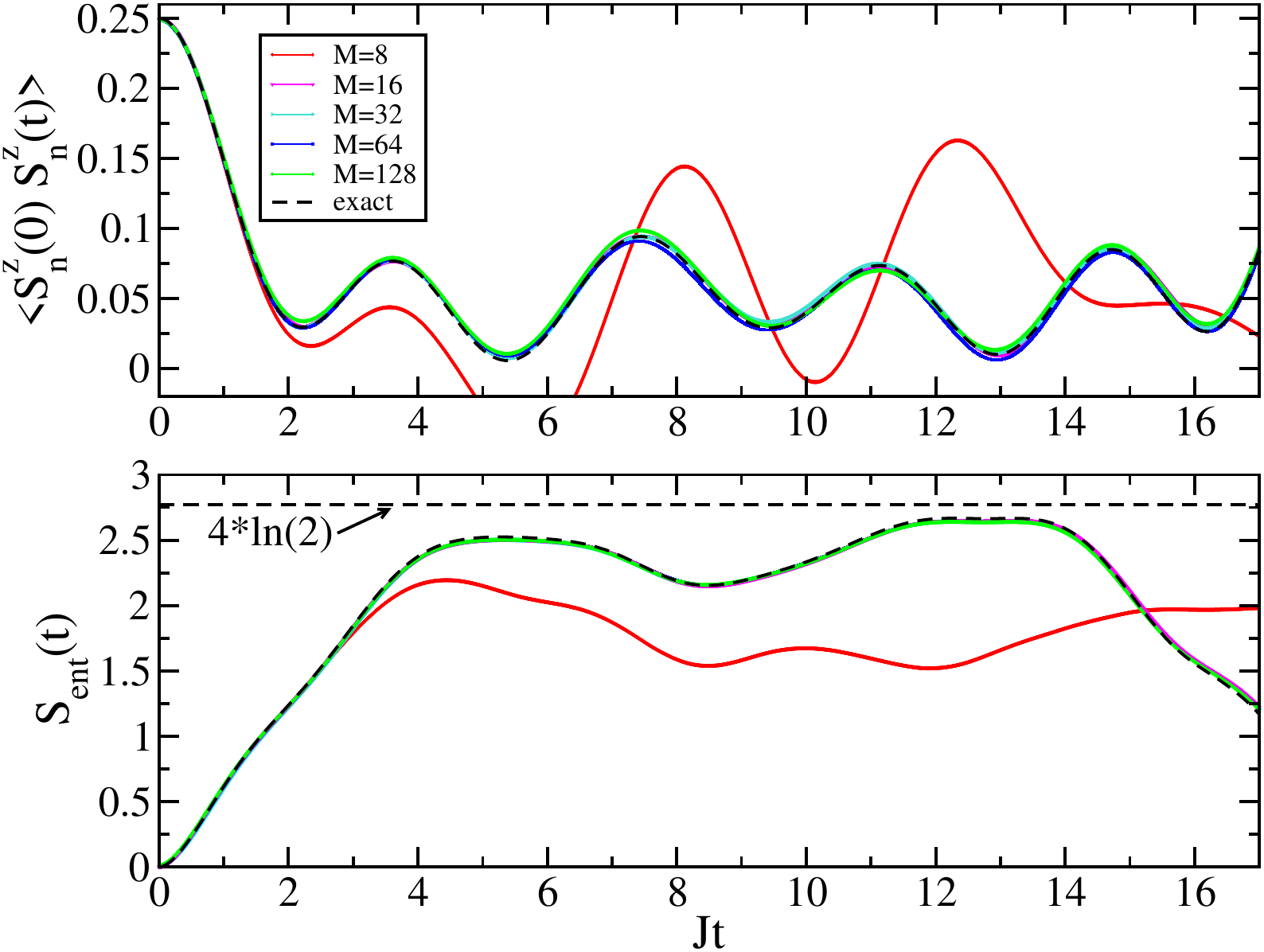}
    \caption{Top: Autocorrelation function $\langle S^z_n(0)S^z_n(t)\rangle$ for $n=N/2$ at infinite temperatures for a Heisenberg chain with OBC and $N=4$ sites. Bottom: Corresponding entanglement entropy for cutting the chain of real and ancilla spins into two equal halves. The RBM results for different $M$ are compared to the exact result.}
    \label{Fig_Realtime1}
\end{figure}

\begin{figure}[!htp]
    \centering
    \includegraphics*[width=0.99\columnwidth]{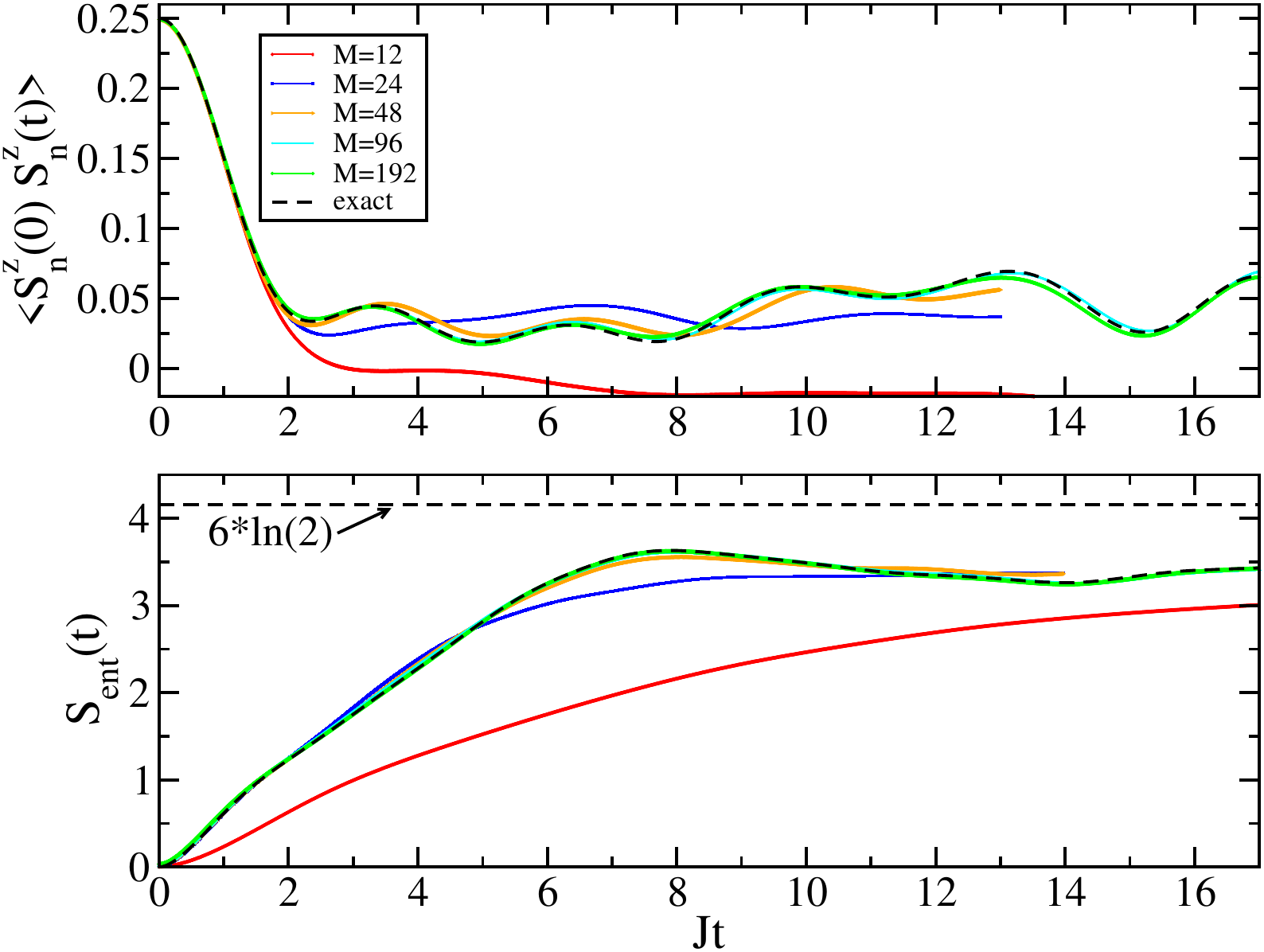}
    \caption{Same as Fig.~\ref{Fig_Realtime1} for a chain with $N=6$ sites. The results for $M=24,48$ are only shown up to $Jt\sim 13$.}
    \label{Fig_Realtime2}
\end{figure}

\begin{figure}[!htp]
    \centering
    \includegraphics*[width=0.99\columnwidth]{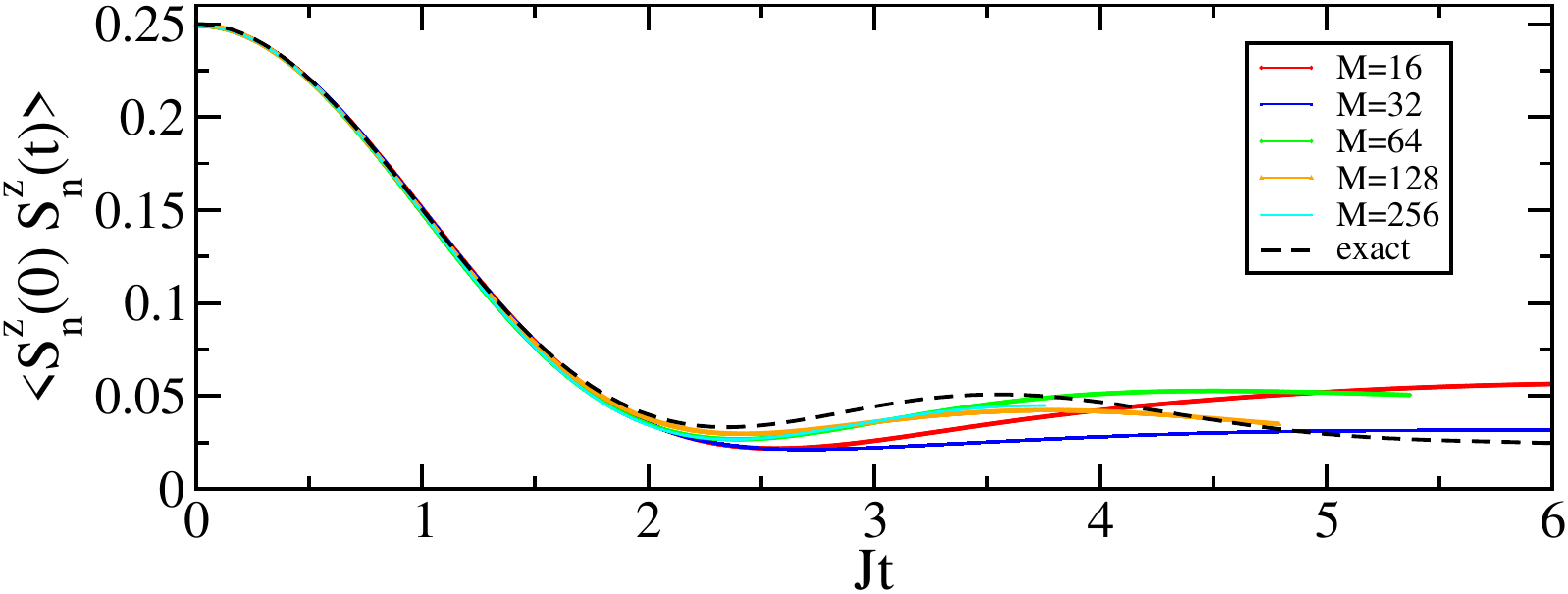}
    \caption{Same as Fig.~\ref{Fig_Realtime1} for a chain with $N=8$ sites. The corresponding entanglement entropies have not been calculated.}
    \label{Fig_Realtime3}
\end{figure}

Among the most interesting questions related to using NNs as approximators for quantum wave functions, is whether or not this approach will lead to major progress in the calculation of dynamical correlation functions. Numerically, real-time evolution is a difficult problem because QMC only gives direct access to imaginary times and analytical continuations of numerical data to real times are not unique. In time-dependent DMRG algorithms, on the other hand, the growth in time of the entanglement entropy often severely limits the accessible simulation times. DMRG is based on a truncation of the Hilbert space. If the reduced density matrix $\rho_A$ has a maximum dimension $\chi$, then the maximal entanglement entropy which can be faithfully represented is $S^{\textrm{max}}_{\textrm{ent}} = -\tr\rho_A\ln\rho_A = \ln\chi$. If the entanglement entropy grows linearly in time---which is the generic case following a quantum quench from a product state and for equilibrium correlations at high temperatures, see below---leading to a volume law $S_{\textrm{ent}}\sim N$ at long times, then one needs $\chi\sim\min\{\exp(t),\exp(N)\}$ states in the truncated Hilbert space. This exponential barrier often prevents to fully address dynamical problems. 
In NN algorithms, on the other hand, the Hilbert space is not explicitly truncated and the hidden units typically are fully connected to all units in the input layer. This means that in principle, states with large entanglement can already be represented by NNs with a small number of hidden units. Indeed, it was shown in Ref.\cite{DengLiDasSarma} that a volume law entangled state can be represented with $M\sim N$ hidden units only.  This, however, does not answer the question whether or not the dynamics of a quantum system can be efficiently simulated in general using a NN representation of the wave function. While it is an interesting theoretical result that specific volume-law entangled states exist which do have an efficient representation in this formalism, much more is required to investigate the long-time dynamics of a quantum system. Here, typically all the eigenstates of the system will contribute with most of them---in a clean, nondisordered system---showing volume-law entanglement. Only if a large fraction of these states can be efficiently represented by a given network can we hope to obtain accurate results for dynamical quantities at long times. \\
In the following, we want to numerically investigate how many hidden units are required to obtain accurate results for the spin-spin autocorrelation function 
\begin{equation}
    \label{Auto}
    \langle S^z_n(0)S^z_n(t)\rangle = \frac{1}{Z}\tr\left\{S^z_n\e^{itH}S^z_n\e^{-itH}\right\}
\end{equation}
at site $n$ of a Heisenberg chain at infinite temperatures using the purification algorithm. In addition, we also calculate the entanglement entropy
\begin{equation}
    \label{Sent}
    S=-\tr\rho_A\ln\rho_A \ ,
\end{equation}
where $\rho_A$ is the reduced density matrix for a subsystem $A$ consisting of half of the chain of real spins and ancilla spins. We begin with results for $N=4$ shown in Fig.~\ref{Fig_Realtime1}.
While the results are numerically stable even for $M=8$ hidden units, we see that we need at least $M=16$ to accurately approximate the exact result. From the panel showing the entanglement entropy, it is clear that while $M=8$ is sufficient to represent certain states with volume-law entanglement \cite{DengLiDasSarma}, it does not capture the full entanglement structure of the time evolved state for times $Jt\gtrsim 2$. \\
Next, we show the same results for $N=6$ in Fig.~\ref{Fig_Realtime2}.
In this case, at least $M=96$ hidden units are required. We note again that already for $M=12,24,48$ the entanglement entropy reaches values at long times of the order of the maximum entanglement in the time evolved state. However, the proper time evolution is only captured for $M\geq 96$. Lastly, we also show in Fig.~\ref{Fig_Realtime3} results for $N=8$.   
Note that, due to the ancillas, this already corresponds to a system with $16$ sites and we do keep the full Hilbert space. Since the calculation of the entanglement entropy is numerically expensive, we only show the correlation function. From the results it is clear that even $M=256$ hidden units are not sufficient. \\
To summarize, the number of hidden units $M$ required to accurately simulate the infinite-temperature autocorrelation \eqref{Auto} increases much faster than linear. Assuming that $M=512$ hidden units give reasonably converged results for $N=8$, the increase follows in the best case scenario a power law $M\sim N^\alpha$ with $\alpha\gtrsim 4$ but is more consistent with an exponential scaling. While a potential power-law increase in the hidden units with system size would make this approach in theory more powerful than time-dependent DMRG, it might be very difficult in practice to benefit from this scaling (if it is not exponential to begin with as seems more likely): while the number of states kept in a time-dependent DMRG calculation is only limited by the available memory and the compute time to perform a single-value decomposition or a diagonalization allowing to reach truncated dimensions of order $\chi\sim 10^5-10^6$, time-dependent NNs for $N\sim 10^2$ with $M\sim 10^8$ hidden units might be difficult to optimize given that the Hilbert space needs to be sampled. We also note that in DMRG the dimension $\chi$ can be adaptively increased with simulation time. Based on the NN results shown in Figs.~\ref{Fig_Realtime1}, \ref{Fig_Realtime2}, \ref{Fig_Realtime3} it seems, on the other hand, that the calculation becomes inaccurate already at short times if an insufficient number of hidden units is used. Furthermore, there does not seem to be an easy way to adaptively increase the number of hidden units during a calculation. Finally, we note that the strong, likely exponential increase of the number of required hidden units is consistent with a recent publication where similar behavior was observed for quench dynamics \cite{LinPollmann}.

\section{Sampling}
\label{sec.Sampling}
To compute the ensemble average of a quantum mechanical operator A (see Eq.~\eqref{Z}) at infinite temperature, one needs to compute the expectation value with respect to every vector of an orthonormal set $\{|\phi_n \rangle \}_{n=1}^{2^N}$
\begin{equation}
    \langle A \rangle_{\beta=0} = \frac{\tr[A]}{2^N} = \frac{1}{2^N}\sum_{n=1}^{2^N} \langle \phi_n |A| \phi_n \rangle \ .
\end{equation}
The idea is now that instead of working with a full orthonormal set we can also sample the trace using random states. These are constructed by choosing $2^N$ random vector components $c_n$ from a symmetric probability distribution (\cite{deRaedt}) to form the infinite-temperature states
\begin{equation}
|\Psi(\beta=0) \rangle = \sum_{n=1}^{2^N} c_n|n\rangle \ .
\label{eq.randstate0}
\end{equation}
It is essential that the vector components are uncorrelated. The ensemble average is then approximated by using $S$ realizations of those states.
\begin{equation}
    \frac{1}{S} \sum_{s=1}^S \langle \Psi_s(\beta=0) | A | \Psi_s(\beta=0) \rangle \approx E(|c_n|^2) \ \tr(A) 
    \label{eq.EVD}
\end{equation}
For normalized states, the expectation value of any $c_n$ is $E(|c_n|^2) = \frac{1}{2^N}$ and thus the correct value of the ensemble average at infinite temperature is obtained. The error goes in general like $\frac{1}{\sqrt{S}}$. An upper bound is computed in Ref.~\cite{deRaedt} in terms of the operator $A$. To include this idea into a variational algorithm using the RBM ansatz, the parameters of the NN must be chosen properly. States that fulfill the above criteria and are efficiently represented by the RBM are, for example, the so called minimally entangled typical thermal states (METTS) \cite{RWhite}. The word \lq typical\rq \ will later be defined and used in Sec.~\ref{sec.Quantum Typicality} in the sense of quantum typicality. To avoid confusion, it is dropped in this paper for the notation of these states.

\subsection{Minimally Entangled Thermal States (METS)}
\label{sec.METS}
The minimally entangled thermal states (METS) were first discussed as a tool to perform the imaginary-time evolution in a standard DMRG algorithm \cite{RWhite}. In Ref.~\cite{RBM-pure}, the approach was then generalized for the NN ansatz. A random METS is defined as
\begin{equation}
|\Psi_0\rangle = \bigotimes_{j=1}^N \left( \frac{\xi_{j,\uparrow} |\uparrow \rangle_j + \xi_{j,\downarrow} |\downarrow \rangle_j } {\sqrt{|\xi_{j,\uparrow}|^2 + |\xi_{j,\downarrow}|^2}} \right),
\label{eq.ProductState}
\end{equation}
where the vector $\vec\xi$ is chosen according to a standard complex normal distribution $\mathbb{C}N(\mu= 0, C = \mathbb{1})$ with $\mu$ the mean and $C$ the covariance matrix of the distribution
\begin{equation}
P_0(\vec\xi) = \left(\frac{1}{\pi} \right)^{2N} \exp \left( -\sum_{j=1}^N \sum_{\sigma=\uparrow, \downarrow} |\xi_{j,\sigma}|^2 \right) \, .
\end{equation}
This is equivalent to choosing the real and imaginary part according to a real normal distribution $N(\mu= 0, \sigma=1/2)$ with $\sigma$ being the variance. A METS state can be obtained in the RBM, Eq.~\eqref{ancilla_ansatz} with $\vec{c}=0$, up to an irrelevant overall constant by setting
\begin{equation}
a_j = -\frac{1}{2} \ln \left( \frac{\xi_{j,\uparrow}}{\xi_{j,\downarrow}} \right) \ ,
\label{eq.METS-a}
\end{equation}
$\vec{W}=0$ and $\vec{b}=0$, see Ref.~\cite{RBM-pure}. These states will be referred to as METS-NN from here on. Numerically, one needs to add some noise to the NN parameters $\vec{W}$ and $\mathbf{b}$ as in the purification method. We decided to choose the real and imaginary parts of both of them from a normal distribution centered around zero with $\sigma_{NN} = 1/N$ so that the variance of the wave function components stays constant for variable system size.

\subsection{Results for the Energy}
\begin{figure}[!htp]
    \centering
   \includegraphics*[width=0.99\columnwidth]{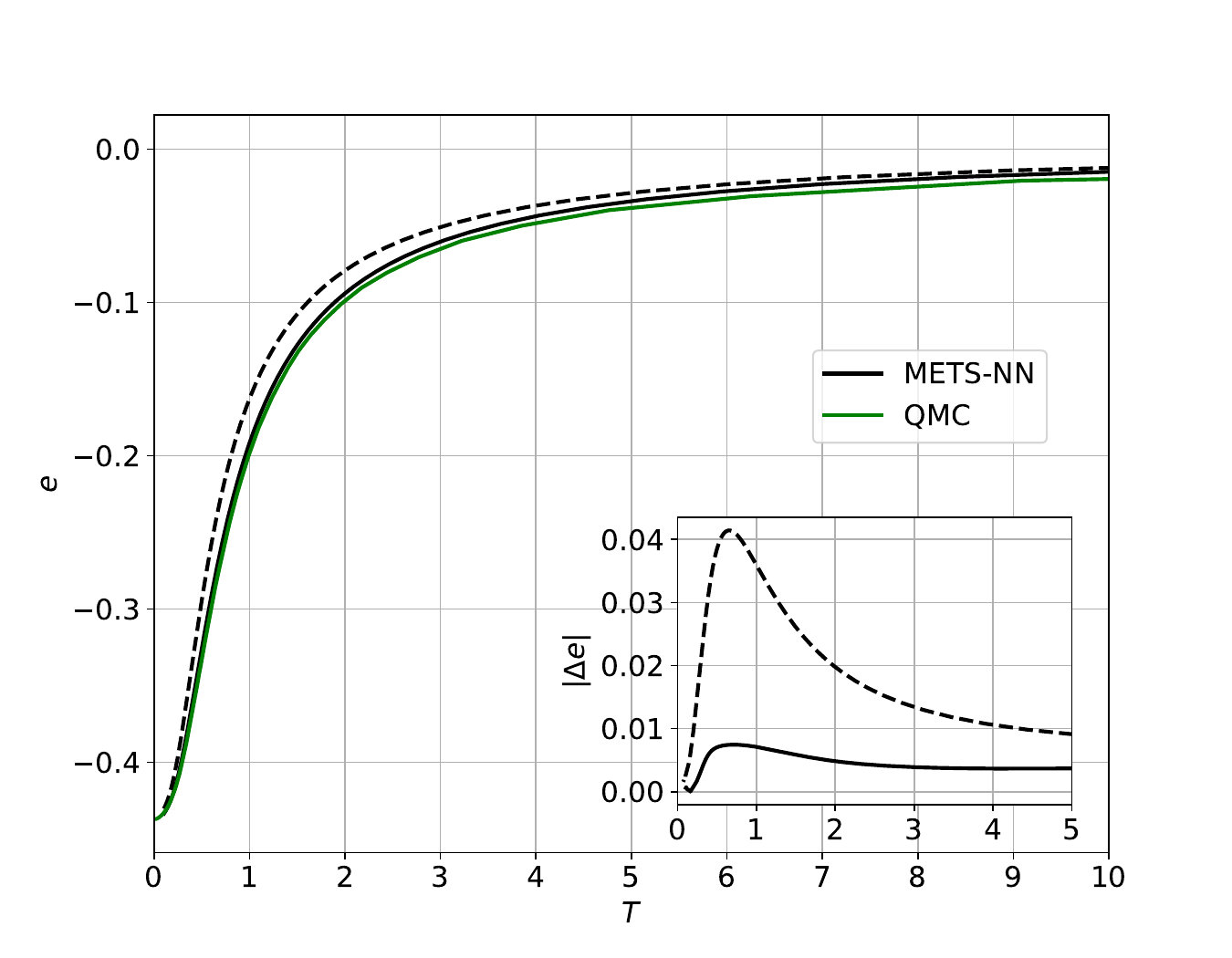}
    \caption{Inner energy of the METS-NN ansatz compared to the QMC solution for an open Heisenberg chain of length $N=30$ using $4\cdot 10^4$ Metropolis samples, $R=2\cdot 10^3$ and $\epsilon = 10^{-6}$. The average is taken over $S=30$ initial states. The dashed lines show the data without $\beta$ modification.}
    \label{fig.ITE-METS-NN}
\end{figure}

\begin{figure}[!htp]
    \centering
   \includegraphics*[width=0.99\columnwidth]{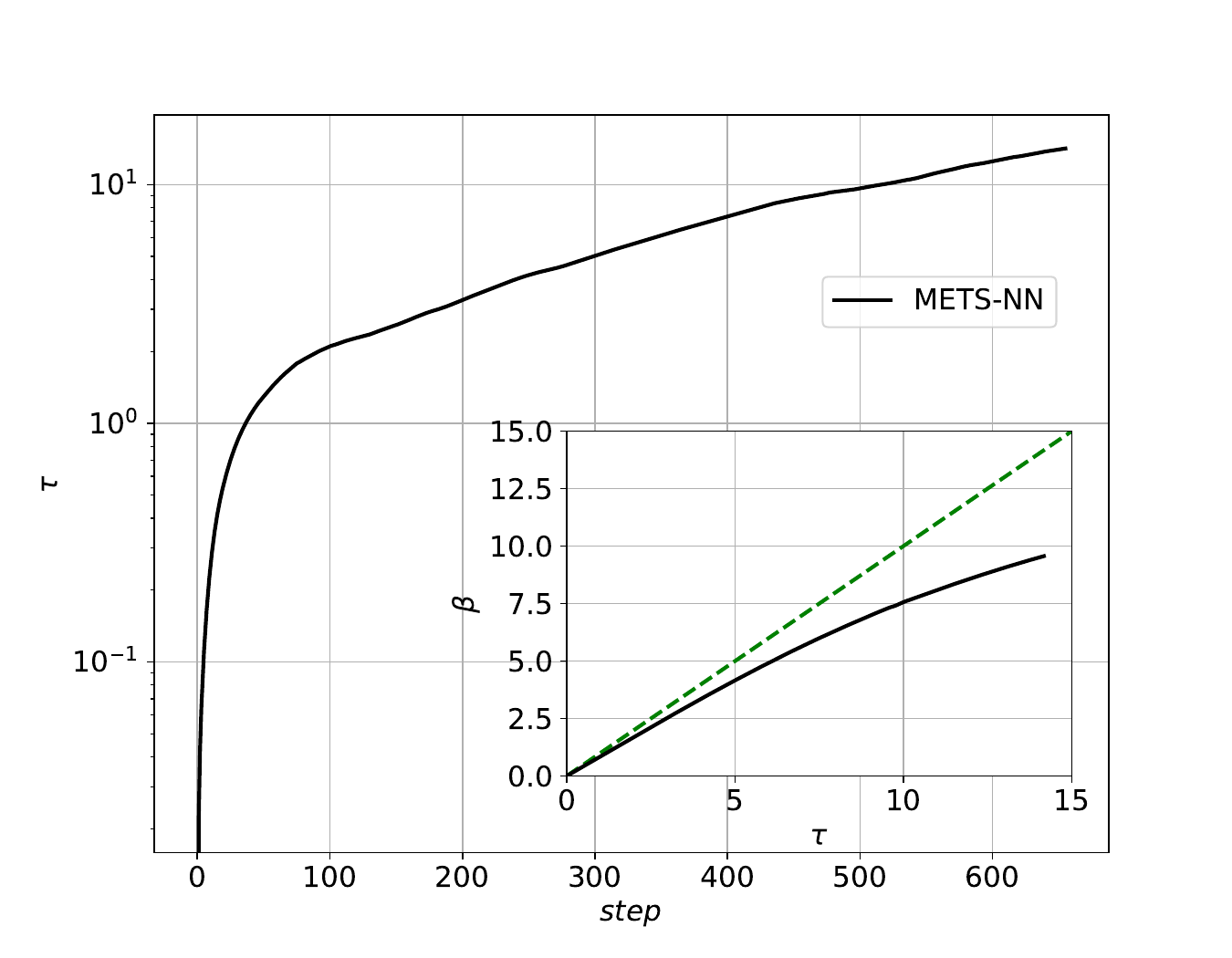}
    \caption{Imaginary time $\tau$ in the Heun algorithm vs number of iterative steps for the METS-NN ansatz for an open Heisenberg chain of length $N=30$ and parameters defined in Fig.~\ref{fig.ITE-METS-NN}. The $\beta$ modification is shown in the inset.}
    \label{fig.ImaginaryTime-METS-NN}
\end{figure}

The set of METS-NN states at infinite temperature is evolved by the adaptive Heun method as explained in Sec.~\ref{subsec.Adaptive Heun Method}. The imaginary time of the algorithm $\tau$ is afterwards modified as explained in Sec.~\ref{subsec.Beta Modification}. 
The results for the energy of a Heisenberg chain of length $N=30$ and $S=30$ states are depicted in Fig.~\ref{fig.ITE-METS-NN}. The accuracy is here also of order $10^{-3}$ which is expected from the sampling error. We also note that the temperatures reached using the purification method are lower than the ones reached here. This is depicted in Fig.~\ref{fig.ImaginaryTime-METS-NN} and due to the already mentioned convergence issues at low temperatures (see Sec.~\ref{subsec.Numerical results for system sizes beyond exact diagonalizations}). Compared to the purification method, this problem is compounded here because one needs to simulate $\sim 30$ states to low temperatures without any such instabilities occurring. Additionally, the state in which these instabilities occur first defines the minimum temperature that can be reached. To allow for the computation of the free energy by integrating the energy results respectively, as it was done for the QMC results in Fig.~\ref{fig.MC-Energy-30}, the evolution must reach lower temperatures. This can only be achieved by increasing the number of hidden units drastically.

\subsection{Correlation Functions}
\begin{figure}[!htp]
    \centering
   \includegraphics*[width=0.99\columnwidth]{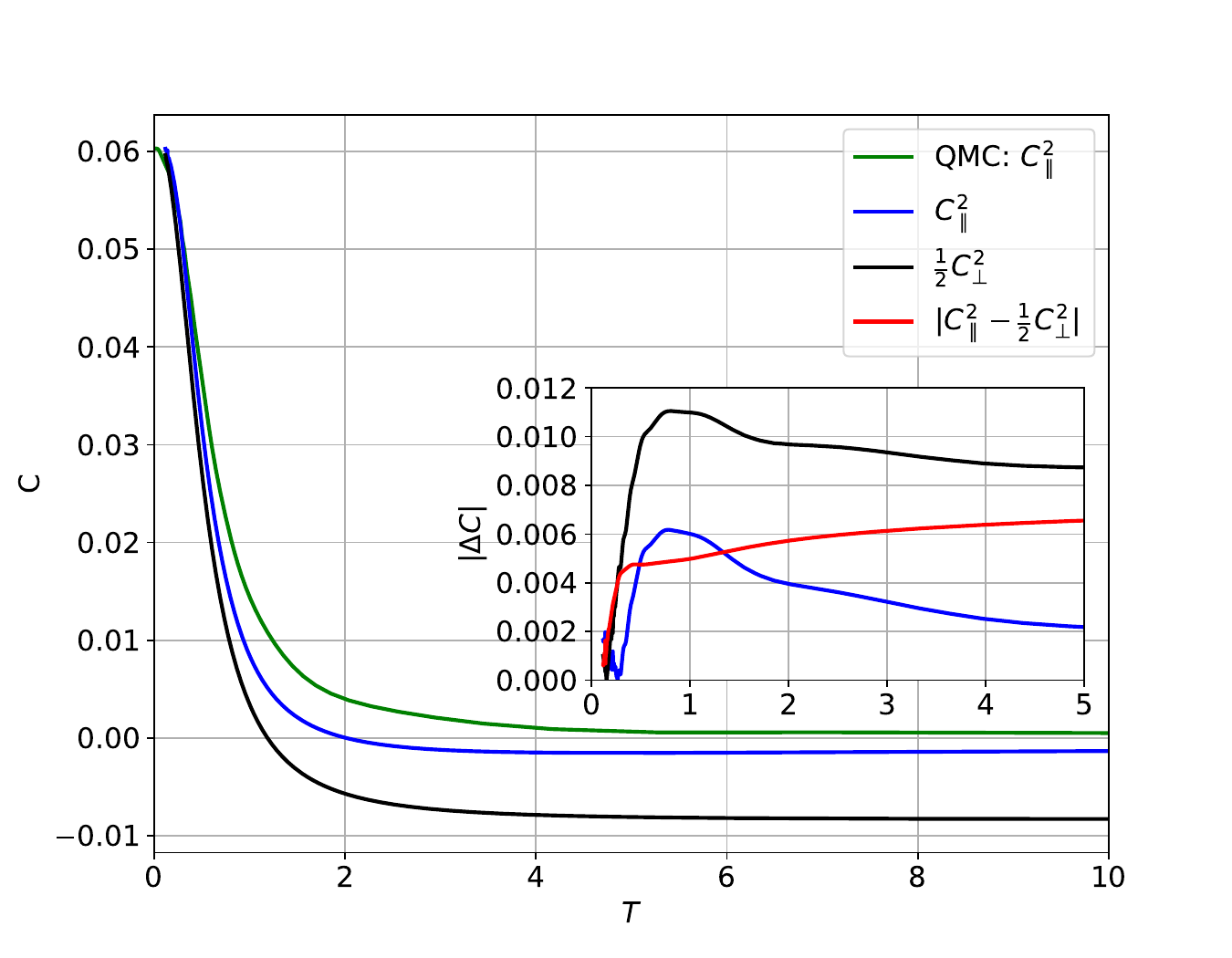}
    \caption{Same as Fig.~\ref{fig.SzSz-AC-30} using the METS-NN states with the parameters specified in Fig.~\ref{fig.ITE-METS-NN}.}
    \label{fig.SzSz-Typ-30}
\end{figure}

For the sampling method, it is checked as well how accurate thermodynamics is approximated beyond the energy by computing spin-spin correlators like $C^d_\parallel$ and $C^d_\perp$ for $k=10$ (see Sec.~\ref{subsec.Numerical results for correlation functions with finite support}). These are not obtained as an average over local operators such as the energy where an average over $N-1$ bond energies is taken. Thus, the METS-NN algorithm results in larger errors for single spin-spin correlators.  In Fig.~\ref{fig.SzSz-Typ-30}, the results for $C^2_\parallel$ and $C^2_\perp$ are compared to the QMC results. It is simply no longer sufficient to use $\sim 30$ states only. Instead, to obtain a result that is consistent within the Monte Carlo sampling error,  $\sim 1000$ states are required. Note that a preselection of states would break the relation in Eq.~\eqref{eq.EVD}. We emphasize that for any specific single state $C^d_\parallel$ and $C^d_\perp$ vary, with neither of them being more accurate in general. The results consequently also violate $\text{SU(2)}$ invariance.

\section{Quantum Typicality}
\label{sec.Quantum Typicality}
The third method investigated in this paper promises to provide one state that is sufficient for the computation of finite temperature expectation values for large system sizes. The definition of thermal pure quantum states (TPQS) or quantum typical states comes without the equal a priori probability postulate of statistical mechanics and replaces the ensemble average by pure state expectation values (see Ref. \cite{Popescu,Popescu2}). Quantum typical states were first discussed for the canonical situation of a small system S coupled to an environment E. In the standard microcanonical formulation, the coupled system is assumed to be in an equiprobable state $\Omega$ corresponding to a definite energy E. In the TPQS formulation, the coupled system is instead assumed to be in a pure state $\rho$. It is shown in Refs.~\cite{Popescu,Popescu2} using Levy's Lemma that the mixed states of the system S---obtained in both cases by tracing out the environment E---are nearly indistinguishable $\rho_S \approx \Omega_S$. This is founded on properties of high-dimensional normed spaces \cite{Milman} and especially the theorem of measure concentration \cite{Matousek}. How this idea can be used to construct a TPQS at arbitrary temperature is discussed in Refs.~\cite{Sugiura,TPQSatFT}, culminating in the following general definition for a TPQS: Let $\{ |n\rangle\}_{n=1}^K$ be an orthonormal basis of the Hilbert space $\mathcal{H}$ and 
\begin{equation}
|\Psi(\beta=0) \rangle = \sum_{n=1}^K c_n|n\rangle
\label{eq.randstate}
\end{equation}
a normalized pure state with components $c_n=\frac{x_n+iy_n}{\sqrt{2}} \in \mathbb{C}$, where the real parameters $x_n$ and $y_n$ are drawn from the unit normal distribution. This state can be evolved in imaginary time 
\begin{equation}
|\Psi(\beta)\rangle = \exp \left(\frac{-\beta H}{2} \right) |\Psi(0)\rangle \ .
\label{eq.Tstate}
\end{equation} 
The ensemble average and the pure state expectation value of an operator $A$ are then defined by 
\begin{equation}
\langle A \rangle_{\beta}^{\textrm{ens}} = \frac{\tr[ A e^{-\beta H}]}{Z} \quad , \quad \langle A \rangle_{\beta} ^{\textrm{TPQS}} = \langle \Psi | A | \Psi \rangle \ ,
\end{equation}
respectively. The operator A is assumed to be a low-degree polynomial ($\text{degree} \leq m$) of local operators \cite{CTPQS}. If $\forall \epsilon>0$ a function $\eta_N(\epsilon)$ exists with $\eta_N(\epsilon) \xrightarrow[]{N\rightarrow \infty} 0$ such that the probability
\begin{equation}
P \left( |\langle A \rangle_{\beta} ^{TPQS} -\langle A \rangle_{\beta} ^{ens}| \geq \epsilon \right) \leq \eta_N(\epsilon) \ ,
\label{eq.TPQS_approx}
\end{equation}
then $|\Psi\rangle$ is called a TPQS \cite{Sugiura}. One can show that the right hand side is for normally distributed $c_n$ bounded from above by
\begin{equation}
\eta_N(\epsilon) \leq \frac{N^{2m} e^{-\mathcal{O}(N)}}{\epsilon^2} \ .
\label{eq.QT}
\end{equation}
This means that $\eta_N(\epsilon)$ is decreasing exponentially in system size $N$. A similar result is obtained in Ref.~\cite{Gemmer2009} and even extended to dynamical correlation functions \cite{SteinigewegGemmer}. For large system sizes, it is therefore sufficient to consider only a few or even a single pure state. In Ref.~\cite{Reimann}, it was investigated thoroughly which minimum requirements the randomly chosen wave function components of the state $|\Psi(\beta=0) \rangle$ need to fulfill for the state to be a TPQS resulting in the following two constraints: 
\begin{itemize}
\item The components $c_n$ are uncorrelated.
\begin{equation}
    \overline{ c_m^* c_n} = \overline{|c_n|^2} \delta_{n,m}
    \label{eq.Corr}
\end{equation}
\item The mixed state density matrix $\rho$ has low purity
\begin{equation}
\tr \rho^2 = \sum_n \rho_n^2 \ll 1 \ ,
\label{eq.2ndcondition}
\end{equation}
where
\begin{equation}
\rho_n = \overline{|c_n|^2/\sqrt{\sum|c_n|^2}} \ .
\end{equation}
This constraint guarantees that averages are not dominated by just a few large $\rho_n$.
\end{itemize} 
Some probability distributions like the normal and the uniform distribution that fulfill these two conditions have been discussed in Refs.~\cite{Sugiura,CTPQS}. Results for normally distributed wave-function components were obtained within the finite-temperature Lanczos method in Refs.~\cite{Schnack1,Schnack2,Schnack3} which include a detailed analysis of the accuracy of this method. In this paper, we are, however, interested in using a variational ansatz like the NN ansatz to compute expectation values of operators at finite temperatures. Unfortunately, a NN ansatz in general and the RBM ansatz 
\begin{equation}
\Psi_\mathbb{W} (\mathbf{s}) = \ \exp \left( -\sum_{i=1}^{N}a_i s_i \right) \prod_{j=1}^{M} \cosh \left( b_j + \sum_{i=1}^N  W_{ij}s_i \right)
\label{eq.RBM}
\end{equation}
in particular are problematic starting points. The non-linear activation function leads, in general, to a small number of amplitudes $|\Psi(s)|$ dominating the wave function, thus violating the second condition stated above for a typical state. 
Additionally, it was proven in Ref. \cite{Page} and discussed in Refs.~\cite{Hackl-Volume-Law,Sugiura-Nature} that TPQS have volume-law entanglement ruling out the METS-NN as possible candidates.

\subsection{Pair-Product (PP) Wave Function}
To fulfill all the mentioned restrictions for a possible variational TPQS, the NN (RBM) is complemented by a Pair-Product (PP) wave function component
\begin{equation}
|\Phi_{PP}\rangle = P_G^{\infty} \left[ \sum_{i,j=0}^{N}\sum_{\sigma, \sigma' = \uparrow, \downarrow} F_{ij}^{\sigma, \sigma'} c_{i, \sigma}^\dagger c_{j, \sigma'}^\dagger \right] ^{\frac{N_e}{2}} |0\rangle \ ,
\label{eq.Pf}
\end{equation}
where $P_G^{\infty} = \prod_i^N (1-n_{i, \uparrow}n_{i, \downarrow} )$ is the Gutzwiller factor prohibiting double occupancy, $c_{i,\sigma}^{\dagger}$ ($c_{i,\sigma}$) the creation (annihilation) operator of a fermion with spin $\sigma$, $F_{ij}^{\sigma,\sigma'}$ are complex parameters, $n_{i, \sigma} = c_{i,\sigma}^{\dagger} c_{i,\sigma}$ the occupation number operator and $N_e$ is the number of electrons which must be even (see Refs.~\cite{Pfaffian1,PP-NN1}). For the Heisenberg model, this translates to $N_e = N$. It should be mentioned that the PP wave function can be generalized to an odd number of electrons $N_e$ \cite{Pfaffian3}, however, since we are interested in expectation values in the thermodynamic limit, it is more convenient to consider spin chains of even length $N$. The PP-ansatz stems from more general considerations of variational wave functions for lattice models in the fermionic formulation \cite{Pfaffian2}. The two approaches are easily combined since $s_{i}^z = \frac{1}{2} (n_{i,\uparrow}- n_{i,\downarrow})$ and there is no double occupation allowed here. This leads to the following PP-NN-ansatz for the overall wave function
\begin{equation}
|\Psi\rangle = \sum_\mathbf{s} \underbrace{\Psi_{\mathbb{W}} (\mathbf{s}) \Phi_{PP}(\mathbf{s})}_{\Psi(s)} |\mathbf{s}\rangle \ .
\label{eq.PP-NN}
\end{equation}
This ansatz was considered in Ref.~\cite{PP-NN1} to improve the accuracy of the ground state of the Heisenberg model and later used to compute the phase diagram of the $J_1-J_2$ square lattice Heisenberg model \cite{PP-NN2}.
The overlap of the PP wave function with a real space configuration $|\boldsymbol{\sigma} \rangle = c_{1,\sigma_1}\cdot \cdot \cdot c_{N,\sigma_N} |0\rangle$ is given by the Pfaffian
\begin{equation}
\Phi_{PP}(\boldsymbol{\sigma}) = (N_e/2)! \ Pf(X[\boldsymbol{\sigma}])
\end{equation}
with the skew-symmetric matrix
\begin{equation}
X_{ij}^{\sigma_i\sigma_j} = F_{ij}^{\sigma_i \sigma_j} - F_{ji}^{\sigma_j \sigma_i} \ ,
\end{equation}
and
\begin{equation}
Pf(X[\boldsymbol{\sigma}]) = \frac{1}{2^{N/2} (N/2)!} \sum_{\alpha \in S_N} \text{sgn}(\alpha) \prod_{i=1}^{N/2} X_{\alpha(2i-1) \alpha(2i)}[\boldsymbol{\sigma}] \ .
\label{eq.Pfaff}
\end{equation}
The real and imaginary parts of the initial variational parameters ${X_{ij}^{\sigma_i \sigma_j} = F_{ij}^{\sigma_i \sigma_j} - F_{ji}^{\sigma_j \sigma_i}}$ are chosen randomly from a normal distribution centered around zero. This means that there are $4$ parameter arrays ($X^{\uparrow \uparrow}$, $X^{\uparrow \downarrow}$, $X^{\downarrow \uparrow}$, $X^{\downarrow \downarrow}$) each containing $(N^2-N)/2$ different parameters and that the initial values for the amplitudes $\Phi_{PP}(\boldsymbol{\sigma})$ are also randomly distributed around zero. We also consider a state, called PP-NN-local in the following, where the projected pairs are limited to nearest neighbors only ($F_{ij}^{\sigma,\sigma'} =0$ if $|i-j|>1$).\\
If one chooses the initial parameters of the RBM to be zero, or at least for numerical convenience so small that the variance of the NN part of the wave function is small compared to the one
of the PP part, the infinite-temperature state is solely described by the PP part. The correlation of the wave functions of two different spin configurations in the PP part is computed to be $\overline{\Phi_{PP}(\boldsymbol{\sigma}_k) \Phi_{PP}(\boldsymbol{\sigma}_m) }=0$ since two configurations differ in every term of the Pfaffian by at least one random variable $X_{ij}^{\sigma_i \sigma_j}$. If the number of parameters is smaller than the Hilbert space dimension, non-zero higher order correlations will exist. However, we expect that such correlations have no influence on the expectation values of local observables. I.e., we expect that the difference in expectation values obtained by an actual TPQS given, for example, by normally distributed $c_n$ and those obtained by the variational PP-NN ansatz in Eq.~\eqref{eq.PP-NN} will be small if a sufficient number of parameters is kept.

\subsection{The Infinite-Temperature TPQS}
\label{subsec.The infinite temperature TPQS}
First, we check if the states chosen according to the PP-NN ansatz, see Eqs.~\eqref{eq.PP-NN}-\eqref{eq.Pfaff}, are good approximations of an infinite-temperature TPQS. As comparison, the normally distributed states chosen according to Eq.~\eqref{eq.randstate} and the METS-NN states chosen according to the NN ansatz, see Eqs.~\eqref{eq.ProductState}-\eqref{eq.METS-a}, are used. Second, the entanglement entropy of those states as a measure of the approximation complexity is also computed.

\subsubsection{Comparison of Typicality Features}
\label{sec.CompareTypicality}
\begin{figure}[!htp]
    \centering
    \includegraphics*[width=0.99\columnwidth]{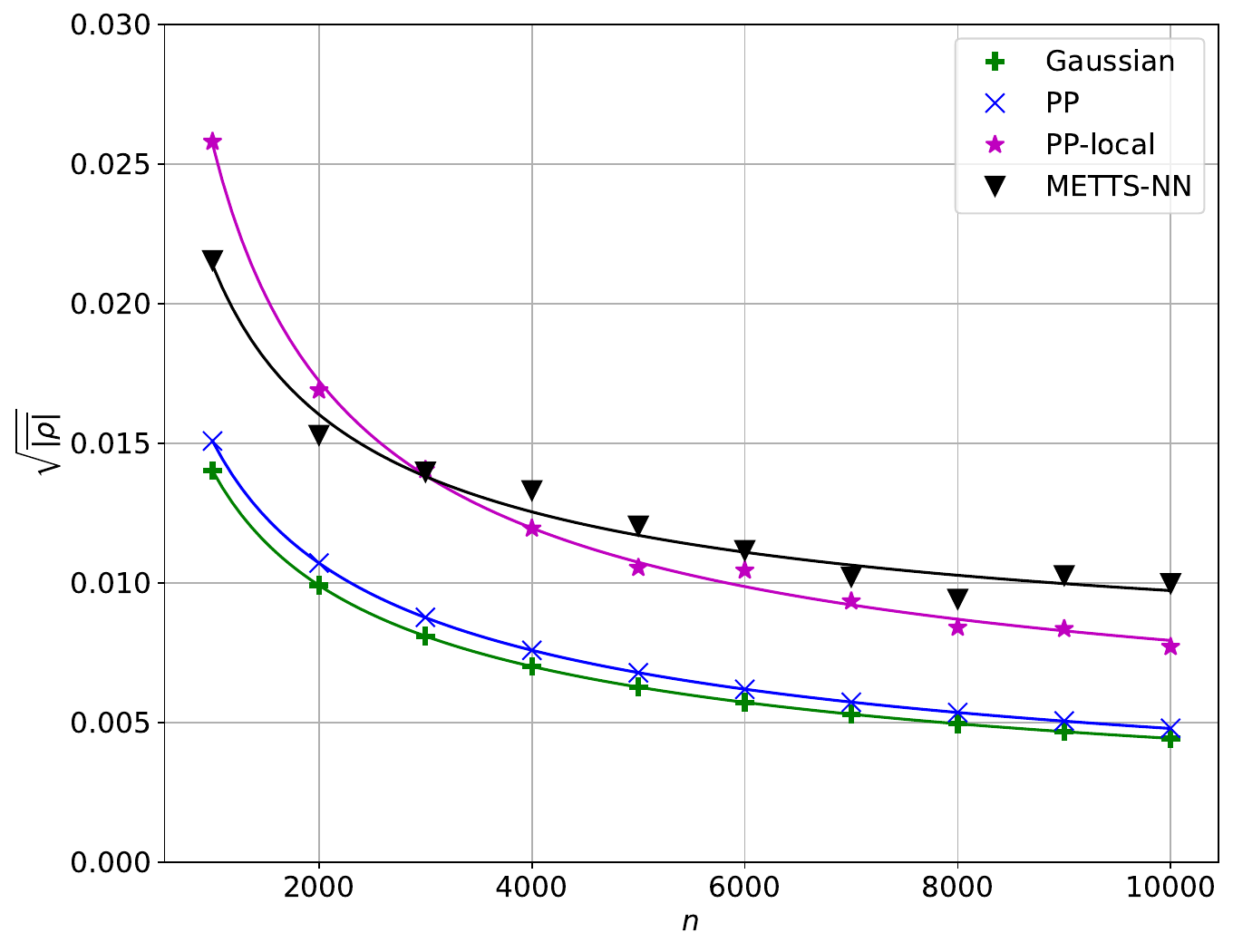}
    \caption{Square root of the mean of the absolute of the pairwise Pearson correlation coefficients of the wave function components for $n$ different sets of parameters and a spin chain of length $N=12$.}
    \label{fig.Correlation}
\end{figure}

\begin{figure}[!htp]
    \centering
    \includegraphics*[width=0.99\columnwidth]{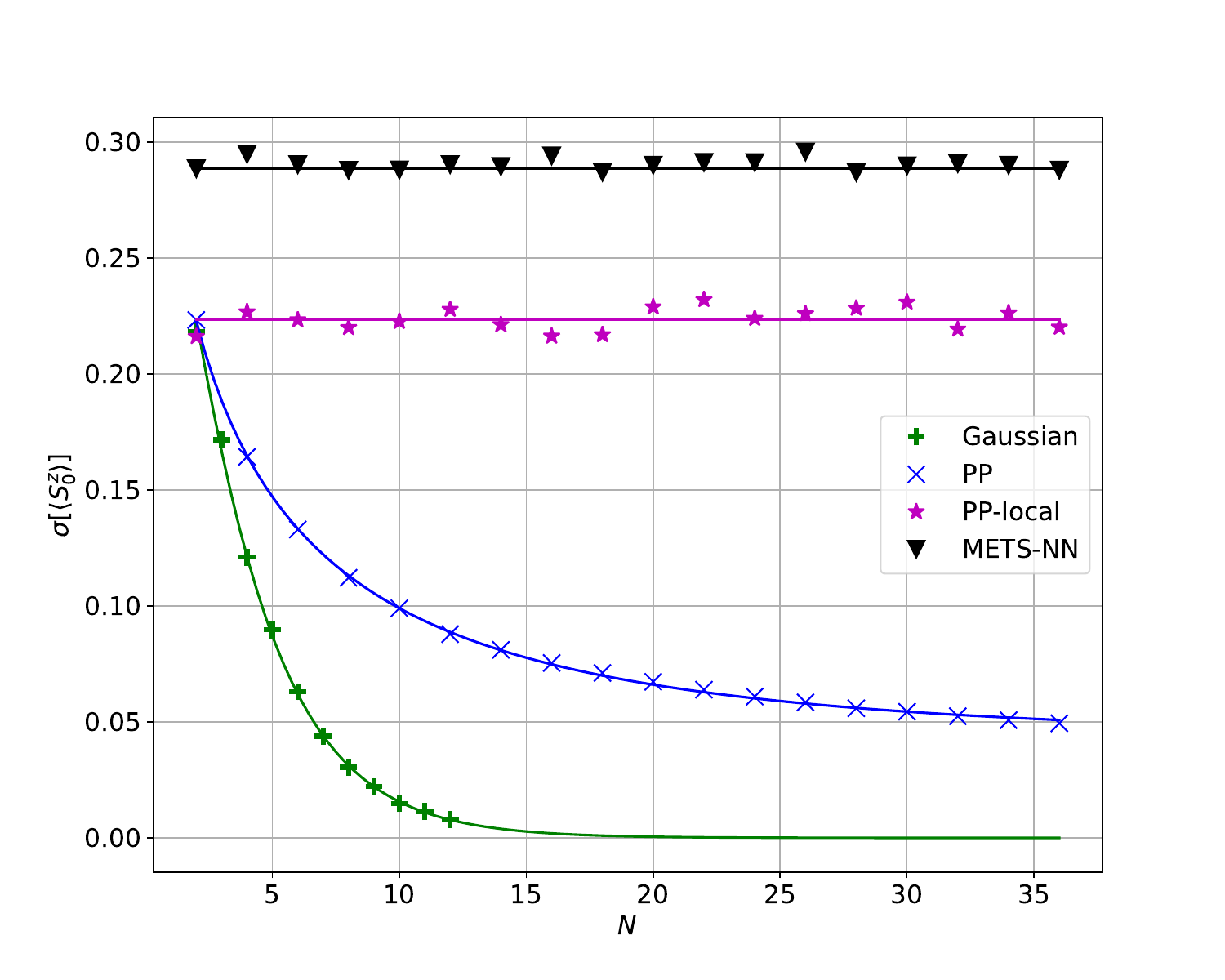}
    \caption{Standard deviation of the operator $\langle S_0^z \rangle$ computed using $10^4$ Metropolis samples and $n=10^3$ different sets of parameters chosen as discussed above.}
    \label{fig.Sz-std}
\end{figure}

It is not a priori clear to which extent the conditions for a TPQS, discussed at the beginning of Sec.~\ref{sec.Quantum Typicality}, are fulfilled by the PP-NN ansatz. Additionally, apart from the entanglement entropy, it is not clear why the METS-NN states do not describe a TPQS.  We investigate first to what extent Eq.~\eqref{eq.Corr} is fulfilled compared to a TPQS with normally distributed random numbers for the wave function components - from here on referred to as the Gaussian case.
It is emphasized that in the PP-NN case the real and imaginary parts of all the NN parameters are not chosen as described in Sec.~\ref{sec.METS} but from a normal distribution centered around zero with $\sigma_{NN} = 1/N$ while those of the PP parameters are always chosen from a normal distribution centered around zero with $\sigma_{PP} = ({N/2})!\cdot \frac{2^{N/2}}{(N!)^{2/N}}$ to keep the variance of both wave function components (see Eq.~\eqref{eq.PP-NN}) constant for growing system size. This way, the NN parameters in the PP-NN ansatz do not contribute to the properties of the infinite-temperature state but are only relevant for the numerics of the imaginary-time evolution. Thus, those parameters can be neglected in the following discussion. To check the validity of Eq.~\eqref{eq.Corr}, the square root of the mean of the absolute value of the pairwise Pearson correlation coefficients is compared in Fig.~\ref{fig.Correlation} for varying number of states $n$. The full Hilbert space is used for a spin chain of length $N=12$. Of course, the diagonal elements are excluded. The results for the Gaussian case show as expected a $n^{-1/2}$ behavior as is checked by a respective fit. While this also holds for the PP ansatz up to an overall constant, the PP-local and the METS-NN ansatz differ significantly from this behavior. \\
To check what this means for the distribution of expectation values, the standard deviation of $\langle S_0^z \rangle$ is computed for $n=10^3$ sets of parameters $\mathbb{W}$ and varying length $N$ in Fig.~\ref{fig.Sz-std}. The expectation values are obtained for every set of parameters by using $10^4$ Metropolis samples. Since the Gaussian case reaches values affected by the sampling error, the full Hilbert space is used. While the Gaussian case shows exponential decay to the infinite-temperature result, the data for the PP ansatz can be fitted very well by a function that is proportional to $e^{-a\sqrt{N}}$, where $a$ is a fitting parameter. However, the METS-NN and the PP-local ansatz show no decay with system size at all since they are product states and thus only depend on the parameters connected to the 1st spin. This is as well as the result for the Gaussian case analytically verified by computing the probability distribution of $\langle S_0^z \rangle$ which is in all those three cases given by
\begin{equation}
    2\langle S_0^z \rangle = \frac{A}{A+B} - \frac{B}{A+B} \ ,
    \label{eq.S0z}
\end{equation}
where
\begin{equation}
    A= \sum_k^{K/2} |z_k^{\uparrow}|^2 \qquad B=\sum_k^{K/2} |z_k^{\downarrow}|^2
\end{equation}
and the real and imaginary part of $z_k^{\uparrow}$ and $z_k^{\downarrow}$ are chosen randomly from a normal distribution. The actual variance of this normal distribution is irrelevant in this case because it cancels out and is thus chosen to be one. Therefore, $A$ and $B$ follow a $\chi_K^2$ distribution. Each term in Eq.\eqref{eq.S0z} is computed to follow a probability distribution $p(y)$ with $y \in [0,1]$. These, the respective value for $K$ and the resulting variances are described in Tab.~\ref{Tab} and plotted as solid lines in Fig.~\ref{fig.Sz-std}.
\begin{table}[!htbp]
\begin{center}
\resizebox{0.99\columnwidth}{!}{
\begin{tabular}{c|c|c|c} 
 Ansatz  &K& $p(y)$ & $\sigma[\langle S_0^z\rangle ]  $ \\ 
  \hline
 Gaussian  & $2^N$& $\frac{\Gamma(2^N)}{\Gamma^2(2^{N-1})} y^{2^{N-1}-1} (1-y)^{2^{N-1}-1}$ & $\sqrt{\frac{2^{N-1}+1}{2^{N+1}+2} -\frac{1}{4}}$  \\  
 PP-local  & 4&  $6y(1-y)$ & $\frac{1}{\sqrt{20}}$\\ 
 METS-NN  & 2&  $1$ & $\frac{1}{\sqrt{12}}$\\ 
\end{tabular}
}
\end{center}
\caption{Probability distributions and variances for the different
ansatz states.}
\label{Tab}
\end{table}
The promising relations for the Gaussian case in Eqs.~\eqref{eq.TPQS_approx} and \eqref{eq.QT} thus do not translate to the discussed variational wave functions with a finite number of parameters. While the exponential decay is weakened for the PP ansatz, there is no decay at all with the METS-NN and the PP-local ansatz. For operators like the Hamiltonian in Eq.\eqref{XXZ}, there is still a decay of the standard deviation in all cases for larger spin chains due to the benefit of averaging over local spin-spin operators along the chain. Naturally, the results for the mean of the expectation value converge in all cases to the infinite temperature value with increasing number of parameter sets (see Ref.\cite{deRaedt} and the discussion in Sec.\ref{sec.METS}). 

\subsubsection{Entanglement Entropy of the Infinite-Temperature TPQS}
\begin{figure}[!htp]
    \centering
   \includegraphics*[width=0.99\columnwidth]{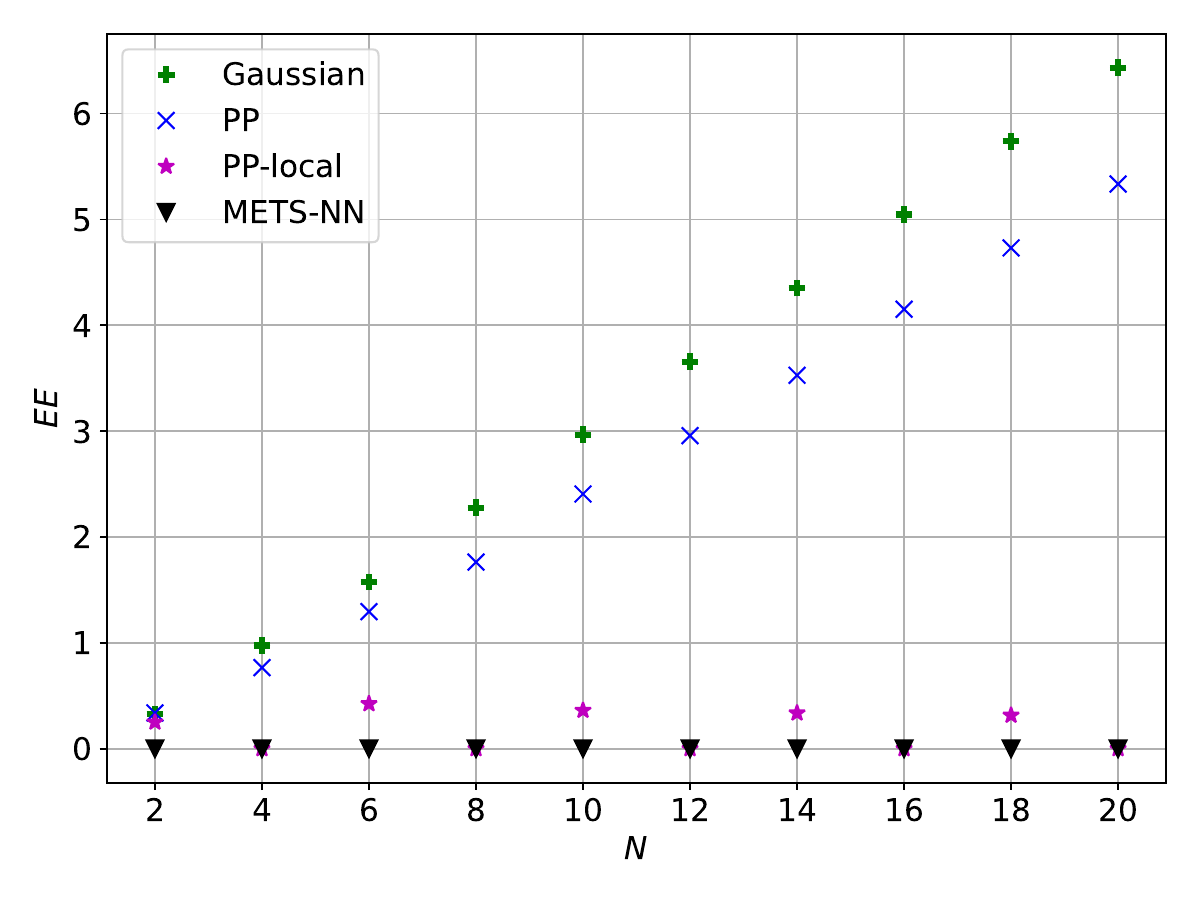}
    \caption{Entanglement entropy of the PP, the PP-local and the METS-NN initial states compared to the Gaussian case for different chain lengths of an open Heisenberg chain.}
    \label{fig.EE}
\end{figure}

Significant differences between the ansatz wave functions become apparent when examining the entanglement entropy in Eq.~\eqref{Sent}, where the system is divided equally in the middle into the subsystems $A$ and $B$. While the PP-NN ansatz shows a volume law scaling due to the PP part of the wave function, the METS-NN ansatz shows no entanglement because the wave function is a product state. The different scaling with the length of the chain is illustrated for the full Hilbert space in Fig.~\ref{fig.EE}. This is very relevant since states with volume law entanglement can in general not be efficiently represented. The RBM is only capable of doing so for some specific states \cite{DengLiDasSarma}. Thus, the METS-NN ansatz defines an initial state that is much easier to approximate. To allow for a similar state in the PP-NN ansatz, the PP-NN-local ansatz is used. This ansatz shows an entanglement entropy that is only nonzero if there is a PP pair that crosses the cut in the middle of the spin chain. This is only the case for chain lengths for which $N/2$ is odd. By increasing the support of the PP ansatz, one can obtain states whose properties lie in between the blue and magenta curves in Figs.~\ref{fig.Sz-std} and \ref{fig.EE}.

\subsection{The variational imaginary-time evolution}
\begin{figure}[!htp]
    \centering
   \includegraphics*[width=0.99\columnwidth]{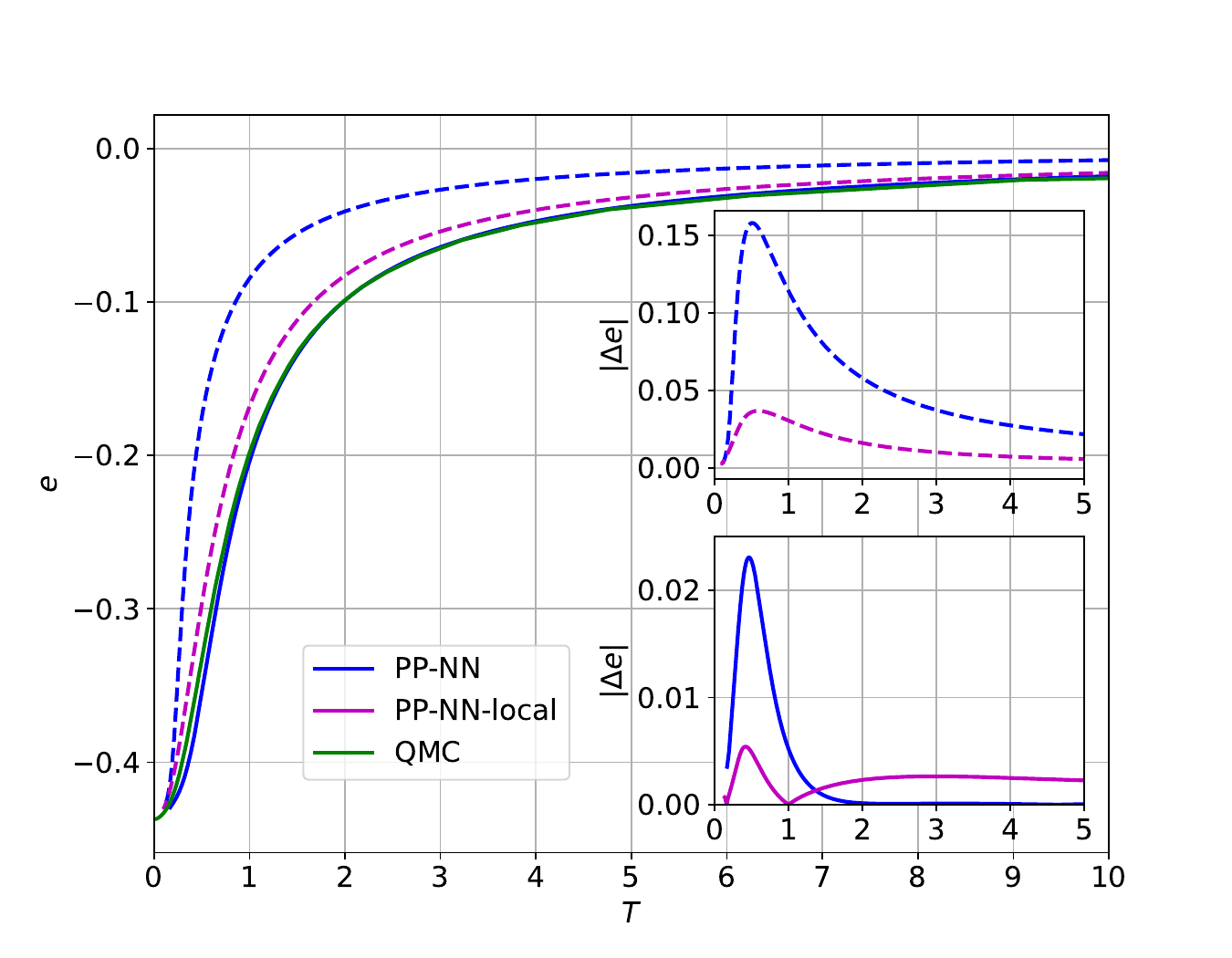}
    \caption{Inner energy of the PP-NN and the PP-NN-local ansatz compared to the QMC solution for an open Heisenberg chain of length $N=30$ using $10^4$ Metropolis samples, $R=10^3$ and $\epsilon = 10^{-6}$. The average is taken over $30$ initial states. The dashed lines show the data without $\beta$ modification. The PP-NN-local results with $\beta$ modification are omitted in the main plot because they are not distinguishable from the QMC results on this scale.}
    \label{fig.ITE}
\end{figure}

\begin{figure}[!htp]
    \centering
   \includegraphics*[width=0.99\columnwidth]{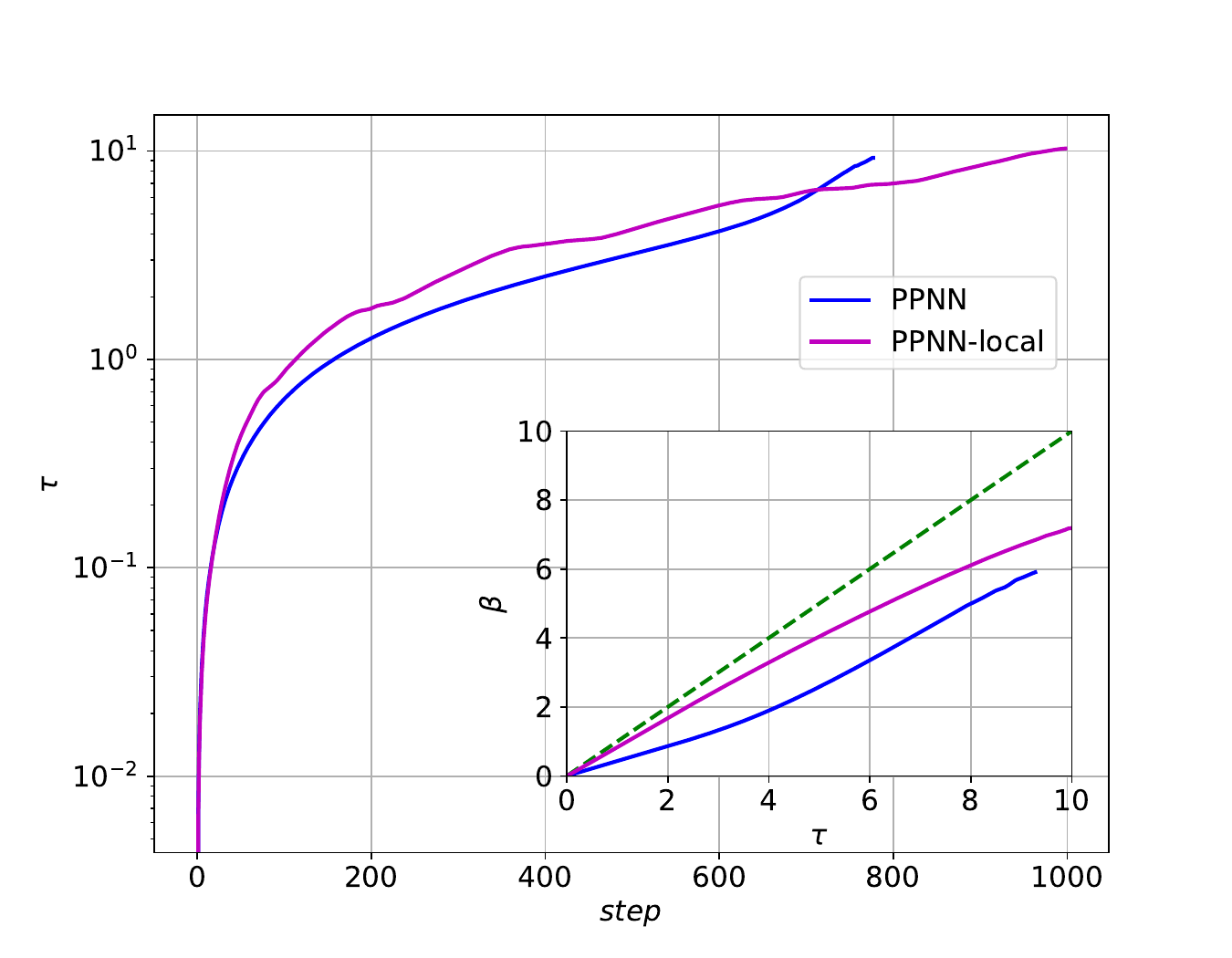}
    \caption{Imaginary time $\tau$ in the Heun algorithm vs number of iterative steps for the PP-NN and the PP-NN-local ansatz for an open Heisenberg chain of length $N=30$ and parameters defined in Fig.~\ref{fig.ITE}. The $\beta$ modification is shown in the inset.}
    \label{fig.ImaginaryTime}
\end{figure}

To compare the PP-NN and the PP-NN-local ansatz to the previous methods, we average in each case over $30$ initial states and time evolve these according to Eq.~\eqref{var_principle3}. The averaging is still useful due to the fact that the error is still dominated by the lack of true quantum typicality over the sampling error (see Fig.~\ref{fig.Sz-std}). We do so again for a Heisenberg chain of length $N=30$. The results are shown in Fig.~\ref{fig.ITE} and compared to the same QMC results as before. The number of samples and the number of iteration steps $R$ is adjusted due to the larger computational cost for the updates of the PP parameters. For the computation of the Pfaffian after a spin flip, an effective update rule based on Cayley's identity is used \cite{Pfaffian1}.
The energy values for high temperatures are more accurate for the PP-NN ansatz, as is expected from the discussion in Sec.~\ref{sec.CompareTypicality}. However, during the course of the evolution, the PP-NN ansatz states tend to deviate further from the QMC results. The maximum deviation is roughly at the point where the heat capacity is maximal. The most elaborate ansatz with the largest number of parameters produces the most inaccurate results. This is a consequence of the large entanglement of the initial states $|\Psi(0)\rangle$ in the PP-NN ansatz. It is therefore more difficult to approximate the typical states and the ones in their vicinity. It is stressed that this is not due to the smaller number of samples, as is evident from the comparison to the PP-NN-local ansatz with zero entanglement entropy. This results in a much better approximation. The difference to the METS-NN ansatz is relatively small. Summarizing, there is no numerical benefit of using TPQS's in the PP-NN ansatz. The theoretical benefit is eaten up by the larger entanglement entropy, leading to a trajectory in the variational manifold that includes states that are harder to represent. It might be possible to find an ansatz that can represent the highly entangled states on the trajectory in special cases, for example for a certain model. While it seems unlikely that this can be done in general, DNNs could possibly lead to a better representation of such states and thus alleviate this issue to some degree \cite{DNN-Efficient}. \\
The modification of the imaginary time shown in Fig.~\ref{fig.ImaginaryTime} illustrates the issue with the variational TPQS in a different way. Especially for high temperatures, the imaginary-time evolution in the PP-NN ansatz shows a very large $\beta$ modification, which means that the proper evolution and the evolution projected onto the variational manifold deviate strongly. Although this feature turns somewhat around for lower temperatures, the PP-NN ansatz states appear not to be a good starting point due to their large initial entanglement. On the other hand, the error of the time evolution after the $\beta$ modification is applied to the PP-NN-local states is of the order $10^{-3}$ which is thus again consistent with the Monte Carlo sampling error.


\section{Conclusions}
\label{Concl}
We have shown that the energy of an open Heisenberg chain can be computed for a large temperature range with an accuracy limited by the sampling error (errors of the order of $10^{-3}$ for an $N=30$ site chain) with  both the purification and the sampling approach. The results for the free energy are more sensitive and can be computed with the purification method with an accuracy which is roughly one order of magnitude worse than for the energy. While we have been concerned with a proof-of-principle investigation of these algorithms here, there are no fundamental obstacles in scaling up both algorithms to larger system sizes.\\
In the quantum typicality approach, there are two competing requirements that could not be fulfilled: The state should be quantum typical and all states on the trajectory defined by the evolution in imaginary time should be efficiently represented by the variational ansatz. While the PP-NN ansatz fulfills the former in a weaker sense, it fails to meet the latter to an extent that the $\beta$ modification cannot fully correct. As discussed in Sec.~\ref{sec.Quantum Typicality}, it seems questionable if deep network structures can fully cure these issues. Additionally, the computational cost of the combined PP-NN ansatz is, of course, considerably higher than the variational wave functions used in the sampling method. Not only because of the much larger number of parameters, but also because of the computation of the spin update for the matrices $X$ in Eq.\eqref{eq.Pfaff} which is, despite using sophisticated updating techniques \cite{Pfaffian1}, more computationally time-consuming than the update of the NN part. \\
The two most promising methods have distinct up- and downsides with respect to the required computation time. In the purification method, it is sufficient, after making sure that the parameter $c$ is within a certain range, to simulate a single state that is part of a Hilbert space of size $2^{2N}$. In the sampling approach, on the other hand, it is necessary to simulate $\gtrsim 30$ states in a much smaller Hilbert space of size $2^{N}$. Using the fact that these states can be time evolved independently of each other, the computational time is significantly lower when using a high-performance cluster. However, the purification method gives a very accurate approximation of the infinite-temperature state and even allows for an accurate computation of correlation functions. In the sampling method, on the other hand, the number of states needs to be increased considerably to also obtain accurate results for local correlation functions.\\
From the attempt to calculate dynamical correlation functions, one can learn that, while certain specific highly entangled states can be efficiently represented as a NN, this is not true for generic highly entangled states. We observe that the number of hidden units must be increased in a way which is consistent with an exponential scaling with system size. It thus appears unlikely that NNs will significantly improve our abilities to calculate dynamical correlation functions, at least not in one dimension. The most fruitful applications of the algorithms discussed in this paper might be to the thermodynamics of two- and three-dimensional frustrated spin models. Here, the standard QMC approach fails and DMRG calculations are limited to small clusters. In these cases, it seems possible that a NN ansatz using either the purification or the sampling method might outperform both QMC and DMRG and might thus provide further insights into the physics of frustrated spin systems.

\section{Acknowledgements}
The authors acknowledge funding by the Deutsche Forschungsgemeinschaft (DFG) via Research Unit FOR 2316. D.W. gratefully acknowledges the PhD fellowship awarded to him by the University of Wuppertal. J.S. acknowledges support by the Natural Sciences and Engineering Research Council (NSERC, Canada).


\begin{thebibliography}{63}%
\makeatletter
\providecommand \@ifxundefined [1]{%
 \@ifx{#1\undefined}
}%
\providecommand \@ifnum [1]{%
 \ifnum #1\expandafter \@firstoftwo
 \else \expandafter \@secondoftwo
 \fi
}%
\providecommand \@ifx [1]{%
 \ifx #1\expandafter \@firstoftwo
 \else \expandafter \@secondoftwo
 \fi
}%
\providecommand \natexlab [1]{#1}%
\providecommand \enquote  [1]{``#1''}%
\providecommand \bibnamefont  [1]{#1}%
\providecommand \bibfnamefont [1]{#1}%
\providecommand \citenamefont [1]{#1}%
\providecommand \href@noop [0]{\@secondoftwo}%
\providecommand \href [0]{\begingroup \@sanitize@url \@href}%
\providecommand \@href[1]{\@@startlink{#1}\@@href}%
\providecommand \@@href[1]{\endgroup#1\@@endlink}%
\providecommand \@sanitize@url [0]{\catcode `\\12\catcode `\$12\catcode
  `\&12\catcode `\#12\catcode `\^12\catcode `\_12\catcode `\%12\relax}%
\providecommand \@@startlink[1]{}%
\providecommand \@@endlink[0]{}%
\providecommand \url  [0]{\begingroup\@sanitize@url \@url }%
\providecommand \@url [1]{\endgroup\@href {#1}{\urlprefix }}%
\providecommand \urlprefix  [0]{URL }%
\providecommand \Eprint [0]{\href }%
\providecommand \doibase [0]{https://doi.org/}%
\providecommand \selectlanguage [0]{\@gobble}%
\providecommand \bibinfo  [0]{\@secondoftwo}%
\providecommand \bibfield  [0]{\@secondoftwo}%
\providecommand \translation [1]{[#1]}%
\providecommand \BibitemOpen [0]{}%
\providecommand \bibitemStop [0]{}%
\providecommand \bibitemNoStop [0]{.\EOS\space}%
\providecommand \EOS [0]{\spacefactor3000\relax}%
\providecommand \BibitemShut  [1]{\csname bibitem#1\endcsname}%
\let\auto@bib@innerbib\@empty
\bibitem [{\citenamefont {Gubernatis}\ \emph {et~al.}(2016)\citenamefont
  {Gubernatis}, \citenamefont {Kawashima},\ and\ \citenamefont
  {Werner}}]{GubernatisQMC}%
  \BibitemOpen
  \bibfield  {author} {\bibinfo {author} {\bibfnamefont {J.}~\bibnamefont
  {Gubernatis}}, \bibinfo {author} {\bibfnamefont {N.}~\bibnamefont
  {Kawashima}},\ and\ \bibinfo {author} {\bibfnamefont {P.}~\bibnamefont
  {Werner}},\ }\href {https://doi.org/10.1017/CBO9780511902581} {\emph
  {\bibinfo {title} {Quantum Monte Carlo Methods: Algorithms for Lattice
  Models}}}\ (\bibinfo  {publisher} {Cambridge University Press},\ \bibinfo
  {year} {2016})\BibitemShut {NoStop}%
\bibitem [{\citenamefont {Sandvik}(2019)}]{SandvikSSE}%
  \BibitemOpen
  \bibfield  {author} {\bibinfo {author} {\bibfnamefont {A.}~\bibnamefont
  {Sandvik}},\ }\href {https://juser.fz-juelich.de/record/864818} {\emph
  {\bibinfo {title} {Many-Body Methods for Real Materials}}},\ edited by\
  \bibinfo {editor} {\bibfnamefont {E.}~\bibnamefont {Pavarini}}, \bibinfo
  {editor} {\bibfnamefont {E.}~\bibnamefont {Koch}},\ and\ \bibinfo {editor}
  {\bibfnamefont {S.}~\bibnamefont {Zhang}}\ (\bibinfo  {publisher}
  {Forschungszentrum Julich},\ \bibinfo {year} {2019})\BibitemShut {NoStop}%
\bibitem [{\citenamefont {Nightingale}\ and\ \citenamefont
  {Umrigar}(1999)}]{Nightingale1999QuantumMC}%
  \BibitemOpen
  \bibfield  {author} {\bibinfo {author} {\bibfnamefont {M.~P.}\ \bibnamefont
  {Nightingale}}\ and\ \bibinfo {author} {\bibfnamefont {C.~J.}\ \bibnamefont
  {Umrigar}},\ }\href
  {https://link.springer.com/book/9780792355519?utm_medium=referral&utm_source=google_books&utm_campaign=3_pier05_buy_print&utm_content=en_08082017}
  {\emph {\bibinfo {title} {Quantum {M}onte {C}arlo methods in {p}hysics and
  {c}hemistry}}}\ (\bibinfo  {publisher} {Springer Dordrecht},\ \bibinfo {year}
  {1999})\BibitemShut {NoStop}%
\bibitem [{\citenamefont {Becca}\ and\ \citenamefont
  {Sorella}(2017)}]{becca_sorella_2017}%
  \BibitemOpen
  \bibfield  {author} {\bibinfo {author} {\bibfnamefont {F.}~\bibnamefont
  {Becca}}\ and\ \bibinfo {author} {\bibfnamefont {S.}~\bibnamefont
  {Sorella}},\ }\href {https://doi.org/10.1017/9781316417041} {\emph {\bibinfo
  {title} {Quantum Monte Carlo Approaches for Correlated Systems}}}\ (\bibinfo
  {publisher} {Cambridge University Press},\ \bibinfo {year}
  {2017})\BibitemShut {NoStop}%
\bibitem [{\citenamefont {Silver}\ \emph {et~al.}(1990)\citenamefont {Silver},
  \citenamefont {Sivia},\ and\ \citenamefont {Gubernatis}}]{SilverSivia}%
  \BibitemOpen
  \bibfield  {author} {\bibinfo {author} {\bibfnamefont {R.~N.}\ \bibnamefont
  {Silver}}, \bibinfo {author} {\bibfnamefont {D.~S.}\ \bibnamefont {Sivia}},\
  and\ \bibinfo {author} {\bibfnamefont {J.~E.}\ \bibnamefont {Gubernatis}},\
  }\bibfield  {title} {\bibinfo {title} {Maximum-entropy method for analytic
  continuation of quantum {M}onte {C}arlo data},\ }\href
  {https://doi.org/10.1103/PhysRevB.41.2380} {\bibfield  {journal} {\bibinfo
  {journal} {Phys. Rev. B}\ }\textbf {\bibinfo {volume} {41}},\ \bibinfo
  {pages} {2380} (\bibinfo {year} {1990})}\BibitemShut {NoStop}%
\bibitem [{\citenamefont {Schollw\"ock}(2005)}]{Schollwock_review}%
  \BibitemOpen
  \bibfield  {author} {\bibinfo {author} {\bibfnamefont {U.}~\bibnamefont
  {Schollw\"ock}},\ }\bibfield  {title} {\bibinfo {title} {The density-matrix
  renormalization group},\ }\href {https://doi.org/10.1103/RevModPhys.77.259}
  {\bibfield  {journal} {\bibinfo  {journal} {Rev. Mod. Phys.}\ }\textbf
  {\bibinfo {volume} {77}},\ \bibinfo {pages} {259} (\bibinfo {year}
  {2005})}\BibitemShut {NoStop}%
\bibitem [{\citenamefont {Schollwöck}(2011)}]{Schollwock_MPS_review}%
  \BibitemOpen
  \bibfield  {author} {\bibinfo {author} {\bibfnamefont {U.}~\bibnamefont
  {Schollwöck}},\ }\bibfield  {title} {\bibinfo {title} {The density-matrix
  renormalization group in the age of matrix product states},\ }\href
  {https://doi.org/https://doi.org/10.1016/j.aop.2010.09.012} {\bibfield
  {journal} {\bibinfo  {journal} {Annals of Physics}\ }\textbf {\bibinfo
  {volume} {326}},\ \bibinfo {pages} {96} (\bibinfo {year} {2011})},\ \bibinfo
  {note} {january 2011 Special Issue}\BibitemShut {NoStop}%
\bibitem [{\citenamefont {Wang}\ and\ \citenamefont {Xiang}(1997)}]{WangXiang}%
  \BibitemOpen
  \bibfield  {author} {\bibinfo {author} {\bibfnamefont {X.}~\bibnamefont
  {Wang}}\ and\ \bibinfo {author} {\bibfnamefont {T.}~\bibnamefont {Xiang}},\
  }\bibfield  {title} {\bibinfo {title} {Transfer-matrix density-matrix
  renormalization-group theory for thermodynamics of one-dimensional quantum
  systems},\ }\href {https://doi.org/10.1103/PhysRevB.56.5061} {\bibfield
  {journal} {\bibinfo  {journal} {Phys. Rev. B}\ }\textbf {\bibinfo {volume}
  {56}},\ \bibinfo {pages} {5061} (\bibinfo {year} {1997})}\BibitemShut
  {NoStop}%
\bibitem [{\citenamefont {Bursill}\ \emph {et~al.}(1996)\citenamefont
  {Bursill}, \citenamefont {Xiang},\ and\ \citenamefont
  {Gehring}}]{BursillXiang}%
  \BibitemOpen
  \bibfield  {author} {\bibinfo {author} {\bibfnamefont {R.~J.}\ \bibnamefont
  {Bursill}}, \bibinfo {author} {\bibfnamefont {T.}~\bibnamefont {Xiang}},\
  and\ \bibinfo {author} {\bibfnamefont {G.~A.}\ \bibnamefont {Gehring}},\
  }\bibfield  {title} {\bibinfo {title} {The density matrix renormalization
  group for a quantum spin chain at non-zero temperature},\ }\href
  {https://doi.org/10.1088/0953-8984/8/40/003} {\bibfield  {journal} {\bibinfo
  {journal} {Journal of Physics: Condensed Matter}\ }\textbf {\bibinfo {volume}
  {8}},\ \bibinfo {pages} {L583} (\bibinfo {year} {1996})}\BibitemShut
  {NoStop}%
\bibitem [{\citenamefont {Sirker}\ and\ \citenamefont
  {Kl\"umper}(2002{\natexlab{a}})}]{SirkerKluemperPRB}%
  \BibitemOpen
  \bibfield  {author} {\bibinfo {author} {\bibfnamefont {J.}~\bibnamefont
  {Sirker}}\ and\ \bibinfo {author} {\bibfnamefont {A.}~\bibnamefont
  {Kl\"umper}},\ }\bibfield  {title} {\bibinfo {title} {Thermodynamics and
  crossover phenomena in the correlation lengths of the one-dimensional t-{J}
  model},\ }\href {https://doi.org/10.1103/PhysRevB.66.245102} {\bibfield
  {journal} {\bibinfo  {journal} {Phys. Rev. B}\ }\textbf {\bibinfo {volume}
  {66}},\ \bibinfo {pages} {245102} (\bibinfo {year}
  {2002}{\natexlab{a}})}\BibitemShut {NoStop}%
\bibitem [{\citenamefont {Sirker}\ and\ \citenamefont
  {Kl\"umper}(2002{\natexlab{b}})}]{SirkerKluemperEPL}%
  \BibitemOpen
  \bibfield  {author} {\bibinfo {author} {\bibfnamefont {J.}~\bibnamefont
  {Sirker}}\ and\ \bibinfo {author} {\bibfnamefont {A.}~\bibnamefont
  {Kl\"umper}},\ }\bibfield  {title} {\bibinfo {title} {Temperature-driven
  crossover phenomena in the correlation lengths of the one-dimensional t-{J}
  model},\ }\href {https://doi.org/10.1209/epl/i2002-00345-2} {\bibfield
  {journal} {\bibinfo  {journal} {Europhys. Lett.}\ }\textbf {\bibinfo {volume}
  {60}},\ \bibinfo {pages} {262} (\bibinfo {year}
  {2002}{\natexlab{b}})}\BibitemShut {NoStop}%
\bibitem [{\citenamefont {Sirker}\ and\ \citenamefont
  {Kl\"umper}(2005)}]{SirkerKluemperDTMRG}%
  \BibitemOpen
  \bibfield  {author} {\bibinfo {author} {\bibfnamefont {J.}~\bibnamefont
  {Sirker}}\ and\ \bibinfo {author} {\bibfnamefont {A.}~\bibnamefont
  {Kl\"umper}},\ }\bibfield  {title} {\bibinfo {title} {Real-time dynamics at
  finite temperature by the density-matrix renormalization group: A
  path-integral approach},\ }\href {https://doi.org/10.1103/PhysRevB.71.241101}
  {\bibfield  {journal} {\bibinfo  {journal} {Phys. Rev. B}\ }\textbf {\bibinfo
  {volume} {71}},\ \bibinfo {pages} {241101} (\bibinfo {year}
  {2005})}\BibitemShut {NoStop}%
\bibitem [{\citenamefont {Enss}\ and\ \citenamefont
  {Sirker}(2012)}]{EnssSirker}%
  \BibitemOpen
  \bibfield  {author} {\bibinfo {author} {\bibfnamefont {T.}~\bibnamefont
  {Enss}}\ and\ \bibinfo {author} {\bibfnamefont {J.}~\bibnamefont {Sirker}},\
  }\bibfield  {title} {\bibinfo {title} {Light cone renormalization and quantum
  quenches in one-dimensional {H}ubbard models},\ }\href
  {https://doi.org/10.1088/1367-2630/14/2/023008} {\bibfield  {journal}
  {\bibinfo  {journal} {New Journal of Physics}\ }\textbf {\bibinfo {volume}
  {14}},\ \bibinfo {pages} {023008} (\bibinfo {year} {2012})}\BibitemShut
  {NoStop}%
\bibitem [{\citenamefont {Feiguin}\ and\ \citenamefont
  {White}(2005)}]{FeiguinWhite}%
  \BibitemOpen
  \bibfield  {author} {\bibinfo {author} {\bibfnamefont {A.~E.}\ \bibnamefont
  {Feiguin}}\ and\ \bibinfo {author} {\bibfnamefont {S.~R.}\ \bibnamefont
  {White}},\ }\bibfield  {title} {\bibinfo {title} {Finite-temperature density
  matrix renormalization using an enlarged {H}ilbert space},\ }\href
  {https://doi.org/10.1103/PhysRevB.72.220401} {\bibfield  {journal} {\bibinfo
  {journal} {Phys. Rev. B}\ }\textbf {\bibinfo {volume} {72}},\ \bibinfo
  {pages} {220401} (\bibinfo {year} {2005})}\BibitemShut {NoStop}%
\bibitem [{\citenamefont {Karrasch}\ \emph {et~al.}(2012)\citenamefont
  {Karrasch}, \citenamefont {Bardarson},\ and\ \citenamefont
  {Moore}}]{KarraschMoore}%
  \BibitemOpen
  \bibfield  {author} {\bibinfo {author} {\bibfnamefont {C.}~\bibnamefont
  {Karrasch}}, \bibinfo {author} {\bibfnamefont {J.~H.}\ \bibnamefont
  {Bardarson}},\ and\ \bibinfo {author} {\bibfnamefont {J.~E.}\ \bibnamefont
  {Moore}},\ }\bibfield  {title} {\bibinfo {title} {Finite-temperature
  dynamical density matrix renormalization group and the {D}rude weight of
  spin-$1/2$ chains},\ }\href {https://doi.org/10.1103/PhysRevLett.108.227206}
  {\bibfield  {journal} {\bibinfo  {journal} {Phys. Rev. Lett.}\ }\textbf
  {\bibinfo {volume} {108}},\ \bibinfo {pages} {227206} (\bibinfo {year}
  {2012})}\BibitemShut {NoStop}%
\bibitem [{\citenamefont {Karrasch}\ \emph {et~al.}(2013)\citenamefont
  {Karrasch}, \citenamefont {Bardarson},\ and\ \citenamefont
  {Moore}}]{KarraschBardarson}%
  \BibitemOpen
  \bibfield  {author} {\bibinfo {author} {\bibfnamefont {C.}~\bibnamefont
  {Karrasch}}, \bibinfo {author} {\bibfnamefont {J.~H.}\ \bibnamefont
  {Bardarson}},\ and\ \bibinfo {author} {\bibfnamefont {J.~E.}\ \bibnamefont
  {Moore}},\ }\bibfield  {title} {\bibinfo {title} {Reducing the numerical
  effort of finite-temperature density matrix renormalization group
  calculations},\ }\href {https://doi.org/10.1088/1367-2630/15/8/083031}
  {\bibfield  {journal} {\bibinfo  {journal} {New Journal of Physics}\ }\textbf
  {\bibinfo {volume} {15}},\ \bibinfo {pages} {083031} (\bibinfo {year}
  {2013})}\BibitemShut {NoStop}%
\bibitem [{\citenamefont {Babenko}\ \emph {et~al.}(2021)\citenamefont
  {Babenko}, \citenamefont {G\"ohmann}, \citenamefont {Kozlowski},
  \citenamefont {Sirker},\ and\ \citenamefont {Suzuki}}]{BabenkoGohmann}%
  \BibitemOpen
  \bibfield  {author} {\bibinfo {author} {\bibfnamefont {C.}~\bibnamefont
  {Babenko}}, \bibinfo {author} {\bibfnamefont {F.}~\bibnamefont {G\"ohmann}},
  \bibinfo {author} {\bibfnamefont {K.~K.}\ \bibnamefont {Kozlowski}}, \bibinfo
  {author} {\bibfnamefont {J.}~\bibnamefont {Sirker}},\ and\ \bibinfo {author}
  {\bibfnamefont {J.}~\bibnamefont {Suzuki}},\ }\bibfield  {title} {\bibinfo
  {title} {Exact real-time longitudinal correlation functions of the massive
  {XXZ} chain},\ }\href {https://doi.org/10.1103/PhysRevLett.126.210602}
  {\bibfield  {journal} {\bibinfo  {journal} {Phys. Rev. Lett.}\ }\textbf
  {\bibinfo {volume} {126}},\ \bibinfo {pages} {210602} (\bibinfo {year}
  {2021})}\BibitemShut {NoStop}%
\bibitem [{\citenamefont {Göhmann}\ \emph {et~al.}(2022)\citenamefont
  {Göhmann}, \citenamefont {Kozlowski}, \citenamefont {Sirker},\ and\
  \citenamefont {Suzuki}}]{GohmannKozlowski}%
  \BibitemOpen
  \bibfield  {author} {\bibinfo {author} {\bibfnamefont {F.}~\bibnamefont
  {Göhmann}}, \bibinfo {author} {\bibfnamefont {K.~K.}\ \bibnamefont
  {Kozlowski}}, \bibinfo {author} {\bibfnamefont {J.}~\bibnamefont {Sirker}},\
  and\ \bibinfo {author} {\bibfnamefont {J.}~\bibnamefont {Suzuki}},\
  }\bibfield  {title} {\bibinfo {title} {{Spin conductivity of the XXZ chain in
  the antiferromagnetic massive regime}},\ }\href
  {https://doi.org/10.21468/SciPostPhys.12.5.158} {\bibfield  {journal}
  {\bibinfo  {journal} {SciPost Phys.}\ }\textbf {\bibinfo {volume} {12}},\
  \bibinfo {pages} {158} (\bibinfo {year} {2022})}\BibitemShut {NoStop}%
\bibitem [{\citenamefont {White}\ and\ \citenamefont
  {Scalapino}(1998)}]{WhiteScalapinoStripes1}%
  \BibitemOpen
  \bibfield  {author} {\bibinfo {author} {\bibfnamefont {S.~R.}\ \bibnamefont
  {White}}\ and\ \bibinfo {author} {\bibfnamefont {D.~J.}\ \bibnamefont
  {Scalapino}},\ }\bibfield  {title} {\bibinfo {title} {Density matrix
  renormalization group study of the striped phase in the 2{D}
  $\mathit{t}\ensuremath{-}\mathit{J}$ model},\ }\href
  {https://doi.org/10.1103/PhysRevLett.80.1272} {\bibfield  {journal} {\bibinfo
   {journal} {Phys. Rev. Lett.}\ }\textbf {\bibinfo {volume} {80}},\ \bibinfo
  {pages} {1272} (\bibinfo {year} {1998})}\BibitemShut {NoStop}%
\bibitem [{\citenamefont {Zheng}\ \emph {et~al.}(2017)\citenamefont {Zheng},
  \citenamefont {Chung}, \citenamefont {Corboz}, \citenamefont {Ehlers},
  \citenamefont {Qin}, \citenamefont {Noack}, \citenamefont {Shi},
  \citenamefont {White}, \citenamefont {Zhang},\ and\ \citenamefont
  {Chan}}]{Zheng1155}%
  \BibitemOpen
  \bibfield  {author} {\bibinfo {author} {\bibfnamefont {B.-X.}\ \bibnamefont
  {Zheng}}, \bibinfo {author} {\bibfnamefont {C.-M.}\ \bibnamefont {Chung}},
  \bibinfo {author} {\bibfnamefont {P.}~\bibnamefont {Corboz}}, \bibinfo
  {author} {\bibfnamefont {G.}~\bibnamefont {Ehlers}}, \bibinfo {author}
  {\bibfnamefont {M.-P.}\ \bibnamefont {Qin}}, \bibinfo {author} {\bibfnamefont
  {R.~M.}\ \bibnamefont {Noack}}, \bibinfo {author} {\bibfnamefont
  {H.}~\bibnamefont {Shi}}, \bibinfo {author} {\bibfnamefont {S.~R.}\
  \bibnamefont {White}}, \bibinfo {author} {\bibfnamefont {S.}~\bibnamefont
  {Zhang}},\ and\ \bibinfo {author} {\bibfnamefont {G.~K.-L.}\ \bibnamefont
  {Chan}},\ }\bibfield  {title} {\bibinfo {title} {Stripe order in the
  underdoped region of the two-dimensional {H}ubbard model},\ }\href
  {https://doi.org/10.1126/science.aam7127} {\bibfield  {journal} {\bibinfo
  {journal} {Science}\ }\textbf {\bibinfo {volume} {358}},\ \bibinfo {pages}
  {1155} (\bibinfo {year} {2017})}\BibitemShut {NoStop}%
\bibitem [{\citenamefont {Carleo}\ and\ \citenamefont
  {Troyer}(2017)}]{TroyerScience}%
  \BibitemOpen
  \bibfield  {author} {\bibinfo {author} {\bibfnamefont {G.}~\bibnamefont
  {Carleo}}\ and\ \bibinfo {author} {\bibfnamefont {M.}~\bibnamefont
  {Troyer}},\ }\bibfield  {title} {\bibinfo {title} {Solving the quantum
  many-body problem with artificial neural networks},\ }\href
  {https://doi.org/10.1126/science.aag2302} {\bibfield  {journal} {\bibinfo
  {journal} {Science}\ }\textbf {\bibinfo {volume} {355}},\ \bibinfo {pages}
  {602} (\bibinfo {year} {2017})}\BibitemShut {NoStop}%
\bibitem [{\citenamefont {Carleo}\ \emph {et~al.}(2019)\citenamefont {Carleo},
  \citenamefont {Cirac}, \citenamefont {Cranmer}, \citenamefont {Daudet},
  \citenamefont {Schuld}, \citenamefont {Tishby}, \citenamefont
  {Vogt-Maranto},\ and\ \citenamefont {Zdeborov\'a}}]{CarleoCirac}%
  \BibitemOpen
  \bibfield  {author} {\bibinfo {author} {\bibfnamefont {G.}~\bibnamefont
  {Carleo}}, \bibinfo {author} {\bibfnamefont {I.}~\bibnamefont {Cirac}},
  \bibinfo {author} {\bibfnamefont {K.}~\bibnamefont {Cranmer}}, \bibinfo
  {author} {\bibfnamefont {L.}~\bibnamefont {Daudet}}, \bibinfo {author}
  {\bibfnamefont {M.}~\bibnamefont {Schuld}}, \bibinfo {author} {\bibfnamefont
  {N.}~\bibnamefont {Tishby}}, \bibinfo {author} {\bibfnamefont
  {L.}~\bibnamefont {Vogt-Maranto}},\ and\ \bibinfo {author} {\bibfnamefont
  {L.}~\bibnamefont {Zdeborov\'a}},\ }\bibfield  {title} {\bibinfo {title}
  {Machine learning and the physical sciences},\ }\href
  {https://doi.org/10.1103/RevModPhys.91.045002} {\bibfield  {journal}
  {\bibinfo  {journal} {Rev. Mod. Phys.}\ }\textbf {\bibinfo {volume} {91}},\
  \bibinfo {pages} {045002} (\bibinfo {year} {2019})}\BibitemShut {NoStop}%
\bibitem [{\citenamefont {Carleo}\ \emph {et~al.}(2018)\citenamefont {Carleo},
  \citenamefont {Nomura},\ and\ \citenamefont {Imada}}]{CarleoNomura}%
  \BibitemOpen
  \bibfield  {author} {\bibinfo {author} {\bibfnamefont {G.}~\bibnamefont
  {Carleo}}, \bibinfo {author} {\bibfnamefont {Y.}~\bibnamefont {Nomura}},\
  and\ \bibinfo {author} {\bibfnamefont {M.}~\bibnamefont {Imada}},\ }\bibfield
   {title} {\bibinfo {title} {Constructing exact representations of quantum
  many-body systems with deep neural networks},\ }\href
  {https://doi.org/10.1038/s41467-018-07520-3} {\bibfield  {journal} {\bibinfo
  {journal} {Nat. Comm.}\ }\textbf {\bibinfo {volume} {9}},\ \bibinfo {pages}
  {5322} (\bibinfo {year} {2018})}\BibitemShut {NoStop}%
\bibitem [{\citenamefont {Schmitt}\ and\ \citenamefont
  {Heyl}(2020)}]{SchmittHeyl}%
  \BibitemOpen
  \bibfield  {author} {\bibinfo {author} {\bibfnamefont {M.}~\bibnamefont
  {Schmitt}}\ and\ \bibinfo {author} {\bibfnamefont {M.}~\bibnamefont {Heyl}},\
  }\bibfield  {title} {\bibinfo {title} {Quantum many-body dynamics in two
  dimensions with artificial neural networks},\ }\href
  {https://doi.org/10.1103/PhysRevLett.125.100503} {\bibfield  {journal}
  {\bibinfo  {journal} {Phys. Rev. Lett.}\ }\textbf {\bibinfo {volume} {125}},\
  \bibinfo {pages} {100503} (\bibinfo {year} {2020})}\BibitemShut {NoStop}%
\bibitem [{\citenamefont {Cybenko}(1989)}]{Cybenko}%
  \BibitemOpen
  \bibfield  {author} {\bibinfo {author} {\bibfnamefont {G.}~\bibnamefont
  {Cybenko}},\ }\bibfield  {title} {\bibinfo {title} {Approximation by
  superpositions of a sigmoidal function.},\ }\href
  {https://doi.org/10.1007/BF02551274} {\bibfield  {journal} {\bibinfo
  {journal} {Math. Control Signal Systems}\ }\textbf {\bibinfo {volume} {2}},\
  \bibinfo {pages} {303} (\bibinfo {year} {1989})}\BibitemShut {NoStop}%
\bibitem [{\citenamefont {Hornik}(1991)}]{Hornik}%
  \BibitemOpen
  \bibfield  {author} {\bibinfo {author} {\bibfnamefont {K.}~\bibnamefont
  {Hornik}},\ }\bibfield  {title} {\bibinfo {title} {Approximation capabilities
  of multilayer feedforward networks},\ }\href
  {https://doi.org/https://doi.org/10.1016/0893-6080(91)90009-T} {\bibfield
  {journal} {\bibinfo  {journal} {Neural Networks}\ }\textbf {\bibinfo {volume}
  {4}},\ \bibinfo {pages} {251} (\bibinfo {year} {1991})}\BibitemShut {NoStop}%
\bibitem [{\citenamefont {Voigtlaender}(2023)}]{Voigtlaender}%
  \BibitemOpen
  \bibfield  {author} {\bibinfo {author} {\bibfnamefont {F.}~\bibnamefont
  {Voigtlaender}},\ }\bibfield  {title} {\bibinfo {title} {The universal
  approximation theorem for complex-valued neural networks},\ }\href
  {https://doi.org/https://doi.org/10.1016/j.acha.2022.12.002} {\bibfield
  {journal} {\bibinfo  {journal} {Applied and Computational Harmonic Analysis}\
  }\textbf {\bibinfo {volume} {64}},\ \bibinfo {pages} {33} (\bibinfo {year}
  {2023})}\BibitemShut {NoStop}%
\bibitem [{\citenamefont {Lieb}\ \emph {et~al.}(1961)\citenamefont {Lieb},
  \citenamefont {Schultz},\ and\ \citenamefont {Mattis}}]{LiebSchultzMattis}%
  \BibitemOpen
  \bibfield  {author} {\bibinfo {author} {\bibfnamefont {E.}~\bibnamefont
  {Lieb}}, \bibinfo {author} {\bibfnamefont {T.}~\bibnamefont {Schultz}},\ and\
  \bibinfo {author} {\bibfnamefont {D.}~\bibnamefont {Mattis}},\ }\bibfield
  {title} {\bibinfo {title} {Two soluble models of an antiferromagnetic
  chain},\ }\href
  {https://doi.org/https://doi.org/10.1016/0003-4916(61)90115-4} {\bibfield
  {journal} {\bibinfo  {journal} {Annals of Physics}\ }\textbf {\bibinfo
  {volume} {16}},\ \bibinfo {pages} {407} (\bibinfo {year} {1961})}\BibitemShut
  {NoStop}%
\bibitem [{\citenamefont {Szab\'o}\ and\ \citenamefont
  {Castelnovo}(2020)}]{SzaboCastelnovo}%
  \BibitemOpen
  \bibfield  {author} {\bibinfo {author} {\bibfnamefont {A.}~\bibnamefont
  {Szab\'o}}\ and\ \bibinfo {author} {\bibfnamefont {C.}~\bibnamefont
  {Castelnovo}},\ }\bibfield  {title} {\bibinfo {title} {Neural network wave
  functions and the sign problem},\ }\href
  {https://doi.org/10.1103/PhysRevResearch.2.033075} {\bibfield  {journal}
  {\bibinfo  {journal} {Phys. Rev. Research}\ }\textbf {\bibinfo {volume}
  {2}},\ \bibinfo {pages} {033075} (\bibinfo {year} {2020})}\BibitemShut
  {NoStop}%
\bibitem [{\citenamefont {Nomura}\ \emph {et~al.}(2017)\citenamefont {Nomura},
  \citenamefont {Darmawan}, \citenamefont {Yamaji},\ and\ \citenamefont
  {Imada}}]{PP-NN1}%
  \BibitemOpen
  \bibfield  {author} {\bibinfo {author} {\bibfnamefont {Y.}~\bibnamefont
  {Nomura}}, \bibinfo {author} {\bibfnamefont {A.}~\bibnamefont {Darmawan}},
  \bibinfo {author} {\bibfnamefont {Y.}~\bibnamefont {Yamaji}},\ and\ \bibinfo
  {author} {\bibfnamefont {M.}~\bibnamefont {Imada}},\ }\bibfield  {title}
  {\bibinfo {title} {Restricted {B}oltzmann machine learning for solving
  strongly correlated quantum systems},\ }\href@noop {} {\bibfield  {journal}
  {\bibinfo  {journal}
  {\href{https://link.aps.org/doi/10.1103/PhysRevB.96.205152}{Phys. Rev. B}}\
  }\textbf {\bibinfo {volume} {96}},\ \bibinfo {pages} {205152} (\bibinfo
  {year} {2017})}\BibitemShut {NoStop}%
\bibitem [{\citenamefont {Nomura}\ and\ \citenamefont {Imada}(2021)}]{PP-NN2}%
  \BibitemOpen
  \bibfield  {author} {\bibinfo {author} {\bibfnamefont {Y.}~\bibnamefont
  {Nomura}}\ and\ \bibinfo {author} {\bibfnamefont {M.}~\bibnamefont {Imada}},\
  }\bibfield  {title} {\bibinfo {title} {Dirac-type nodal spin liquid revealed
  by refined quantum many-body solver using neural-network wave function,
  correlation ratio, and level spectroscopy},\ }\href
  {https://doi.org/10.1103/PhysRevX.11.031034} {\bibfield  {journal} {\bibinfo
  {journal} {Phys. Rev. X}\ }\textbf {\bibinfo {volume} {11}},\ \bibinfo
  {pages} {031034} (\bibinfo {year} {2021})}\BibitemShut {NoStop}%
\bibitem [{\citenamefont {Hams}\ and\ \citenamefont {Raedt}(2000)}]{deRaedt}%
  \BibitemOpen
  \bibfield  {author} {\bibinfo {author} {\bibfnamefont {A.}~\bibnamefont
  {Hams}}\ and\ \bibinfo {author} {\bibfnamefont {H.~D.}\ \bibnamefont
  {Raedt}},\ }\bibfield  {title} {\bibinfo {title} {Fast algorithm for finding
  the eigenvalue distribution of very large matrices},\ }\href
  {https://doi.org/10.1103/PhysRevE.62.4365} {\bibfield  {journal} {\bibinfo
  {journal} {\href{https://link.aps.org/doi/10.1103/PhysRevE.62.4365}{Phys.
  Rev. E}}\ }\textbf {\bibinfo {volume} {62}},\ \bibinfo {pages} {4365}
  (\bibinfo {year} {2000})}\BibitemShut {NoStop}%
\bibitem [{\citenamefont {Gemmer}\ \emph {et~al.}(2009)\citenamefont {Gemmer},
  \citenamefont {Michel},\ and\ \citenamefont {Mahler}}]{Gemmer2009}%
  \BibitemOpen
  \bibfield  {author} {\bibinfo {author} {\bibfnamefont {J.}~\bibnamefont
  {Gemmer}}, \bibinfo {author} {\bibfnamefont {M.}~\bibnamefont {Michel}},\
  and\ \bibinfo {author} {\bibfnamefont {G.}~\bibnamefont {Mahler}},\ }\bibinfo
  {title} {Typicality of observables and states},\ in\ \href
  {https://doi.org/10.1007/978-3-540-70510-9_8} {\emph {\bibinfo {booktitle}
  {Quantum Thermodynamics: Emergence of Thermodynamic Behavior Within Composite
  Quantum Systems}}}\ (\bibinfo  {publisher} {Springer Berlin Heidelberg},\
  \bibinfo {year} {2009})\ pp.\ \bibinfo {pages} {85--93}\BibitemShut {NoStop}%
\bibitem [{\citenamefont {Steinigeweg}\ \emph {et~al.}(2014)\citenamefont
  {Steinigeweg}, \citenamefont {Gemmer},\ and\ \citenamefont
  {Brenig}}]{SteinigewegGemmer}%
  \BibitemOpen
  \bibfield  {author} {\bibinfo {author} {\bibfnamefont {R.}~\bibnamefont
  {Steinigeweg}}, \bibinfo {author} {\bibfnamefont {J.}~\bibnamefont
  {Gemmer}},\ and\ \bibinfo {author} {\bibfnamefont {W.}~\bibnamefont
  {Brenig}},\ }\bibfield  {title} {\bibinfo {title} {Spin-current
  autocorrelations from single pure-state propagation},\ }\href
  {https://doi.org/10.1103/PhysRevLett.112.120601} {\bibfield  {journal}
  {\bibinfo  {journal} {Phys. Rev. Lett.}\ }\textbf {\bibinfo {volume} {112}},\
  \bibinfo {pages} {120601} (\bibinfo {year} {2014})}\BibitemShut {NoStop}%
\bibitem [{\citenamefont {Sugiura}(2017)}]{Sugiura}%
  \BibitemOpen
  \bibfield  {author} {\bibinfo {author} {\bibfnamefont {S.}~\bibnamefont
  {Sugiura}},\ }\href@noop {} {\emph {\bibinfo {title} {Formulation of
  Statistical Mechanics Based on Thermal Pure Quantum States}}}\ (\bibinfo
  {publisher} {\href{https://www.springer.com/de/book/9789811015052}{Springer
  Singapore}},\ \bibinfo {year} {2017})\BibitemShut {NoStop}%
\bibitem [{\citenamefont {Nys}\ \emph {et~al.}(2023)\citenamefont {Nys},
  \citenamefont {Denis},\ and\ \citenamefont {Carleo}}]{nys2023realtime}%
  \BibitemOpen
  \bibfield  {author} {\bibinfo {author} {\bibfnamefont {J.}~\bibnamefont
  {Nys}}, \bibinfo {author} {\bibfnamefont {Z.}~\bibnamefont {Denis}},\ and\
  \bibinfo {author} {\bibfnamefont {G.}~\bibnamefont {Carleo}},\ }\href@noop {}
  {\bibinfo {title} {Real-time quantum dynamics of thermal states with neural
  thermofields}} (\bibinfo {year} {2023}),\ \Eprint
  {https://arxiv.org/abs/2309.07063} {arXiv:2309.07063 [quant-ph]} \BibitemShut
  {NoStop}%
\bibitem [{\citenamefont {Nomura}\ \emph {et~al.}(2021)\citenamefont {Nomura},
  \citenamefont {Yoshioka},\ and\ \citenamefont {Nori}}]{PurifyingDBM}%
  \BibitemOpen
  \bibfield  {author} {\bibinfo {author} {\bibfnamefont {Y.}~\bibnamefont
  {Nomura}}, \bibinfo {author} {\bibfnamefont {N.}~\bibnamefont {Yoshioka}},\
  and\ \bibinfo {author} {\bibfnamefont {F.}~\bibnamefont {Nori}},\ }\bibfield
  {title} {\bibinfo {title} {Purifying deep {B}oltzmann machines for thermal
  quantum states},\ }\href {https://doi.org/10.1103/PhysRevLett.127.060601}
  {\bibfield  {journal} {\bibinfo  {journal} {Phys. Rev. Lett.}\ }\textbf
  {\bibinfo {volume} {127}},\ \bibinfo {pages} {060601} (\bibinfo {year}
  {2021})}\BibitemShut {NoStop}%
\bibitem [{\citenamefont {Hendry}\ \emph {et~al.}(2022)\citenamefont {Hendry},
  \citenamefont {Chen},\ and\ \citenamefont {Feiguin}}]{RBM-pure}%
  \BibitemOpen
  \bibfield  {author} {\bibinfo {author} {\bibfnamefont {D.}~\bibnamefont
  {Hendry}}, \bibinfo {author} {\bibfnamefont {H.}~\bibnamefont {Chen}},\ and\
  \bibinfo {author} {\bibfnamefont {A.}~\bibnamefont {Feiguin}},\ }\bibfield
  {title} {\bibinfo {title} {Neural network representation for minimally
  entangled typical thermal states},\ }\href
  {https://doi.org/10.1103/PhysRevB.106.165111} {\bibfield  {journal} {\bibinfo
   {journal} {Phys. Rev. B}\ }\textbf {\bibinfo {volume} {106}},\ \bibinfo
  {pages} {165111} (\bibinfo {year} {2022})}\BibitemShut {NoStop}%
\bibitem [{\citenamefont {Hackl}\ \emph {et~al.}(2020)\citenamefont {Hackl},
  \citenamefont {Guaita}, \citenamefont {Shi}, \citenamefont {Haegeman},
  \citenamefont {Demler},\ and\ \citenamefont {Cirac}}]{Geometry}%
  \BibitemOpen
  \bibfield  {author} {\bibinfo {author} {\bibfnamefont {L.}~\bibnamefont
  {Hackl}}, \bibinfo {author} {\bibfnamefont {T.}~\bibnamefont {Guaita}},
  \bibinfo {author} {\bibfnamefont {T.}~\bibnamefont {Shi}}, \bibinfo {author}
  {\bibfnamefont {J.}~\bibnamefont {Haegeman}}, \bibinfo {author}
  {\bibfnamefont {E.}~\bibnamefont {Demler}},\ and\ \bibinfo {author}
  {\bibfnamefont {J.~I.}\ \bibnamefont {Cirac}},\ }\bibfield  {title} {\bibinfo
  {title} {{Geometry of variational methods: dynamics of closed quantum
  systems}},\ }\href {https://doi.org/10.21468/SciPostPhys.9.4.048} {\bibfield
  {journal} {\bibinfo  {journal} {SciPost Phys.}\ }\textbf {\bibinfo {volume}
  {9}},\ \bibinfo {pages} {48} (\bibinfo {year} {2020})}\BibitemShut {NoStop}%
\bibitem [{\citenamefont {{Sandvik, A. W.}}(2012)}]{Sandvik-Webpage}%
  \BibitemOpen
  \bibfield  {author} {\bibinfo {author} {\bibnamefont {{Sandvik, A. W.}}},\
  }\href@noop {} {\bibinfo {title} {Quantum {M}onte {C}arlo methods at work for
  novel phases of matter}} (\bibinfo {year} {2012}),\ \bibinfo {note}
  {\url{http://physics.bu.edu/~sandvik/trieste12/index.html}, Last accessed on
  2023-10-13}\BibitemShut {NoStop}%
\bibitem [{\citenamefont {Sandvik}(2010)}]{Sandvik}%
  \BibitemOpen
  \bibfield  {author} {\bibinfo {author} {\bibfnamefont {A.~W.}\ \bibnamefont
  {Sandvik}},\ }\bibfield  {title} {\bibinfo {title} {Computational studies of
  quantum spin systems},\ }\href {https://doi.org/10.1063/1.3518900} {\bibfield
   {journal} {\bibinfo  {journal} {AIP Conference Proceedings}\ }\textbf
  {\bibinfo {volume} {1297}},\ \bibinfo {pages} {135} (\bibinfo {year}
  {2010})}\BibitemShut {NoStop}%
\bibitem [{\citenamefont {Zhang}\ \emph {et~al.}(2023)\citenamefont {Zhang},
  \citenamefont {Webber}, \citenamefont {Lindsey}, \citenamefont {Berkelbach},\
  and\ \citenamefont {Weare}}]{SpuriousModes}%
  \BibitemOpen
  \bibfield  {author} {\bibinfo {author} {\bibfnamefont {H.}~\bibnamefont
  {Zhang}}, \bibinfo {author} {\bibfnamefont {R.~J.}\ \bibnamefont {Webber}},
  \bibinfo {author} {\bibfnamefont {M.}~\bibnamefont {Lindsey}}, \bibinfo
  {author} {\bibfnamefont {T.~C.}\ \bibnamefont {Berkelbach}},\ and\ \bibinfo
  {author} {\bibfnamefont {J.}~\bibnamefont {Weare}},\ }\bibfield  {title}
  {\bibinfo {title} {Understanding and eliminating spurious modes in
  variational {M}onte {C}arlo using collective variables},\ }\href
  {https://doi.org/10.1103/PhysRevResearch.5.023101} {\bibfield  {journal}
  {\bibinfo  {journal} {Phys. Rev. Res.}\ }\textbf {\bibinfo {volume} {5}},\
  \bibinfo {pages} {023101} (\bibinfo {year} {2023})}\BibitemShut {NoStop}%
\bibitem [{\citenamefont {Sinibaldi}\ \emph {et~al.}(2023)\citenamefont
  {Sinibaldi}, \citenamefont {Giuliani}, \citenamefont {Carleo},\ and\
  \citenamefont {Vicentini}}]{Sinibaldi2023unbiasingtime}%
  \BibitemOpen
  \bibfield  {author} {\bibinfo {author} {\bibfnamefont {A.}~\bibnamefont
  {Sinibaldi}}, \bibinfo {author} {\bibfnamefont {C.}~\bibnamefont {Giuliani}},
  \bibinfo {author} {\bibfnamefont {G.}~\bibnamefont {Carleo}},\ and\ \bibinfo
  {author} {\bibfnamefont {F.}~\bibnamefont {Vicentini}},\ }\bibfield  {title}
  {\bibinfo {title} {Unbiasing time-dependent variational {M}onte {C}arlo by
  projected quantum evolution},\ }\href
  {https://doi.org/10.22331/q-2023-10-10-1131} {\bibfield  {journal} {\bibinfo
  {journal} {{Quantum}}\ }\textbf {\bibinfo {volume} {7}},\ \bibinfo {pages}
  {1131} (\bibinfo {year} {2023})}\BibitemShut {NoStop}%
\bibitem [{\citenamefont {Deng}\ \emph {et~al.}(2017)\citenamefont {Deng},
  \citenamefont {Li},\ and\ \citenamefont {Das~Sarma}}]{DengLiDasSarma}%
  \BibitemOpen
  \bibfield  {author} {\bibinfo {author} {\bibfnamefont {D.-L.}\ \bibnamefont
  {Deng}}, \bibinfo {author} {\bibfnamefont {X.}~\bibnamefont {Li}},\ and\
  \bibinfo {author} {\bibfnamefont {S.}~\bibnamefont {Das~Sarma}},\ }\bibfield
  {title} {\bibinfo {title} {Quantum entanglement in neural network states},\
  }\href {https://doi.org/10.1103/PhysRevX.7.021021} {\bibfield  {journal}
  {\bibinfo  {journal} {Phys. Rev. X}\ }\textbf {\bibinfo {volume} {7}},\
  \bibinfo {pages} {021021} (\bibinfo {year} {2017})}\BibitemShut {NoStop}%
\bibitem [{\citenamefont {Lin}\ and\ \citenamefont
  {Pollmann}(2022)}]{LinPollmann}%
  \BibitemOpen
  \bibfield  {author} {\bibinfo {author} {\bibfnamefont {S.-H.}\ \bibnamefont
  {Lin}}\ and\ \bibinfo {author} {\bibfnamefont {F.}~\bibnamefont {Pollmann}},\
  }\bibfield  {title} {\bibinfo {title} {Scaling of neural-network quantum
  states for time evolution},\ }\href
  {https://doi.org/https://doi.org/10.1002/pssb.202100172} {\bibfield
  {journal} {\bibinfo  {journal} {physica status solidi (b)}\ }\textbf
  {\bibinfo {volume} {259}},\ \bibinfo {pages} {2100172} (\bibinfo {year}
  {2022})}\BibitemShut {NoStop}%
\bibitem [{\citenamefont {White}(2009)}]{RWhite}%
  \BibitemOpen
  \bibfield  {author} {\bibinfo {author} {\bibfnamefont {S.~R.}\ \bibnamefont
  {White}},\ }\bibfield  {title} {\bibinfo {title} {Minimally entangled typical
  quantum states at finite temperature},\ }\href
  {https://doi.org/10.1103/PhysRevLett.102.190601} {\bibfield  {journal}
  {\bibinfo  {journal} {Phys. Rev. Lett.}\ }\textbf {\bibinfo {volume} {102}},\
  \bibinfo {pages} {190601} (\bibinfo {year} {2009})}\BibitemShut {NoStop}%
\bibitem [{\citenamefont {Short}\ and\ \citenamefont {Winter}(2006)}]{Popescu}%
  \BibitemOpen
  \bibfield  {author} {\bibinfo {author} {\bibfnamefont {S.~P.~A.}\
  \bibnamefont {Short}}\ and\ \bibinfo {author} {\bibfnamefont
  {A.}~\bibnamefont {Winter}},\ }\bibfield  {title} {\bibinfo {title}
  {Entanglement and the foundations of statistical mechanics},\ }\href
  {https://doi.org/10.1038/nphys444} {\bibfield  {journal} {\bibinfo  {journal}
  {Nature Phys}\ }\textbf {\bibinfo {volume} {2}},\ \bibinfo {pages}
  {754} (\bibinfo {year} {2006})}\BibitemShut {NoStop}%
\bibitem [{\citenamefont {Popescu}\ \emph {et~al.}(2005)\citenamefont
  {Popescu}, \citenamefont {Short},\ and\ \citenamefont {Winter}}]{Popescu2}%
  \BibitemOpen
  \bibfield  {author} {\bibinfo {author} {\bibfnamefont {S.}~\bibnamefont
  {Popescu}}, \bibinfo {author} {\bibfnamefont {A.~J.}\ \bibnamefont {Short}},\
  and\ \bibinfo {author} {\bibfnamefont {A.}~\bibnamefont {Winter}},\
  }\bibfield  {title} {\bibinfo {title} {The foundations of statistical
  mechanics from entanglement: Individual states vs. averages},\ }\href@noop {}
  {\bibfield  {journal} {\bibinfo  {journal}
  {\href{https://arxiv.org/abs/quant-ph/0511225}{arXiv:quant-ph/0511225}}\ }
  (\bibinfo {year} {2005})}\BibitemShut {NoStop}%
\bibitem [{\citenamefont {Milman}\ and\ \citenamefont
  {Schechtman}(1986)}]{Milman}%
  \BibitemOpen
  \bibfield  {author} {\bibinfo {author} {\bibfnamefont {V.}~\bibnamefont
  {Milman}}\ and\ \bibinfo {author} {\bibfnamefont {G.}~\bibnamefont
  {Schechtman}},\ }\href@noop {} {\emph {\bibinfo {title} {Asymptotic Theory of
  Finite Dimensional Normed Spaces}}}\ (\bibinfo  {publisher}
  {\href{https://link.springer.com/book/10.1007/978-3-540-38822-7}{Springer
  Berlin, Heidelberg}},\ \bibinfo {year} {1986})\BibitemShut {NoStop}%
\bibitem [{\citenamefont {Matoušek}(2002)}]{Matousek}%
  \BibitemOpen
  \bibfield  {author} {\bibinfo {author} {\bibfnamefont {J.}~\bibnamefont
  {Matoušek}},\ }\href@noop {} {\emph {\bibinfo {title} {Lectures on Discrete
  Geometry}}}\ (\bibinfo  {publisher}
  {\href{https://doi.org/10.1007/978-1-4613-0039-7}{Springer New York}},\
  \bibinfo {year} {2002})\BibitemShut {NoStop}%
\bibitem [{\citenamefont {Sugiura}\ and\ \citenamefont
  {Shimizu}(2012)}]{TPQSatFT}%
  \BibitemOpen
  \bibfield  {author} {\bibinfo {author} {\bibfnamefont {S.}~\bibnamefont
  {Sugiura}}\ and\ \bibinfo {author} {\bibfnamefont {A.}~\bibnamefont
  {Shimizu}},\ }\bibfield  {title} {\bibinfo {title} {Thermal pure quantum
  states at finite temperature},\ }\href
  {https://doi.org/10.1103/PhysRevLett.111.010401} {\bibfield  {journal}
  {\bibinfo  {journal}
  {\href{https://link.aps.org/doi/10.1103/PhysRevLett.108.240401}{Phys. Rev.
  Lett.}}\ }\textbf {\bibinfo {volume} {108}},\ \bibinfo {pages} {240401}
  (\bibinfo {year} {2012})}\BibitemShut {NoStop}%
\bibitem [{\citenamefont {Sugiura}\ and\ \citenamefont
  {Shimizu}(2013)}]{CTPQS}%
  \BibitemOpen
  \bibfield  {author} {\bibinfo {author} {\bibfnamefont {S.}~\bibnamefont
  {Sugiura}}\ and\ \bibinfo {author} {\bibfnamefont {A.}~\bibnamefont
  {Shimizu}},\ }\bibfield  {title} {\bibinfo {title} {Canonical thermal pure
  quantum state},\ }\href {https://doi.org/10.1103/PhysRevLett.111.010401}
  {\bibfield  {journal} {\bibinfo  {journal}
  {\href{https://link.aps.org/doi/10.1103/PhysRevLett.111.010401}{Phys. Rev.
  Lett.}}\ }\textbf {\bibinfo {volume} {111}},\ \bibinfo {pages} {010401}
  (\bibinfo {year} {2013})}\BibitemShut {NoStop}%
\bibitem [{\citenamefont {P.Reimann}(2007)}]{Reimann}%
  \BibitemOpen
  \bibfield  {author} {\bibinfo {author} {\bibnamefont {P.Reimann}},\
  }\bibfield  {title} {\bibinfo {title} {Typicality for generalized
  microcanonical ensembles},\ }\href
  {https://doi.org/10.1103/PhysRevLett.111.010401} {\bibfield  {journal}
  {\bibinfo  {journal}
  {\href{https://link.aps.org/doi/10.1103/PhysRevLett.99.160404}{Phys. Rev.
  Lett.}}\ }\textbf {\bibinfo {volume} {99}},\ \bibinfo {pages} {160404}
  (\bibinfo {year} {2007})}\BibitemShut {NoStop}%
\bibitem [{\citenamefont {Schnack}\ \emph
  {et~al.}(2020{\natexlab{a}})\citenamefont {Schnack}, \citenamefont {Richter},
  \citenamefont {Heitmann}, \citenamefont {Richter},\ and\ \citenamefont
  {Steinigeweg}}]{Schnack1}%
  \BibitemOpen
  \bibfield  {author} {\bibinfo {author} {\bibfnamefont {J.}~\bibnamefont
  {Schnack}}, \bibinfo {author} {\bibfnamefont {J.}~\bibnamefont {Richter}},
  \bibinfo {author} {\bibfnamefont {T.}~\bibnamefont {Heitmann}}, \bibinfo
  {author} {\bibfnamefont {J.}~\bibnamefont {Richter}},\ and\ \bibinfo {author}
  {\bibfnamefont {R.}~\bibnamefont {Steinigeweg}},\ }\bibfield  {title}
  {\bibinfo {title} {Finite-size scaling of typicality-based estimates},\
  }\href {https://doi.org/doi:10.1515/zna-2020-0031} {\bibfield  {journal}
  {\bibinfo  {journal} {Zeitschrift für Naturforschung A}\ }\textbf {\bibinfo
  {volume} {75}},\ \bibinfo {pages} {465} (\bibinfo {year}
  {2020}{\natexlab{a}})}\BibitemShut {NoStop}%
\bibitem [{\citenamefont {Schnack}\ \emph
  {et~al.}(2020{\natexlab{b}})\citenamefont {Schnack}, \citenamefont
  {Richter},\ and\ \citenamefont {Steinigeweg}}]{Schnack2}%
  \BibitemOpen
  \bibfield  {author} {\bibinfo {author} {\bibfnamefont {J.}~\bibnamefont
  {Schnack}}, \bibinfo {author} {\bibfnamefont {J.}~\bibnamefont {Richter}},\
  and\ \bibinfo {author} {\bibfnamefont {R.}~\bibnamefont {Steinigeweg}},\
  }\bibfield  {title} {\bibinfo {title} {Accuracy of the finite-temperature
  {L}anczos method compared to simple typicality-based estimates},\ }\href
  {https://doi.org/10.1103/PhysRevResearch.2.013186} {\bibfield  {journal}
  {\bibinfo  {journal} {Phys. Rev. Res.}\ }\textbf {\bibinfo {volume} {2}},\
  \bibinfo {pages} {013186} (\bibinfo {year} {2020}{\natexlab{b}})}\BibitemShut
  {NoStop}%
\bibitem [{\citenamefont {Schlüter}\ \emph {et~al.}(2021)\citenamefont
  {Schlüter}, \citenamefont {Gayk}, \citenamefont {Schmidt}, \citenamefont
  {Honecker},\ and\ \citenamefont {Schnack}}]{Schnack3}%
  \BibitemOpen
  \bibfield  {author} {\bibinfo {author} {\bibfnamefont {H.}~\bibnamefont
  {Schlüter}}, \bibinfo {author} {\bibfnamefont {F.}~\bibnamefont {Gayk}},
  \bibinfo {author} {\bibfnamefont {H.-J.}\ \bibnamefont {Schmidt}}, \bibinfo
  {author} {\bibfnamefont {A.}~\bibnamefont {Honecker}},\ and\ \bibinfo
  {author} {\bibfnamefont {J.}~\bibnamefont {Schnack}},\ }\bibfield  {title}
  {\bibinfo {title} {Accuracy of the typicality approach using {C}hebyshev
  polynomials},\ }\href {https://doi.org/doi:10.1515/zna-2021-0116} {\bibfield
  {journal} {\bibinfo  {journal} {Zeitschrift für Naturforschung A}\ }\textbf
  {\bibinfo {volume} {76}},\ \bibinfo {pages} {823} (\bibinfo {year}
  {2021})}\BibitemShut {NoStop}%
\bibitem [{\citenamefont {Page}(1993)}]{Page}%
  \BibitemOpen
  \bibfield  {author} {\bibinfo {author} {\bibfnamefont {D.~N.}\ \bibnamefont
  {Page}},\ }\bibfield  {title} {\bibinfo {title} {Average entropy of a
  subsystem},\ }\href {https://doi.org/10.1103/PhysRevLett.71.1291} {\bibfield
  {journal} {\bibinfo  {journal}
  {\href{https://link.aps.org/doi/10.1103/PhysRevLett.71.1291}{Phys. Rev.
  Lett.}}\ }\textbf {\bibinfo {volume} {71}},\ \bibinfo {pages} {1291}
  (\bibinfo {year} {1993})}\BibitemShut {NoStop}%
\bibitem [{\citenamefont {Bianchi}\ \emph {et~al.}(2022)\citenamefont
  {Bianchi}, \citenamefont {Hackl}, \citenamefont {Kieburg}, \citenamefont
  {Rigol},\ and\ \citenamefont {Vidmar}}]{Hackl-Volume-Law}%
  \BibitemOpen
  \bibfield  {author} {\bibinfo {author} {\bibfnamefont {E.}~\bibnamefont
  {Bianchi}}, \bibinfo {author} {\bibfnamefont {L.}~\bibnamefont {Hackl}},
  \bibinfo {author} {\bibfnamefont {M.}~\bibnamefont {Kieburg}}, \bibinfo
  {author} {\bibfnamefont {M.}~\bibnamefont {Rigol}},\ and\ \bibinfo {author}
  {\bibfnamefont {L.}~\bibnamefont {Vidmar}},\ }\bibfield  {title} {\bibinfo
  {title} {Volume-law entanglement entropy of typical pure quantum states},\
  }\href {https://doi.org/10.1103/PRXQuantum.3.030201} {\bibfield  {journal}
  {\bibinfo  {journal}
  {\href{https://link.aps.org/doi/10.1103/PRXQuantum.3.030201}{PRX Quantum}}\
  }\textbf {\bibinfo {volume} {3}},\ \bibinfo {pages} {030201} (\bibinfo {year}
  {2022})}\BibitemShut {NoStop}%
\bibitem [{\citenamefont {Nakagawa}\ \emph {et~al.}(2018)\citenamefont
  {Nakagawa}, \citenamefont {Watanabe}, \citenamefont {Fujita},\ and\
  \citenamefont {Sugiura}}]{Sugiura-Nature}%
  \BibitemOpen
  \bibfield  {author} {\bibinfo {author} {\bibfnamefont {Y.~O.}\ \bibnamefont
  {Nakagawa}}, \bibinfo {author} {\bibfnamefont {M.}~\bibnamefont {Watanabe}},
  \bibinfo {author} {\bibfnamefont {H.}~\bibnamefont {Fujita}},\ and\ \bibinfo
  {author} {\bibfnamefont {S.}~\bibnamefont {Sugiura}},\ }\bibfield  {title}
  {\bibinfo {title} {Universality in volume-law entanglement of scrambled pure
  quantum states},\ }\href {https://doi.org/10.1038/s41467-018-03883-9}
  {\bibfield  {journal} {\bibinfo  {journal}
  {\href{https://doi.org/10.1038/s41467-018-03883-9}{Nature Communications}}\
  }\textbf {\bibinfo {volume} {9}},\ \bibinfo {pages} {1635} (\bibinfo {year}
  {2018})}\BibitemShut {NoStop}%
\bibitem [{\citenamefont {Tahara}\ and\ \citenamefont
  {Imada}(2008)}]{Pfaffian1}%
  \BibitemOpen
  \bibfield  {author} {\bibinfo {author} {\bibfnamefont {D.}~\bibnamefont
  {Tahara}}\ and\ \bibinfo {author} {\bibfnamefont {M.}~\bibnamefont {Imada}},\
  }\bibfield  {title} {\bibinfo {title} {Variational {M}onte {C}arlo method
  combined with quantum-number projection and multi-variable optimization},\
  }\href@noop {} {\bibfield  {journal} {\bibinfo  {journal}
  {\href{https://doi.org/10.1143/JPSJ.77.114701}{J. Phys. Soc. Jpn.}}\ }\textbf
  {\bibinfo {volume} {77}} (\bibinfo {year} {2008})}\BibitemShut {NoStop}%
\bibitem [{\citenamefont {Bajdich}\ \emph {et~al.}(2006)\citenamefont
  {Bajdich}, \citenamefont {Mitas}, \citenamefont {Drobn\'y},\ and\
  \citenamefont {Wagner}}]{Pfaffian3}%
  \BibitemOpen
  \bibfield  {author} {\bibinfo {author} {\bibfnamefont {M.}~\bibnamefont
  {Bajdich}}, \bibinfo {author} {\bibfnamefont {L.}~\bibnamefont {Mitas}},
  \bibinfo {author} {\bibfnamefont {G.}~\bibnamefont {Drobn\'y}},\ and\
  \bibinfo {author} {\bibfnamefont {L.~K.}\ \bibnamefont {Wagner}},\ }\bibfield
   {title} {\bibinfo {title} {Pfaffian pairing wave functions in
  electronic-structure quantum {M}onte {C}arlo simulations},\ }\href
  {https://doi.org/10.1103/PhysRevLett.96.130201} {\bibfield  {journal}
  {\bibinfo  {journal}
  {\href{https://link.aps.org/doi/10.1103/PhysRevLett.96.130201}{Phys. Rev.
  Lett.}}\ }\textbf {\bibinfo {volume} {96}},\ \bibinfo {pages} {130201}
  (\bibinfo {year} {2006})}\BibitemShut {NoStop}%
\bibitem [{\citenamefont {Misawa}\ \emph {et~al.}(2019)\citenamefont {Misawa},
  \citenamefont {Morita}, \citenamefont {Yoshimi}, \citenamefont {Kawamura},
  \citenamefont {Motoyama}, \citenamefont {Ido}, \citenamefont {Ohgoe},
  \citenamefont {Imada},\ and\ \citenamefont {Kato}}]{Pfaffian2}%
  \BibitemOpen
  \bibfield  {author} {\bibinfo {author} {\bibfnamefont {T.}~\bibnamefont
  {Misawa}}, \bibinfo {author} {\bibfnamefont {S.}~\bibnamefont {Morita}},
  \bibinfo {author} {\bibfnamefont {K.}~\bibnamefont {Yoshimi}}, \bibinfo
  {author} {\bibfnamefont {M.}~\bibnamefont {Kawamura}}, \bibinfo {author}
  {\bibfnamefont {Y.}~\bibnamefont {Motoyama}}, \bibinfo {author}
  {\bibfnamefont {K.}~\bibnamefont {Ido}}, \bibinfo {author} {\bibfnamefont
  {T.}~\bibnamefont {Ohgoe}}, \bibinfo {author} {\bibfnamefont
  {M.}~\bibnamefont {Imada}},\ and\ \bibinfo {author} {\bibfnamefont
  {T.}~\bibnamefont {Kato}},\ }\bibfield  {title} {\bibinfo {title}
  {mvmc---open-source software for many-variable variational {M}onte {C}arlo
  method},\ }\href@noop {} {\bibfield  {journal} {\bibinfo  {journal}
  {\href{https://doi.org/10.1016/j.cpc.2018.08.014}{Computer Physics
  Communications}}\ }\textbf {\bibinfo {volume} {235}},\ \bibinfo {pages} {447}
  (\bibinfo {year} {2019})}\BibitemShut {NoStop}%
\bibitem [{\citenamefont {Gao}\ and\ \citenamefont
  {Duan}(2017)}]{DNN-Efficient}%
  \BibitemOpen
  \bibfield  {author} {\bibinfo {author} {\bibfnamefont {X.}~\bibnamefont
  {Gao}}\ and\ \bibinfo {author} {\bibfnamefont {L.-M.}\ \bibnamefont {Duan}},\
  }\bibfield  {title} {\bibinfo {title} {Efficient representation of quantum
  many-body states with deep neural networks},\ }\href
  {https://doi.org/10.1038/s41467-017-00705-2} {\bibfield  {journal} {\bibinfo
  {journal} {\href{https://doi.org/10.1038/s41467-017-00705-2}{Nature
  Communications}}\ }\textbf {\bibinfo {volume} {8}},\ \bibinfo {pages} {662}
  (\bibinfo {year} {2017})}\BibitemShut {NoStop}%
\end{thebibliography}
%

\end{document}